%% file: nimrec.tex
\documentclass{elsart}
\pdfoutput=1

 \usepackage{graphicx}

\usepackage{amssymb}
\usepackage{afterpage}

\input{boonesym}
\def\vecx  {\ensuremath{\mathbf{x}}}
\def\like    {\ensuremath{\mathcal{L}}}
\def\likeemu    {\ensuremath{\mathcal{R}_{e/\mu}}}
\def\likeepi    {\ensuremath{\mathcal{R}_{e/\pi}}}

\def\hit     {\ensuremath{\mbox{hit}}}
\def\sunhit {\ensuremath{\mbox{\scriptsize{unhit}}}}
\def\shit     {\ensuremath{\mbox{\scriptsize{hit}}}}

\def\ROmega{\ensuremath{\mathrm{\Omega}}}
\def\point     {\ensuremath{\mbox{\scriptsize{point}}}}
\def\iso     {\ensuremath{\mbox{\scriptsize{iso}}}}
\def\sci     {\ensuremath{\mbox{\scriptsize{sci}}}}
\def\cer     {\ensuremath{\mbox{\scriptsize{Ch}}}}
\def\direct     {\ensuremath{\mbox{\scriptsize{direct}}}}
\def\indirect   {\ensuremath{\mbox{\scriptsize{indirect}}}}
\def\smid      {\ensuremath{\mbox{\scriptsize{mid}}}} 
\def\tc        {\ensuremath t^c}
\def\sprompt     {\ensuremath{\mbox{\scriptsize{prompt}}}}
\def\slate  {\ensuremath{\mbox{\scriptsize{late}}}}
\def\sn     {\ensuremath{\mbox{\scriptsize{n}}}}
\def\sf  {\ensuremath{\mbox{\scriptsize{f}}}}
\def\sl  {\ensuremath{\mbox{\scriptsize{l}}}}
\def\stot  {\ensuremath{\mbox{\scriptsize{tot}}}}

\def\mgg  {\ensuremath{M_{\gamma\gamma}}}
\def\nearsim {\ensuremath{\mathord{\sim}}}

 \usepackage{lineno}

\begin{document}

\begin{frontmatter} 



\title{The Extended-Track Event Reconstruction for MiniBooNE}

\author[Princeton]			{R.~B.~Patterson$\;^1$}\thanks{Now at the California Institute of Technology, Physics Department 103-33, Pasadena, CA 91125, USA.},
\author[Princeton]			{E.~M.~Laird},
\author[Alabama]			{Y.~Liu},
\author[Princeton]			{P.~D.~Meyers},
\author[Alabama]			{I.~Stancu},
\author[Princeton]			{H.~A.~Tanaka$\;^2$}\thanks{Now at the University of British Columbia, Department of Physics and Astronomy, Vancouver, BC V6T 1Z1, Canada.}

\address[Alabama]{University of Alabama,  Department of Physics and Astronomy, Tuscaloosa, AL, 35487 USA}
\address[Princeton]{Princeton University, Department of Physics, Joseph Henry Laboratories,  Princeton, NJ 08544  USA}


\begin{abstract}
The Booster Neutrino Experiment (MiniBooNE) searches for $\num\to\nue$ oscillations using the $\nearsim1 \gev$ neutrino beam produced by the FNAL Booster synchrotron.  The array of photomultiplier tubes (PMTs) lining the MiniBooNE detector records Cherenkov and scintillation photons from the charged particles produced in neutrino interactions.  We describe a maximum likelihood fitting algorithm used to reconstruct the basic properties (position, direction, energy) of these particles from the charges and times measured by the PMTs.  The likelihoods returned from fitting an event to different particle hypotheses are used to categorize it as a signal $\nue$ event or as one of the background $\num$ processes, in particular charged current quasi-elastic scattering and neutral current $\piz$ production. The reconstruction and event selection techniques described here can be applied to current and future neutrino experiments using similar Cherenkov-based detection.
\end{abstract}

\end{frontmatter}

\input{introduction}
\input{nuint}
\input{detector}
\input{onetrack}
\input{twotrack} 
\input{performance} 
\input{pid}




\end{document}

%% file: introduction.tex
\section{Introduction}
\label{sec:intro}
\indent Cherenkov detectors of $\mathord{\gtrsim}1~\mbox{kiloton}$ have played a key role in the 
establishment of the phenomenon of neutrino oscillations~\cite{kamioka,superk,sno}. These detectors
typically consist of a large volume of a homogeneous transparent medium (water or mineral oil)
with a high index of refraction surrounded by an array of photomultiplier tubes (PMTs). 
Cherenkov photons produced by charged particles emerging from the neutrino interactions and traversing the medium with $\beta>1/n$
(where $n$ is the index of refraction and $\beta = v/c$)  are detected
by the PMTs. The photons are emitted at an angle $\theta_C$ relative to the
track direction, where $\cos\theta_C=1/n\beta$.
The radiation is azimuthally symmetric about the track direction, resulting in a ring-like 
pattern that can be identified on the PMT array. The quantity, spatial distribution, and arrival times of these photons provide information 
on the location, direction, and energy of the particle.  We refer to the extraction of such information from the charge and time measurements of the PMTs  as ``reconstruction.''
An analysis of the ring profile can also provide information on the identity of the particle and, if multiple rings are detected, the number of particles.

\indent  In this paper, we discuss the event reconstruction algorithms used in the Booster Neutrino
Experiment (MiniBooNE), which searches for an excess of $\nue$ interactions indicative of $\num\to\nue$ oscillations using the Fermilab $\nearsim 1\ \textrm{GeV}$ neutrino beam.  The MiniBooNE detector is a sphere of radius 610.6 cm filled with Marcol~7 mineral oil ($n\approx 1.47$) and divided into two optically isolated regions by an opaque shell of radius 575 cm, concentric with the sphere.  The surface of the inner ``main'' region is instrumented with an array of 1280 inward-facing 8~in.~PMTs which detect the light produced by the neutrino interactions.  The outer ``veto'' region is instrumented with 240 PMTs and is used to tag charged particles that enter the detector from outside ({\em e.g.}, cosmic muons) or that exit the main region.  Though no scintillator was added to the Marcol 7, it scintillates weakly, 
resulting in the production of delayed isotropic light for particles with sub- and super-Cherenkov
velocities.  A more detailed description of the experiment can be found in Refs.~\cite{prl} and~\cite{detnim}. 



%% file: nuint.tex
\section{Neutrino Interactions at MiniBooNE}
\label{sec:nuint}
The $\nearsim 1 \gev$ neutrino beam at MiniBooNE results in interactions with relatively low
outgoing particle multiplicity.  The largest interaction channel is the charged current quasi-elastic (CCQE) process:
\begin{equation}\label{equ:ccqe}
\nu_\ell + n \to  \ell^- + p
\end{equation}
which accounts for $\nearsim 40\%$ of all neutrino interactions in the MiniBooNE detector.  Since the recoil proton of Eq.~(\ref{equ:ccqe}) is typically below Cherenkov threshold, only the outgoing lepton produces significant light.  In modeling such events within a reconstruction algorithm, one can therefore consider only the Cherenkov and scintillation light produced by the outgoing lepton. While
the recoiling nucleon can produce scintillation light, this additional source of light is not considered
in the reconstruction.

A muon in the MiniBooNE detector exhibits minimum-ionizing behavior through most of its path with little chance of radiative energy loss.  In contrast, electrons typically induce electromagnetic showers, resulting in additional electrons and positrons that emit Cherenkov light.  Thus, muons and electrons have significantly different Cherenkov ring patterns.  This difference is the basis for distinguishing
a muon track from an electron track, and the reconstruction algorithm has a model for each.

The next most abundant process in MiniBooNE is single pion production, which occurs
primarily via $\Delta$ resonance or coherent scattering:
\begin{equation}
\begin{array}{lcll}
\nu_\ell + N  & \to &\ell^- + N' + \pi & \ \ \ \ \ (\mbox{CC } \pi) \\ 
\nu_\ell + N  & \to & \nu_\ell + N'  + \pi & \ \ \ \ \ (\mbox{NC } \pi)  
\end{array}
\end{equation} 
where $N^{(')}$ denotes a nucleon in the case of resonance production and
a nucleus in the case of coherent production.  Of particular interest is neutral current (NC) $\piz$ production.  The two photons from $\piz\to\gamma\gamma$ decay induce electromagnetic showers indistinguishable in MiniBooNE from those induced by electrons.  Since the recoil nucleon/nucleus is typically below Cherenkov threshold, these NC $\piz$ events are well-described in the reconstruction algorithm by two electron-like tracks pointing back to a common vertex.  The presence of two distinct Cherenkov rings allows for the separation of $\piz$ events from electron ($\nue$ CCQE) events.  However, $\piz$ misidentification (for example, due to a large energy asymmetry between the two photons or due to a small photon opening angle which leaves the Cherenkov rings overlapping) accounts for most of the reducible background in the $\num\to\nue$ oscillation search.  Thus, successful reconstruction and identification of $\piz$ events is critical.

While other channels can be accommodated by the reconstruction algorithm, the $\nue$ appearance oscillation analysis uses only four event models: single electron track, single muon track, two $\gamma$ tracks, and two $\gamma$ tracks with a $\piz$ invariant mass.

%% file: detector.tex
\section{The Detector Response}
\label{sec:detector}
\begin{figure}[htb]
\centering
\includegraphics[width=65mm]{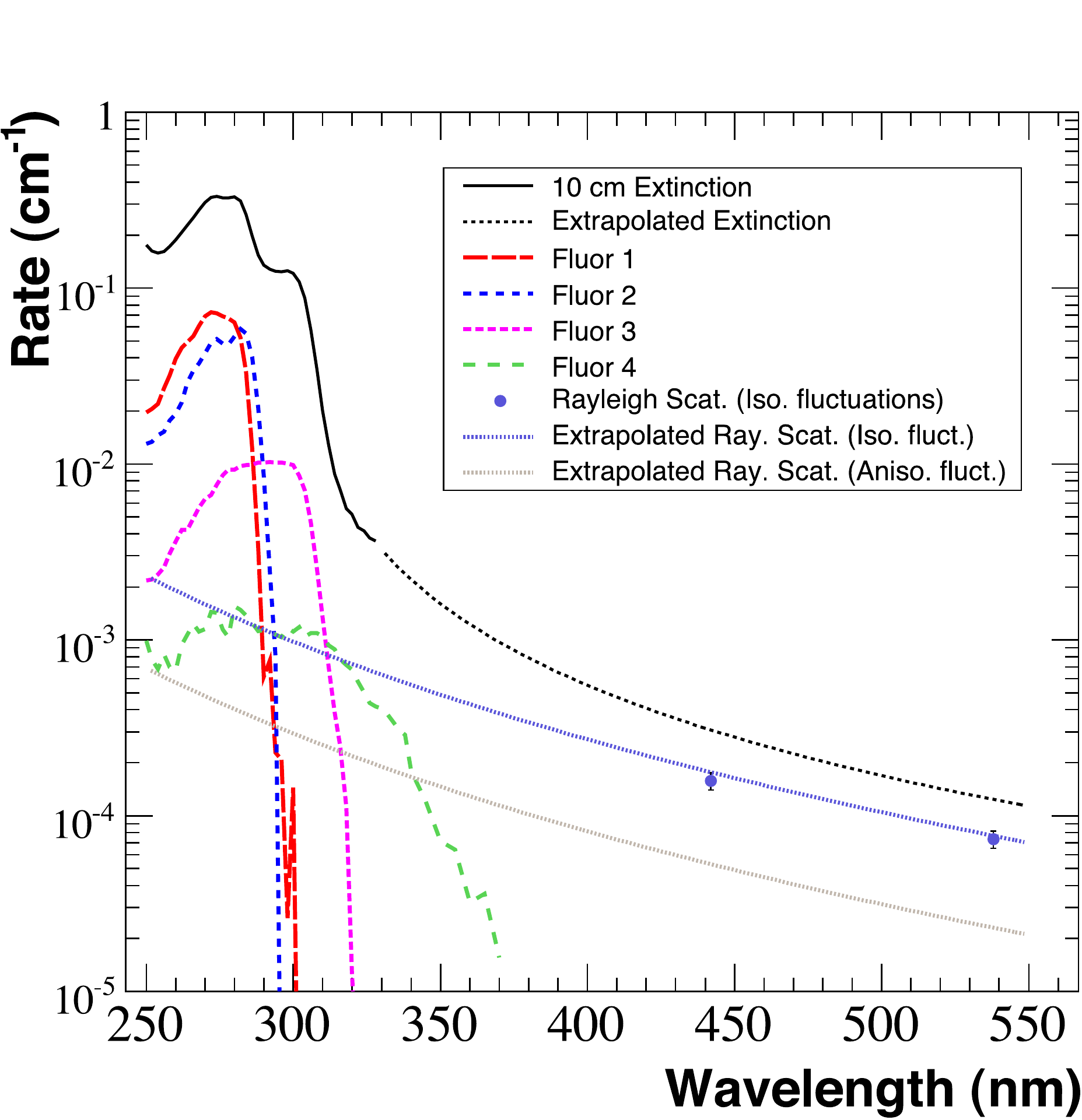}
\includegraphics[width=65mm]{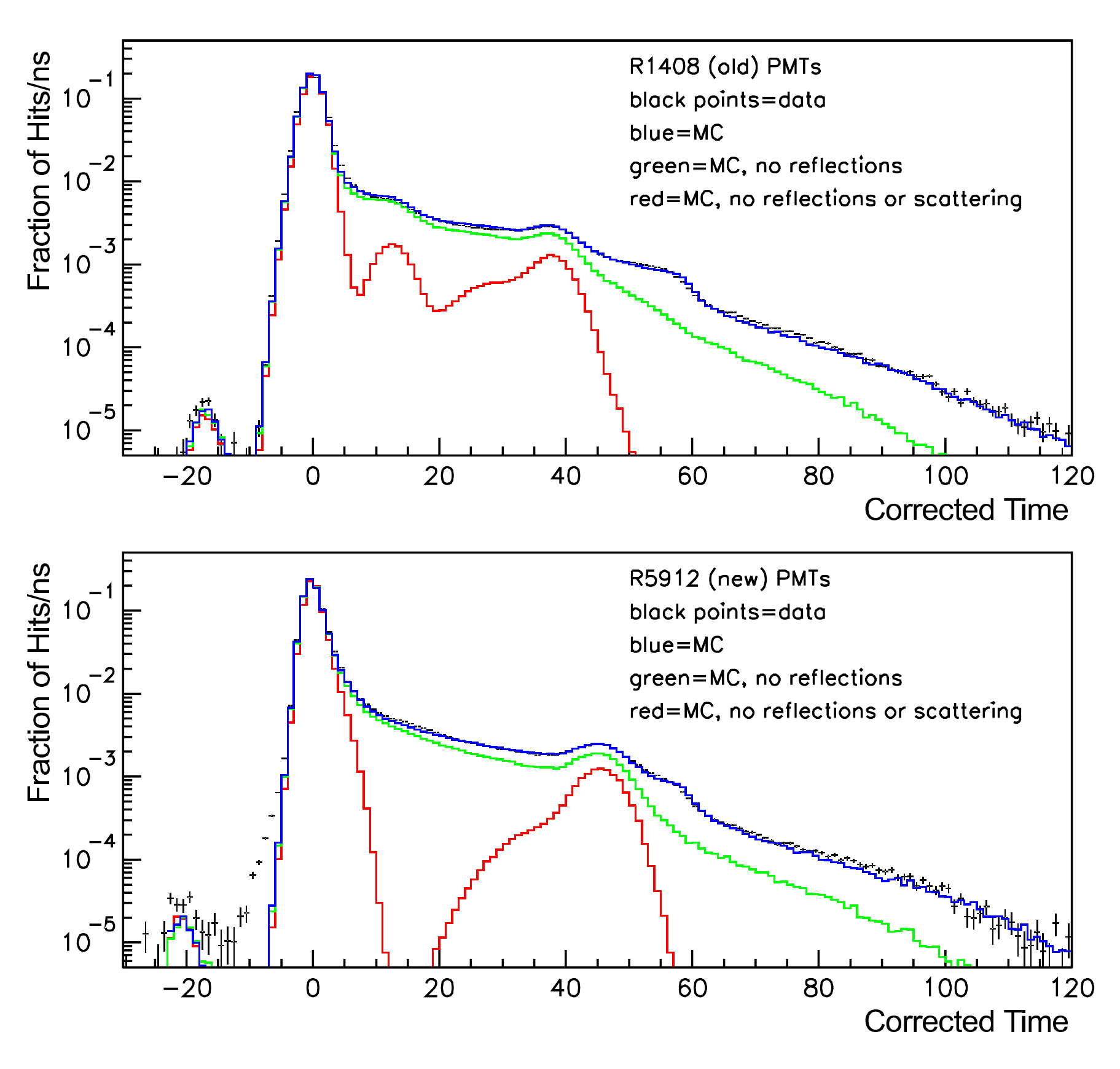}
\caption{\label{fig:optical} Left: Rates of optical processes in Marcol 7  as a function of wavelength. The solid black line is the overall extinction rate obtained from laboratory measurements. The dashed black line is the extrapolated rate based on {\em in situ} data. The curves labeled ``Fluor $n$'' are the 
excitation rates for the four identified fluorescence processes. The light blue points represent the 
measured rates of Rayleigh scattering, and the dashed light blue and gray lines represent extrapolated rates. 
Right: Reconstructed photon arrival times for R1408 (top) and R5912 (bottom) PMTs
for 397 nm light flashed from the center of the detector.  The black histogram is the distribution from data and the blue is the complete Monte Carlo simulation.  The green (red) shows the simulation with reflections (and scattering) turned off.}
\end{figure}

The Marcol~7 mineral oil used in MiniBooNE exhibits a rich array of optical phenomena despite its $\nearsim20 $~m extinction length near the peak of PMT sensitivity ($\nearsim 400$~nm).  Cherenkov and scintillation light production is accompanied by photon absorption, fluorescence (with several possible excitation/emission spectra and lifetimes), and Rayleigh and Raman scattering.  The left plot in Figure \ref{fig:optical} summarizes the rates of various processes as a function of wavelength.  The cumulative extinction rate is shown as the black line.  In the near-ultraviolet region ($\mathord{<}320\nm$), fluorescence processes dominate the extinction rate.   The fluorescence lifetimes range from $1-35\,\ns$.  In the visible region ($\mathord{>}320\nm$), the dominant processes are Rayleigh scattering and absorption.

The 1520~8~in.~PMTs in MiniBooNE are of two types: 322 model R5912 PMTs and 1198 model R1408 PMTs, both from Hamamatsu.  All of the R5912 PMTs are located in the main PMT array.  The PMT time response (particularly the late-pulsing behavior) and the variation of PMT efficiency with incident angle have been characterized in external measurements using a pulsed LED~\cite{pmt}.  The PMTs {\em in situ} have been studied using a laser calibration system that produces sub-nanosecond bursts of nearly isotropic light in the detector.

The right panel of Figure \ref{fig:optical} shows the times of PMT hits recorded in laser calibration events relative to their expected times.  The colored histograms come from Monte Carlo simulation, with the red curve showing PMT and electronics timing effects (including pre- and late-pulsing), the green curve including photon scattering, and the blue curve adding photon reflections from the PMTs and the surface of the main detector region.  The resulting time structure matches well with that seen in data (black points).  We note that only the earliest photoelectron's time is reported when multiple photoelectrons are present in a hit.  (A fresh hit can be seen after an electronics dead time of $\nearsim 300$~ns.)  The reported charge, however, does account for additional photoelectrons.

Absorption, scattering, reflections, and fluorescence processes influence both the topology and the time structure of hits.  Additionally, scintillation light, emitted with a lifetime of $\nearsim 35 \ns$, adds intrinsically delayed light to the prompt Cherenkov photons.  The event models within the reconstruction algorithm account for all these phenomena.  In the discussion, we refer to Cherenkov and scintillation light that travels undisturbed from the particle track to the PMT as ``direct'' light.  Light that undergoes fluorescence, scattering, or reflection we term ``indirect'' light.

A detailed GEANT3-based~\cite{geant3} Monte Carlo simulation of the MiniBooNE detector serves as the central tool for developing the reconstruction algorithm's predictive models.  The accuracy of the models influences performance through resolution and biases in extracted track parameters and through the particle separation capabilities of the likelihood-based hypothesis testing.  To balance the need for likelihoods that are as accurate as possible with the desire to contain the computational demands, detector and track model information extracted from the Monte Carlo simulation is tabulated, approximated, or parametrized wherever possible while minimizing any sacrifice in performance.  Also, thanks to the excellent stability of the detector's optical properties, a single configuration of the reconstruction is sufficient for use on all MiniBooNE physics data.

%% file: onetrack.tex
\section{The Single Track Model}
\label{sec:onetrack}
The reconstruction of a single track in the detector, whether an electron or muon, is based
on a model with seven parameters:
\begin{itemize}
\item the starting point: $x_0, y_0, z_0$ 
\item the starting time: $t_0$
\item the direction: $\theta_0$, $\phi_0$
\item the kinetic energy: $E_0$.
\end{itemize}
The starting point and time are referred to as the ``vertex'' of the event.  The
direction is described in terms of the polar angle $\theta_0$ defined with respect 
to the direction of the neutrino beam and the azimuthal angle $\phi_0$ about this beam axis. 
For the electron model, the stochastic variations
in the electromagnetic showers are modeled in average -- no attempt is made to account for event-by-event variations in the shower profile.
This is also true of the much smaller variations in the muon propagation ({\em e.g.}, straggling).  This simplification allows the track to be fully specified from its initial
properties summarized in the seven parameters above. We refer to this
vector of parameters as $\vecx$. 

The observables of an event are the measurements from the 1280 PMTs in
the main region of the detector. For each PMT, we have:
\begin{itemize}
\item a bit indicating whether the tube registered a hit
\item if the tube was hit, the measured charge of the hit
\item if the tube was hit, the measured time of the hit.
\end{itemize}
Assuming that the PMTs behave independently, the likelihood for an observed  set of 
PMT measurements given track parameters $\vecx$ can be expressed as:
\begin{equation}
\like(\vecx) = \displaystyle{\prod_{\sunhit}} (1-P(i\;\hit;\vecx)) \times  
\displaystyle{\prod_{\shit}} \;P(i\;\hit;\vecx)\; f_q(q_i;\vecx)\; f_t(t_i;\vecx)
\label{eq:likex}
\end{equation}
where the products are taken over all unhit and hit PMTs and:
\begin{itemize}
\item $P(i\;\hit;\vecx)$ is the probability that the $i$th tube is hit, given parameters $\vecx$.
\item $q_i$, $t_i$ are the measured charge and time on the $i$th PMT.
\item $f_q(q_i;\vecx)$ is the probability distribution function (PDF) for the measured charge
on the $i$th PMT, given $\vecx$, evaluated at $q_i$.
\item $f_t(t_i;\vecx)$ is the PDF for the measured time given $\vecx$, evaluated
at $t_i$.
\end{itemize}
It is convenient to work with the negative logarithm of $\like$, and since the charge- and time-related
portions of the likelihood are distinct, it is helpful to define:
\begin{equation}
F(\vecx) \equiv -\log\like(\vecx) \equiv F_q(\vecx) + F_t(\vecx)
\label{eq:likefact}
\end{equation}
where
\begin{equation}
\begin{array}{ll}
F_q(\vecx) = - \displaystyle{\sum_{\sunhit}} \log(1-P(i\;\hit;\vecx)) & - \displaystyle{\sum_{\shit}} 
\log\left(P(i\;\hit;\vecx)f_q(q_i;\vecx) \right)\\ 
F_t(\vecx) = -\displaystyle{\sum_{\shit}} \log\left(f_t(t_i;\vecx)\right)
\end{array}
\label{eq:qtlike}
\end{equation}
For brevity, we refer to $F_q$ and $F_t$ as the charge and time likelihoods, respectively, although they are
the negative logarithms of the likelihoods.

\subsection{The Charge Likelihood}
If the number of photoelectrons $n$ produced in a PMT is known, then $f_q(q)$ for the observed charge
$q$ is fully specified in terms of the response properties of the PMT (assumed fixed) without any reference to any other property of the detector. Further, $n$ is a Poisson variable whose mean $\mu$
is a function of the track parameters $\vecx$.  These
considerations motivate a two-step approach to calculating the charge likelihood.

In the first step, using the known optical photon and particle propagation properties of the detector, one determines for a given particle type (electron/muon) and set of track parameters $\vecx$ the average number of photoelectrons that a particular PMT should observe. This quantity is referred to as $\mu_i$, the ``predicted charge'' for the $i$th PMT. In calculating the predicted charge, one considers the quantity and angular distribution of light produced by the track, the absorption,
 scattering, and fluorescence of light as it traverses the mineral oil, the acceptance of the PMT, and 
anything else that influences the mapping $\vecx\mapsto\{\mu_i\}$.

\begin{figure}[tb]
\centering
\includegraphics[width=12 cm] {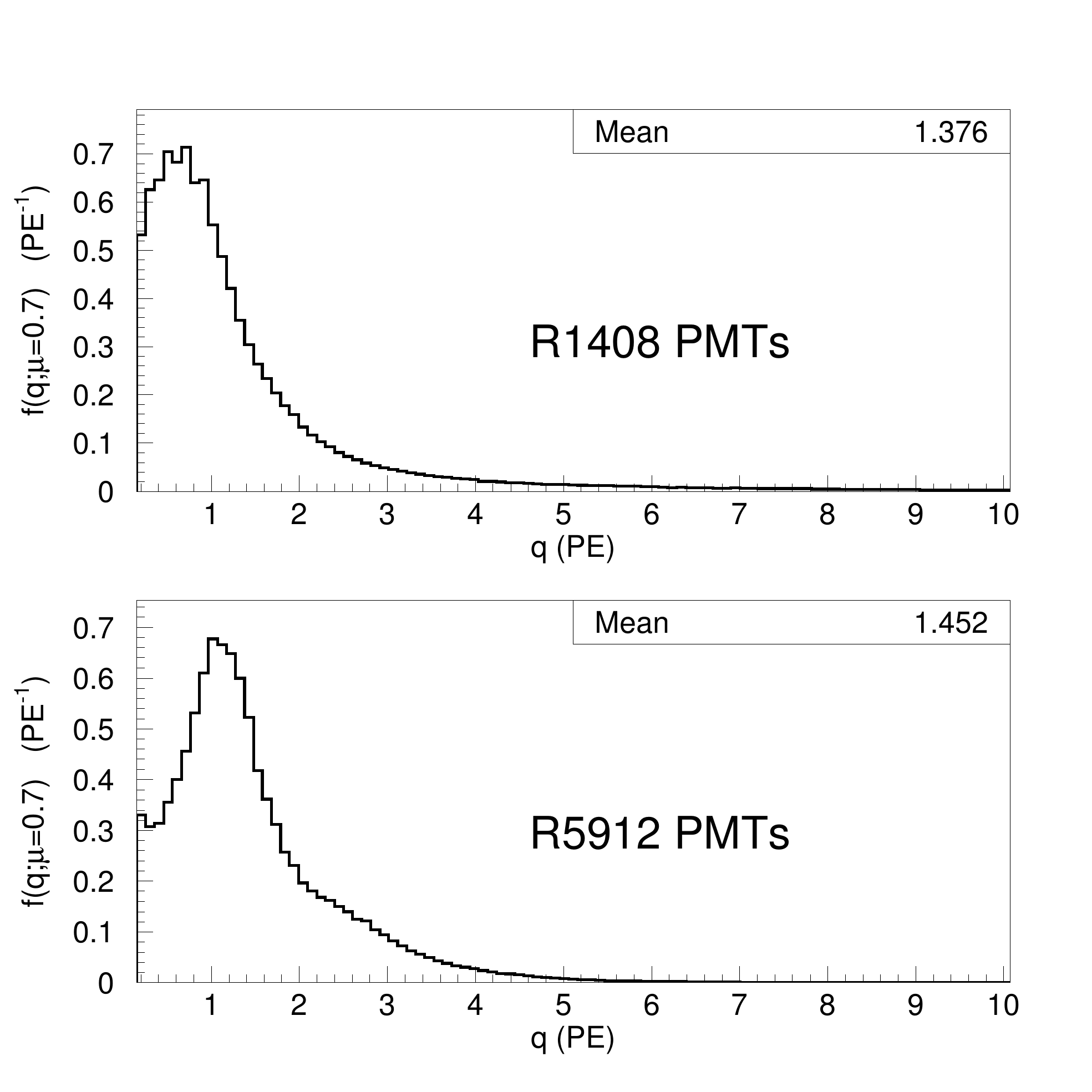}
\caption{\label{fig:qprob} Examples of $f(q;\mu)$ for $\mu=0.7$ photoelectrons.  Note that the distribution only includes cases where the PMT registers a hit.}
\end{figure}

For the second step, $F_q$ given in Eq.~(\ref{eq:qtlike}) can be rewritten in terms of the predicted mean charges $\mu_i$:
\begin{equation}
F_q(\vecx) \equiv -\displaystyle{\sum_{\sunhit} } \log(1-P(i\;\hit;\mu_i)) - \displaystyle{\sum_{\shit} } \log(P(i\;\hit;\mu_i) f_q(q_i;\mu_i))
\label{eq:mui}
\end{equation}
where $P$ and $f_{q}$ are now functions of the $\left\{ \mu_i\right\}$, with the mapping $\vecx\mapsto\{\mu_i\}$ from the first step left implicit.  These two functions depend only on the properties of the PMT response and are, in particular, independent of the happenings within the detector volume. This stepwise approach decouples the track and optical photon model from the PMT response model.  The latter is established once using laser calibration data.  The laser system provides a source of approximately isotropic and prompt light with controllable intensity.  The value of $\mu$ for a particular laser intensity can be determined from the occupancy of the PMTs.  The function $f_q(q;\mu)$ for each PMT type can be determined by obtaining the distribution of measured charges $q$ seen at various intensities $\mu$.  These distributions incorporate all processes associated with the charge response, such as the cascade of charge down the dynode chain, the amplification and digitization of the anode signal, and the conversion of the digital readouts into calibrated charge values.  Examples of $f_q(q;\mu)$ obtained in this way are shown in Figure \ref{fig:qprob}.

\subsection{Calculating the Predicted Charge $\mu$}
We begin our description of the predicted charge calculation with a simple
scenario involving a fictitious point source of isotropic direct light.  This is then generalized
to a line source of isotropic light, appropriate for the scintillation production of a track.  We then consider a line source in which the emission has a non-trivial (but azimuthally symmetric) angular distribution, representative of Cherenkov radiation.  Finally,
we account for the contribution of indirect light ({\em i.e.}, scattering, reflections, etc.) from
both the scintillation and the Cherenkov emission.  In order to simplify notation and without loss of generality, the discussion involves fixed but arbitrary PMT location and track parameters $\vecx$.

One needs only two parameters to describe the geometric relationship between an isotropic point source and a PMT.  We choose (1)~the 
distance $r$ between the source and the PMT and (2)~the angle of incidence $\eta$ at the PMT, with 
$\eta=0$ corresponding to incidence parallel to the PMT axis.  The predicted charge on the PMT can be written
\begin{equation}
\mu_{\point,\sci} = \Phi_{\sci}\; \ROmega(r) \; T_{\sci}(r) \; \epsilon(\eta)
\label{eq:isopoint}
\end{equation}
where $\Phi_{\sci}$ is an event-energy-dependent scintillation light yield ({\em i.e.}, $\Phi_{\sci}$ $\mathord{=}$ $\Phi_{\sci}(E_0)$\,), $\ROmega(r)$ is a $r$-dependent solid angle factor, $T_{\sci}(r)$ is the transmission of the oil and PMT glass as a function of light propagation distance, and $\epsilon(\eta)$ is the acceptance of the PMT as a function of the angle of incidence. The transmission depends on the wavelength spectrum of the light source and, thus, is specific in this case to scintillation.
The wavelength-dependent PMT photocathode efficiency is also included in $T(r)$.  The solid angle factor and PMT angular acceptance are purely geometric, independent of the particulars of the light production.

\begin{figure}[tb]
\centering
\includegraphics[width=6 cm] {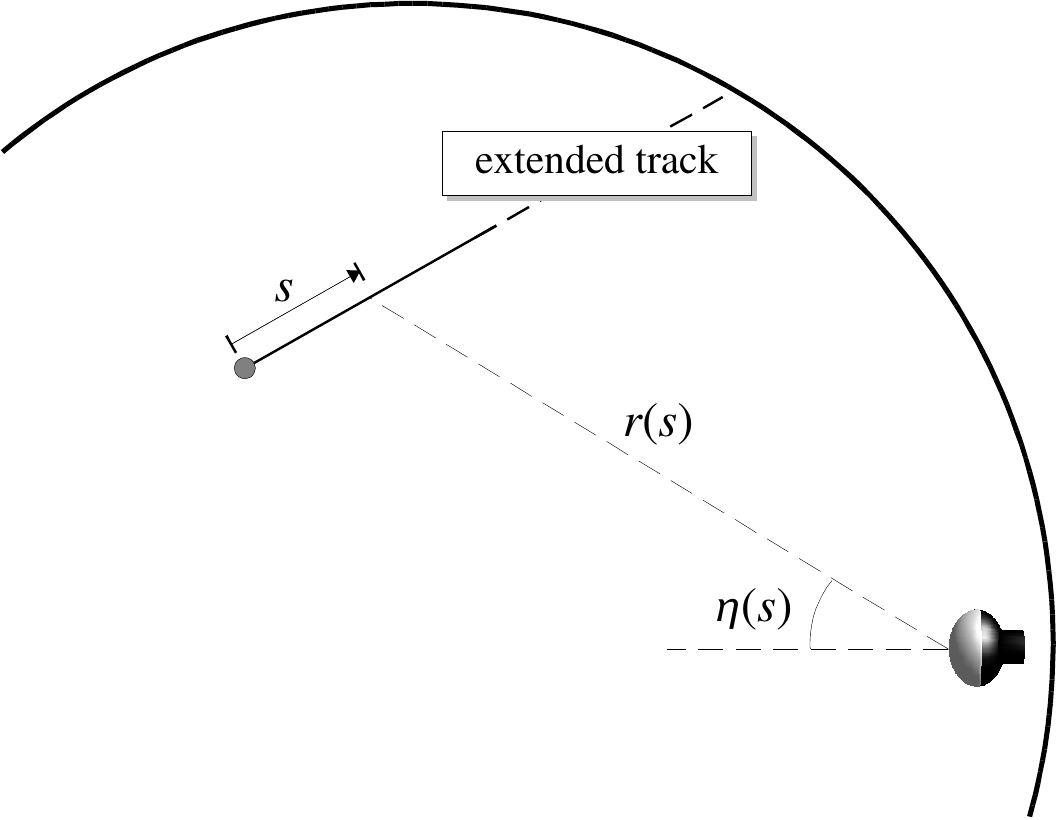}\hspace{1.00 cm}
\includegraphics[width=6 cm] {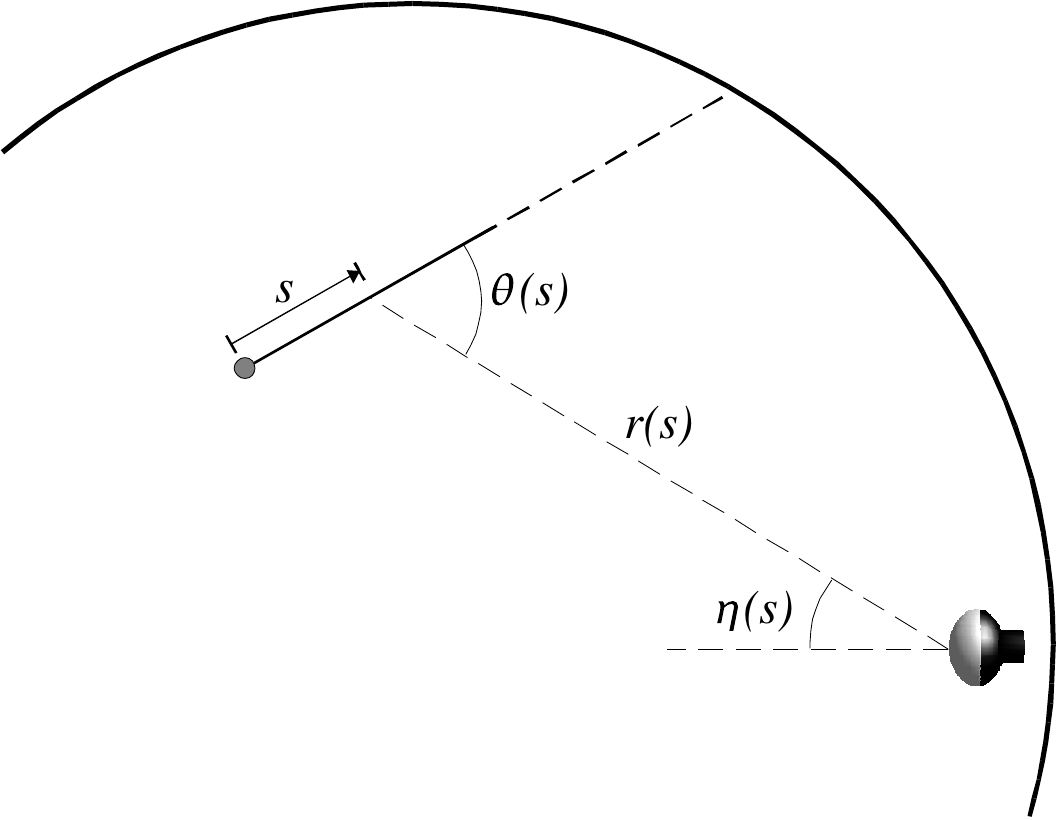} 
\caption{\label{fig:geometry} Geometry of a line source of isotropic scintillation light (left) and directional Cherenkov light (right). In each case $s$ is the distance from the origin of the track, $r(s)$ is the distance to the PMT from the point along the track, and $\eta(s)$ is the angle of PMT incidence for light emitted from this point.  The presence of Cherenkov light requires the additional parameter $\theta(s)$ describing the light emission angle.}
\end{figure}

\begin{figure}[tb]
\centering
\includegraphics[width=12 cm] {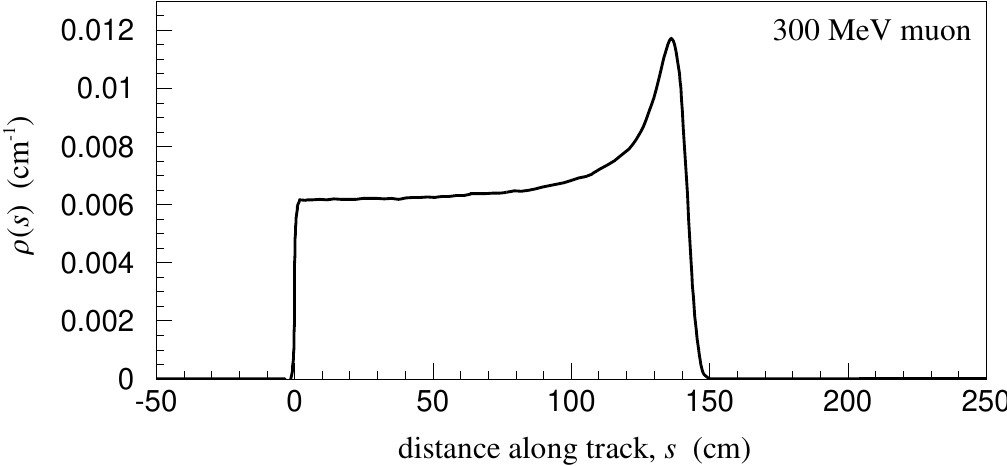}
\caption{\label{fig:rhoscie} Scintillation emission profile for $300\mev$ muons as a function
of $s$, the distance along the track.}
\vspace{0.50 cm}
\includegraphics[width=12 cm] {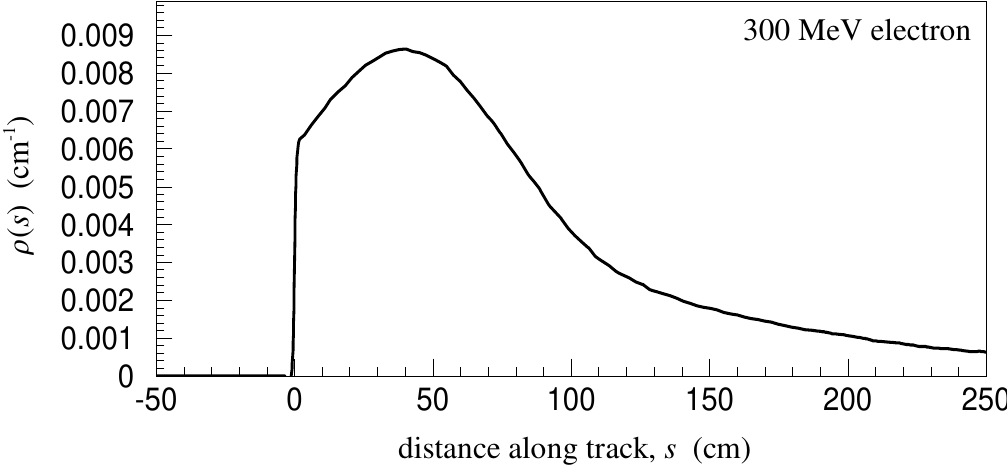}
\caption{\label{fig:rhoscimu} Scintillation emission profile for $300\mev$ electrons as a function
of $s$, the distance along the track.}
\end{figure}

To generalize to an extended source of scintillation light, as depicted in Figure \ref{fig:geometry}(left), the latter three factors in Eq.~(\ref{eq:isopoint}) become functions of $s$, the distance along the track from its origin.   We must also include an emission profile $\rho_{\sci}(s)$ describing the distribution of scintillation production along the track.  This profile satisfies $\int_{-\infty}^{\infty} \rho(s) ds = 1$.  The predicted charge can now be obtained by integrating along the track's axis:
\begin{equation}\label{equ:muscidir}
\mu_{\sci} = \Phi_{\sci} \int_{-\infty}^{\infty} ds\; \rho_{\sci}(s) \, \ROmega(s) \, T_{\sci}(s) \, \epsilon(s)\ .
\label{eq:musci}
\end{equation}
The dependences of $\ROmega$, $T_{\sci}$, and $\epsilon$ on $r$ and $\eta$ have been recast as dependences
on $s$, as the relationship $s\mapsto(r, \eta)$ is fixed for given track parameters and PMT location. 
Figures \ref{fig:rhoscie} and \ref{fig:rhoscimu} show $\rho_{\sci}(s)$ for 300 $\mev$ muons
and electrons, obtained via Monte Carlo simulation. The distribution is relatively flat for muons until the end of the track, where the ionization rate per unit track length becomes larger.  (Saturation effects are taken into account in the simulation.)  For electrons,
the distribution reflects the showering behavior of the track, with $\rho_{\sci}(s)$ rising and falling
with the number of shower particles.

 We  now consider an extended track emitting light with a non-trivial angular distribution, as is the case for Cherenkov radiation. We introduce $\theta$, the angle to the PMT from the track, as shown in Figure \ref{fig:geometry}(right), and express the predicted
charge $\mu_{\cer}$ due to Cherenkov radiation as:
\begin{equation}
\mu_{\cer} = \Phi_{\cer}  \int_{-\infty}^{\infty} ds \rho_{\cer}(s) \, \ROmega(s) \, T_{\cer}(s) \, \epsilon(s) \, g(\cos\theta(s); s)
\label{eq:mucer}
\end{equation}

This expression differs from its scintillation counterpart by the presence of an angular emission
profile $g(\cos\theta(s);s)$.  Note that $g$ depends on $s$ in two distinct ways: the angular profile of the light changes  as the track propagates and loses energy, and the angle $\theta$ to the PMT  changes depending on which part of the track we are considering.

\begin{figure}[t]
\centering
\includegraphics[width=10.1 cm] {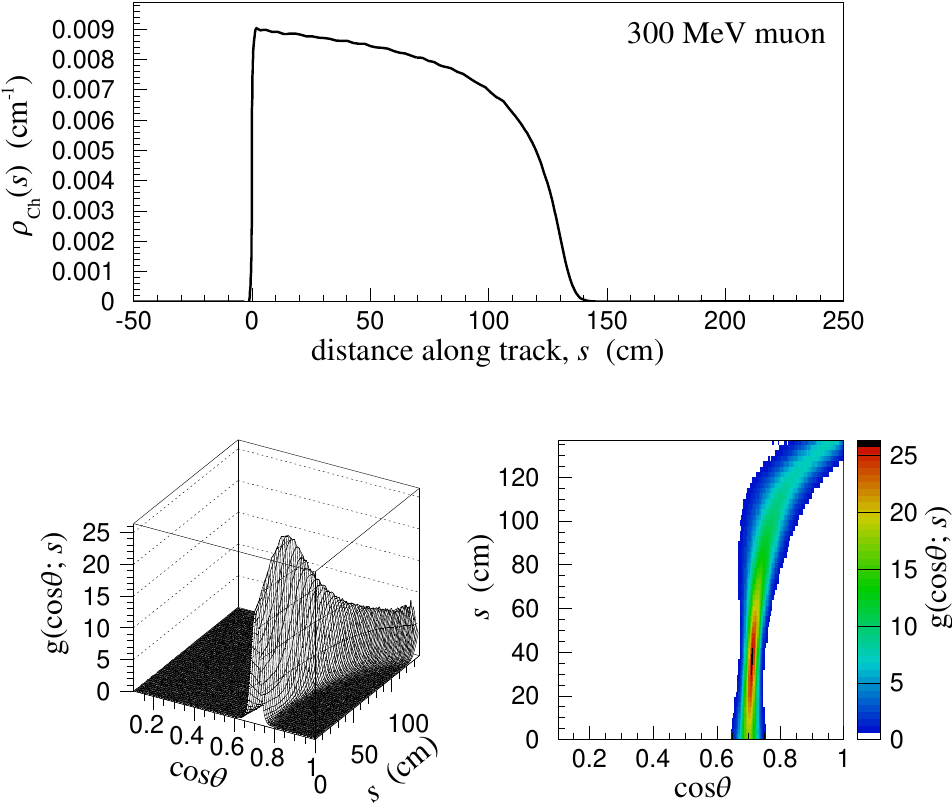}
\caption{\label{fig:rhochermu} Top: Cherenkov emission profile for $300\mev$ muons as
a function of $s$, the distance along the track. Bottom: Two-dimensional emission
profile showing the angular distribution of Cherenkov light as a function of $s$
for a $300\mev$ muon.}
\vspace{0.8cm}
\includegraphics[width=10.1 cm] {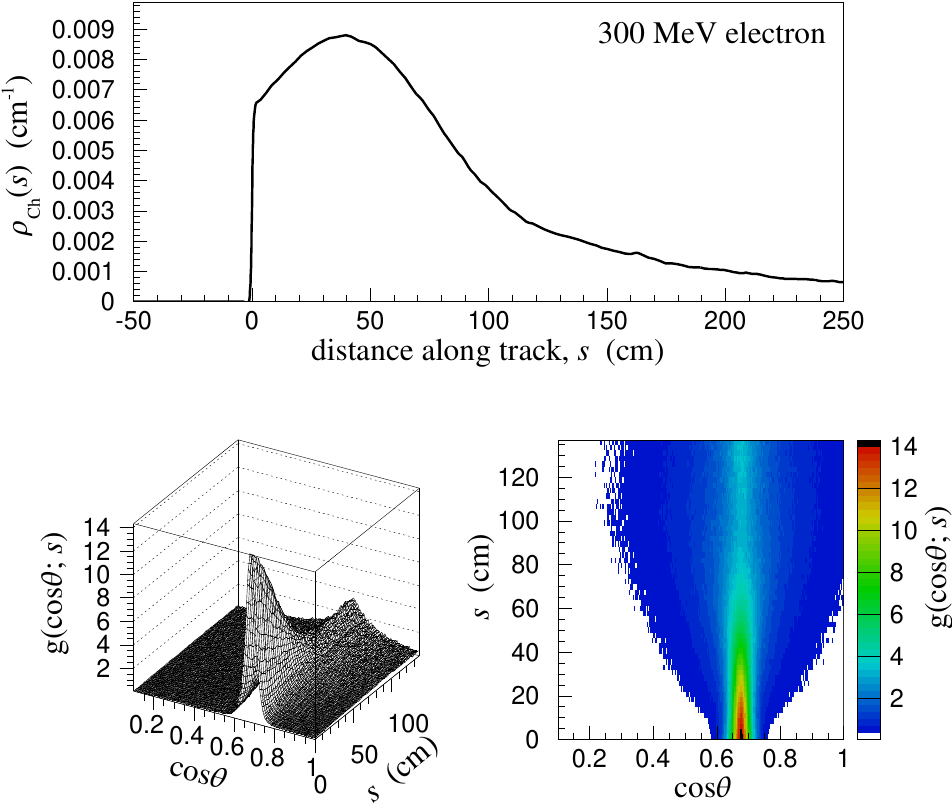}
\caption{\label{fig:rhochere} Top: Cherenkov emission profile for $300\mev$ electrons as
a function of $s$, the distance along the track. Bottom: Two-dimensional emission
profile showing the angular distribution of Cherenkov light as a function of $s$
for a $300\mev$ electron.}
\end{figure}

Figures \ref{fig:rhochermu} and \ref{fig:rhochere} show  $\rho_{\cer}(s)$ and $g(\cos\theta;s)$  for simulated 300 $\mev$ muons and electrons.
As the muon propagates, loses energy, and approaches the Cherenkov threshold, the rate of
Cherenkov radiation per unit track length decreases and the Cherenkov angle becomes smaller.
The scattering of the muon, which causes deviations of the track from its original direction, has been
included in the Monte Carlo simulation and results in the spread of the angular distribution about the nominal $\cos\theta_C$. For electrons, $\rho_{\cer}(s)$ follows the shower profile, like $\rho_{\sci}(s)$. The presence of the shower particles is readily apparent in the
$g(\cos\theta;s)$ distribution, which becomes substantially wider as the ``track'' propagates.
\afterpage{\clearpage}

In writing Eq.~(\ref{eq:mucer}), a few simplifications are made.  A more rigorous treatment would involve $\frac{d^2P(\cos\theta,\phi)}{d\cos\theta d\phi}$, the differential probability of sending an emitted Cherenkov photon in the direction $(\cos\theta,\phi)$, which would be integrated over the solid angle subtended by the PMT. By assuming that $\frac{d^2P(\cos\theta,\phi)}{d\cos\theta d\phi}$ is constant over the PMT face, the integration can be effected by simply multiplying the differential probability by the solid angle $\ROmega$. Also, the azimuthal symmetry of the emission reduces the 
two-dimensional PDF to a one-dimensional expression, namely $g(\cos\theta)$. We take $g(\cos\theta)$
to satisfy
\begin{equation}
\int_{-1}^{1} d\cos\theta\, g(\cos\theta;s)  =1
\label{eq:lightnorm}
\end{equation}
for all values of $s$, with the result that a factor of $(4\pi)^{-1}$ is absorbed into the definition 
$\Phi_{\cer}$.

\subsection{Indirect Light (scattering, fluorescence, reflections)}
The above formalism determines the predicted charge for light arriving at the PMTs directly from the
track without any redirection. However, the detector has sources of indirect light from
scattering, fluorescence, and reflection as discussed in Section \ref{sec:detector}.

\begin{figure}[t]
\centering
\includegraphics[width=6 cm] {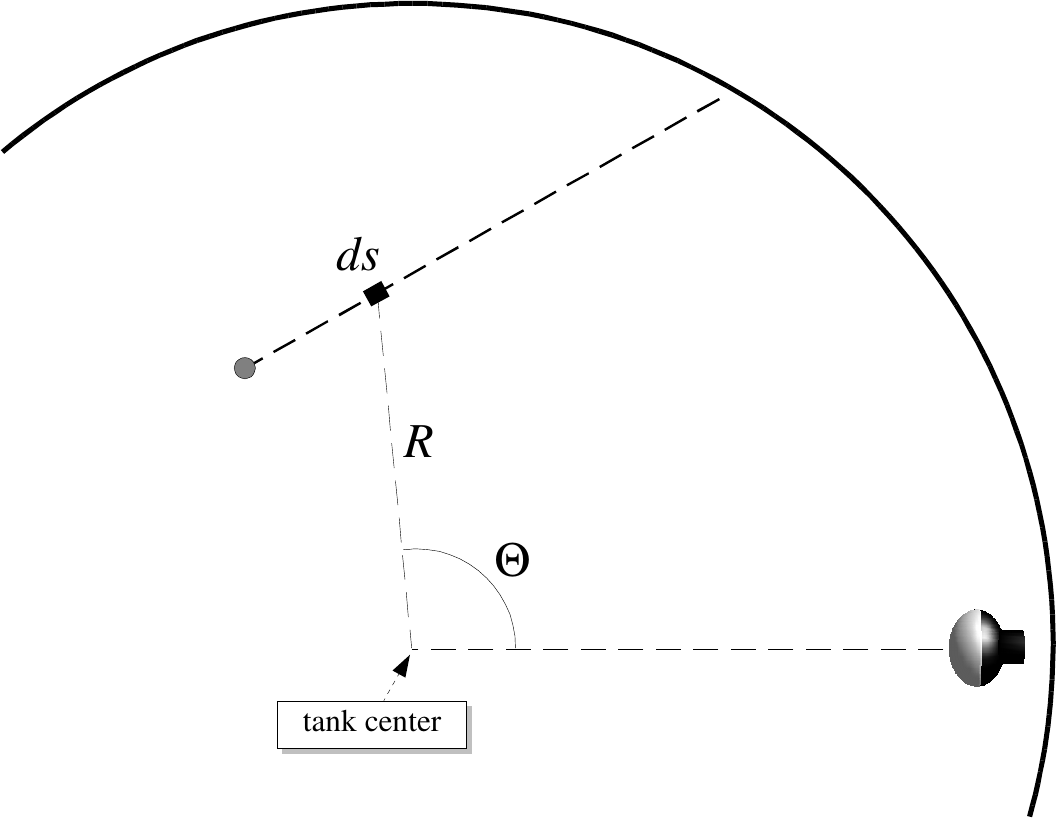}\hspace{1.00 cm}
\includegraphics[width=6 cm] {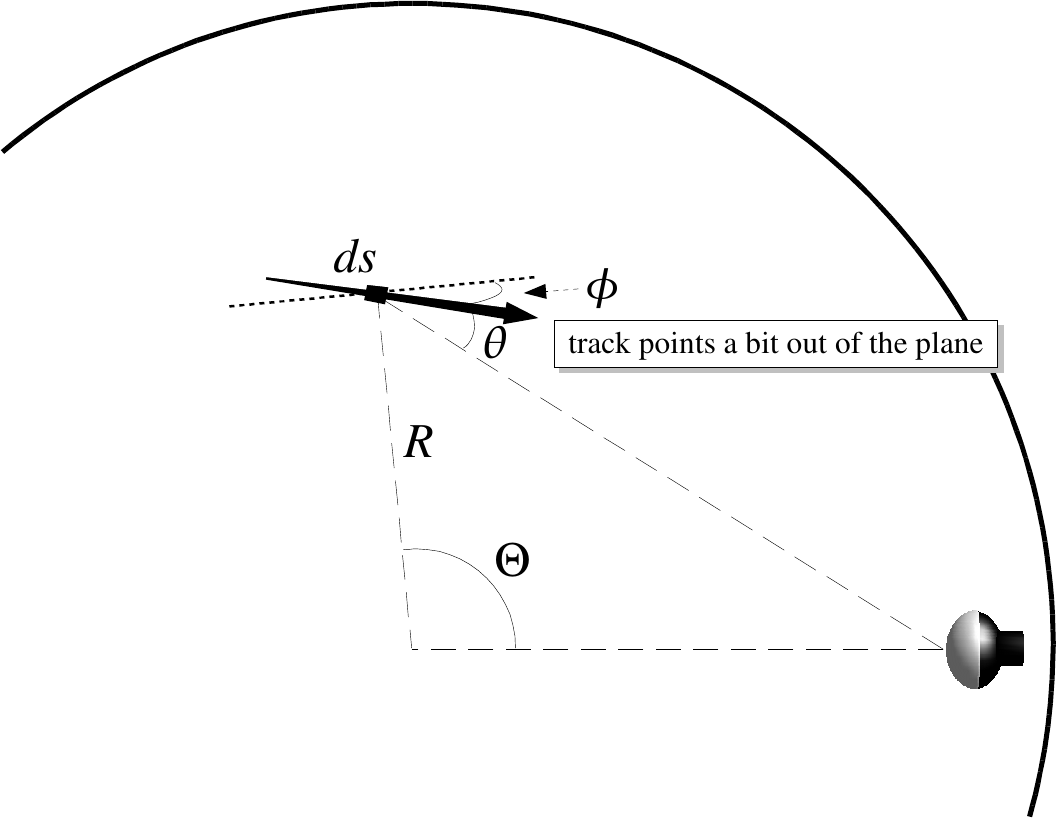} \\
\caption{\label{fig:indirect} Geometry of indirect light from an isotropic (left) and directional (right) source of light.}
\end{figure}

The geometry for indirect light given scintillation (isotropic) emission is shown in Figure \ref{fig:indirect}(left), where an infinitesimal element $ds$ along the track is  situated at radius $R$ from the center of the detector and at angle $\Theta$ relative to the position of the PMT.  The direct light from this track element is simply the integrand of Eq.~(\ref{equ:muscidir}):
\begin{equation}
d\mu_{\sci}^{\direct} =  ds\; \Phi_{\sci} \, \rho_{\sci}(s) \, \ROmega(s) \, T_{\sci}(s) \, \epsilon(s)\ .
\label{eq:isoindirect}
\end{equation}
An analytic expression for the {\em indirect light} would involve an elaborate
integral over emission angles and scattering points throughout the tank. Rather than attempt this, we observe
that the value of such an integral must be proportional to the source strength and must otherwise
depend only on the topological variables $R$ and $\Theta$. The source strength 
dependence can be eliminated by forming a ratio of the indirect and direct light predictions:
\begin{equation}
A_{\sci}(R,\cos\Theta) \equiv \frac{d\mu^{\indirect}_{\sci}}{d\mu^{\direct}_{\sci}}\ .
\label{eq:asci}
\end{equation}
$A_{\sci}$, which we refer to as the {\em scattering table} and which we build via the detector simulation, is a property only of the detector optics and the $(R, \cos\Theta)$ of the track element. With this table, the indirect light contribution can be immediately incorporated into the expression for predicted charge:
\begin{equation}
\mu_{\sci} = \Phi_{\sci} \int_{-\infty}^{\infty} ds \,  \rho_{\sci}(s) \, \ROmega(s) T_{\sci}(s) \epsilon(s) 
\left[ 1 + A_{\sci}\left(R(s), \cos\Theta(s)\right) \right]
\label{eq:asci2}
\end{equation}
where the dependence of $R$ and $\Theta$ on $s$ has been made explicit.

For Cherenkov light, the situation is more complex since the light emission is anisotropic. Two additional
variables are needed to specify the direction of a {\em vector} source relative to the PMT position and the
tank center. We make the following non-unique choice for these two variables, as shown in Figure \ref{fig:indirect}(right):
\begin{itemize}
\item $\theta$: the angle between (1) the source direction vector and (2) the source-to-PMT ray (the same $\theta$ as defined elsewhere).
\item $\phi$: The angle between (1) the plane containing the tank center, the PMT, and the source, and (2) the
plane containing the track and the tank center.
\end{itemize}

The intensity of the indirect Cherenkov light is normalized with respect to a fictitious {\em isotropic} Cherenkov source with predicted charge $d\mu_{\cer}^{\direct,\iso}$, so that:
\begin{equation}
A_{\cer}(R,\cos\Theta, \cos\theta, \phi)\equiv \frac{d\mu_{\cer}^{\indirect}}{d\mu_{\cer}^{\direct,\iso}}
\label{eq:acer}
\end{equation}
The total Cherenkov contribution to the mean predicted charge, including both direct and indirect components
is
\begin{equation}
\mu_{\cer} = \mu^{\direct}_{\cer} + \mu^{\indirect}_{\cer}
\label{eq:mucertot}
\end{equation}
where $\mu^{\direct}_{\cer}$ is given by Eq.~(\ref{eq:mucer}) and $\mu^{\indirect}_{\cer}$
is given by
\begin{equation}
\begin{array}{lll}
\mu^{\indirect}_{\cer} &  =   \Phi_{\cer} & {\displaystyle  \int_{-\infty}^{\infty}}\! ds \; [ \rho_{\cer}(s) \, \ROmega(s) \, T_{\cer}(s) \, \epsilon(s) \\
 & & ~~~~~~~\times  A_{\cer} \left(R(s), \cos\Theta(s), \cos\theta(s), \phi(s) \right) ]
\end{array}
\label{eq:mucerindirec}
\end{equation}
Implicit in this expression is the assumption that the Cherenkov angular emission profile does not change as the track propagates (and loses energy).  This simplification has negligible impact, as the processes that result in indirect light tend to destroy the initial emission pattern anyway.

We now have a complete charge prediction, where the total predicted charge for the PMT is given by:
\begin{equation}
\mu = \mu_{\cer} + \mu_{\sci}
\label{eq:mutotal}
\end{equation}

\subsection{Computation of Integrals}
The predicted charge integrals above would be impractical to evaluate
with sufficient spatial granularity within a maximum likelihood fit, where they must be evaluated
multiple times for every PMT in every event.  To mitigate this, all integrations are 
performed and tabulated beforehand as follows.

\begin{figure}[t]
\centering
\includegraphics[width=4 cm,angle=90] {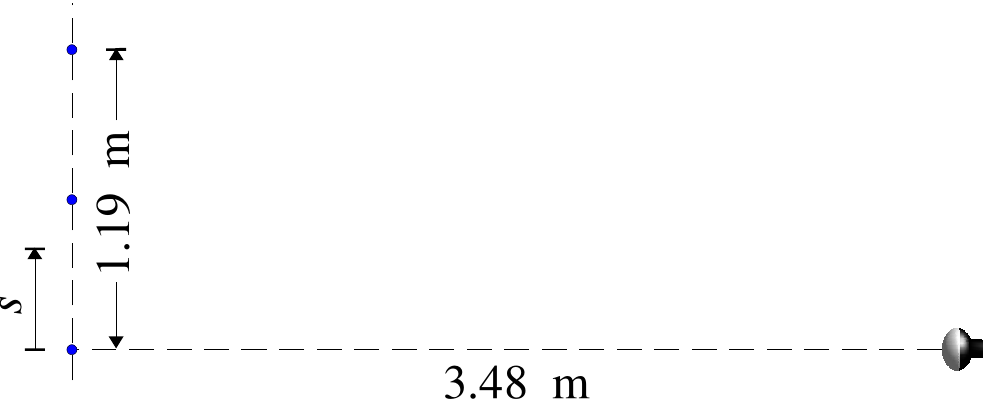} \\
\includegraphics[width=12 cm] {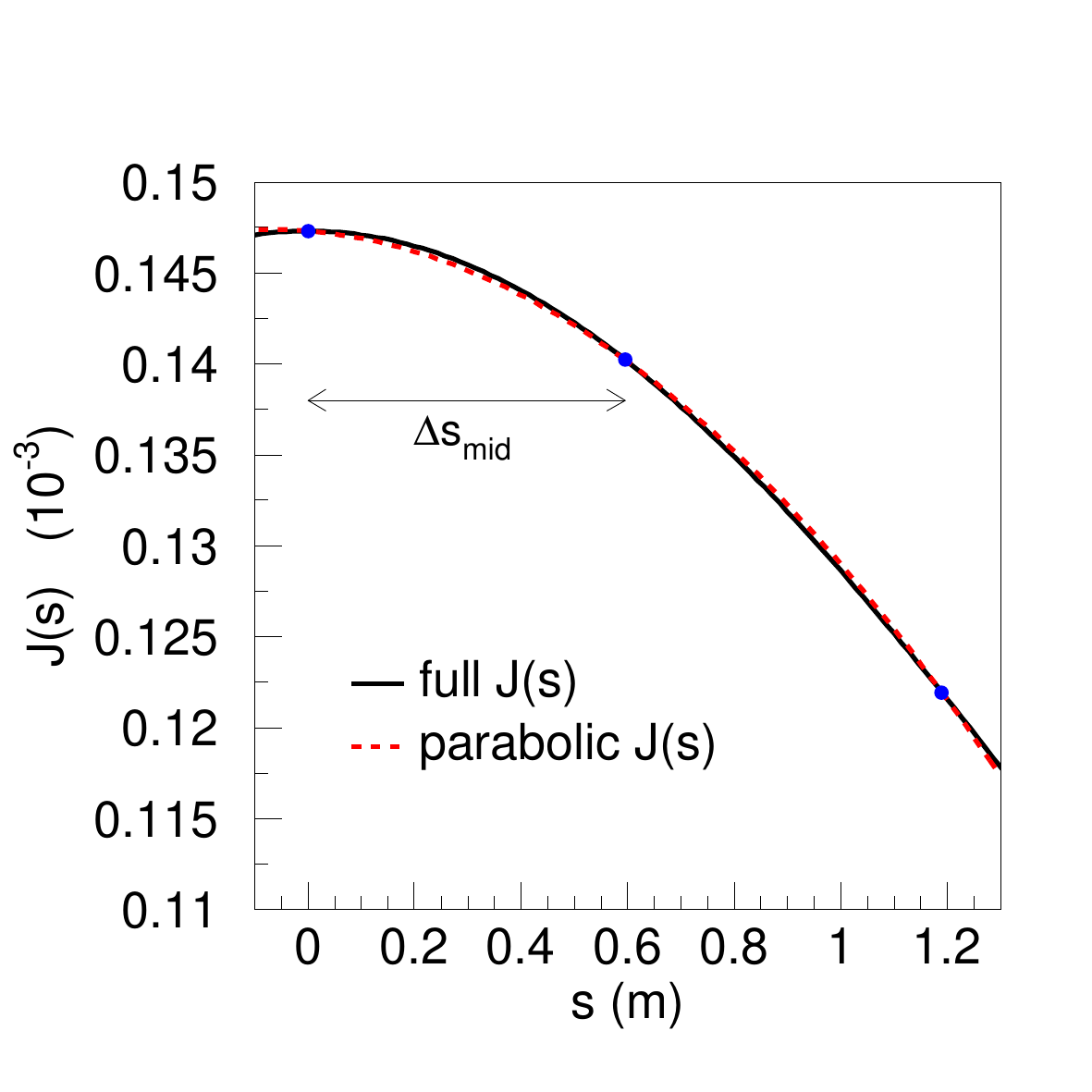} 
\caption{\label{fig:parabola} Example of $J(s)$ for the track geometry shown in
the top figure, where a muon 3.48 meters from a PMT  moves perpendicular
to the PMT axis. $J(s)$ is evaluated at three points located along the track, corresponding
to the three points in the figure. In the bottom panel, the parabolic approximation for $J(s)$
based on the three evaluated points (dashed red) is compared to the exact form (solid black).}
\end{figure}

The integrand of Eq.~(\ref{eq:musci}), which pertains to direct scintillation light, is the product of a ``source'' factor $\Phi_{\sci}\rho_{\sci}(s)$ and an ``acceptance'' factor
\begin{equation}
J(s) \equiv \ROmega(s) T_{\sci}(s) \epsilon(s)\ .
\label{eq:js}
\end{equation}
As in Eq.~(\ref{eq:musci}), the dependence on $r$ and $\eta$ is recast through $s$.
If $J(s)$ varies gradually enough, it can be
approximated with a parabolic form:
\begin{equation}
J(s) = j_0 + j_1\, s + j_2\, s^2\ .
\label{eq:jquad}
\end{equation}
An example of this is shown in Figure \ref{fig:parabola}.  The predicted charge then becomes
\begin{equation}
\mu_{\sci}^{\direct} = \Phi_{\sci} \left[  j_0\! \int_{-\infty}^{\infty}\! ds\;  \rho_{\sci}(s) +
                                           j_1\! \int_{-\infty}^{\infty}\! ds\;  \rho_{\sci}(s)\, s +
                                           j_2\! \int_{-\infty}^{\infty}\! ds\;  \rho_{\sci}(s)\, s^2
 \right]\ .
\label{eq:muscij}
\end{equation}
The first integral is identically unity.  The remaining two depend on the energy $E_0$ via $\rho_{\sci}(s)$, but they depend on no other track parameters.  This allows one to tabulate the
integrals beforehand as a function of $E_0$, eliminating their evaluation from the actual likelihood calculation. 

The parabolic coefficients $\left\{ j_i \right\}$ are obtained by evaluating Eq.~(\ref{eq:js}) at three points
along the track: the start of the track ($s\,\mathord{=}\,0$),  the midpoint of the Cherenkov  emission profile ($s\,\mathord{=}\,\Delta s_{\smid}$)
and twice this distance from the start ($s\,\mathord{=}\,2\Delta s_{\smid}$). The choice of where to evaluate
$J(s)$ is somewhat arbitrary; one could choose any three points that sample a large fraction of the range over which light is produced. The extension to indirect isotropic light is straightforward with
a redefinition of $J(s)$ that incorporates the scattering table:
\begin{equation}
J_{\sci}^{\indirect}(s) \equiv \ROmega(s)\, T_{\sci}(s)\, \epsilon(s)\, A_{\sci}\left( R(s), \cos\Theta(s) \right)\ .
 \label{eq:jsci}
\end{equation}

An analogous three-integral expression can be formed for the Cherenkov predicted charge:
\begin{equation}
\displaystyle{
\begin{array}{lcl}
\mu_{\cer}^{\direct} &  = \Phi_{\cer}  &\left[\;\; j_0 \displaystyle{\int_{-\infty}^{\infty}}\! ds\; \rho_{\cer}(s) g(\cos\theta(s);s)\right. \\
                                   & &  +\;  j_1 \displaystyle{\int_{-\infty}^{\infty}}\! ds\;  \rho_{\cer}(s)\, g(\cos\theta(s);s) s \\
                                  & & \left.+\; j_2 \displaystyle{\int_{-\infty}^{\infty}}\! ds\;  \rho_{\cer}(s)\, g(\cos\theta(s);s) s^2\;\right]\ .
\end{array} 
}                                 
\label{eq:mucerj}
\end{equation}
As in the scintillation case, the integrals depend on the energy of the track. However, they also depend on two
parameters which define the PMT-track geometry. We choose these parameters to be the vertex-to-PMT
distance, $r(0)$; and the cosine of the angle-to-PMT viewed from the vertex, $\cos\theta(0)$. Labeling the
integrals $\mathcal{I}^{\cer}_i$, we obtain
\begin{equation}
\mu_{\cer}^{\direct} = \Phi_{\cer}\left( j_0 \mathcal{I}^{\cer}_0 + j_1 \mathcal{I}^{\cer}_1 + j_2 \mathcal{I}^{\cer}_2\right )\ .
\label{eq:mucerj2}
\end{equation}
Although the integrals $\left\{I_i^{\cer} \right\}$ depend on three parameters, it is still feasible
to tabulate their values ahead of time, eliminating their explicit evaluation within the
likelihood calculation.

For indirect Cherenkov light, recall that the scattering table $A_{\cer}$ is normalized to
a fictitious isotropic Cherenkov source. Thus, the parabolic method used for indirect scintillation 
can be applied here to give
\begin{equation}
J_{\cer}^{\indirect}(s) \equiv \ROmega(s)\, T_{\cer}(s)\, \epsilon(s)\,\, A_{\cer}\left( R(s), \cos\Theta(s) \right)\ ,
\label{eq:jcersindir}
\end{equation}
with two energy-dependent integrals  $I^{\cer}_{i=1,2}=\int\! ds \rho_{\cer}(s) s^{i}$ to be tabulated.

\subsection{The Time Likelihood}
The time portion of the likelihood involves the PDF $f_t(t;\vecx)$ for the measured PMT hit time $t$ given
track parameters $\vecx$. Some of the dependence on the track parameters can be eliminated by using
the ``corrected time'' $t^c$:
\begin{equation}
t^c = t-t_0 - \frac{r(\Delta s_{\smid}\!(E_0))}{c_n} - \frac{\Delta s_{\smid}\!(E_0)}{c}\ .
\label{eq:tc}
\end{equation}
The expression removes from $t$ three terms, namely the starting time of the track $t_0$, the expected time for light to propagate from the track midpoint to the PMT (with the speed of light in mineral oil given by $c_n$), and the time for the particle to propagate from its starting point to the midpoint.
The dependence of $\Delta s_{\smid}$ on the track energy $E_0$ is made explicit. 
Defining $t^c$ with respect to the track midpoint (as opposed to, say, the start of the track) improves
the validity of the simplifications we make below.

Given the $t\mapsto\tc$ substitution, we seek the PDF $f_{\tc}(t^c_i;\vecx)$ for the corrected time given arbitrary PMT-track
configurations.  The space of configurations is five dimensional.  However, producing tables
of $f_{\tc}(\tc)$ as a function of five parameters is impractical.
To reduce the task, we make the assumption that the corrected time PDF depends only on the track energy,
the predicted ``prompt'' charge, and the predicted ``late'' charge, where ``prompt'' corresponds
roughly to light arriving at the PMT directly from the track without delays induced by emission lifetimes,
scattering, {\em etc.}  Loosely, the assumption is that the shape of the corrected time spectrum
is dominated by the physical extent of the track (characterized by its energy), and by the amount of
prompt and late light reaching the PMT. The extent of the track affects the spread of possible hit times,
since there is a spread of photon production times, while the amounts of prompt and late light affect 
the proportion of peaked ``prompt'' response to the tail of ``late'' response.

This assumption reduces the configuration to three dimensions. We make one further simplification
by assuming that for  a given energy $E_0$, $f_{\tc}(\tc)$ can be modeled by separate primitive prompt and late
distributions, indexed by $\mu_{\sprompt}$ and $\mu_{\slate}$, respectively, from which we can
construct the full PDF. As a result,
the PDF is indexed not by the fundamentally two-dimensional space of $(\mu_{\sprompt}, \mu_{\slate})$, but
by $\mu_{\sprompt}$ and $\mu_{\slate}$ separately, with the full PDF calculated on-the-fly as described below.  In practice, Cherenkov and scintillation primitive distributions are
created and used as proxies for the desired prompt and late distributions, respectively.

\begin{figure}[t]
\centering
\includegraphics[width=12 cm] {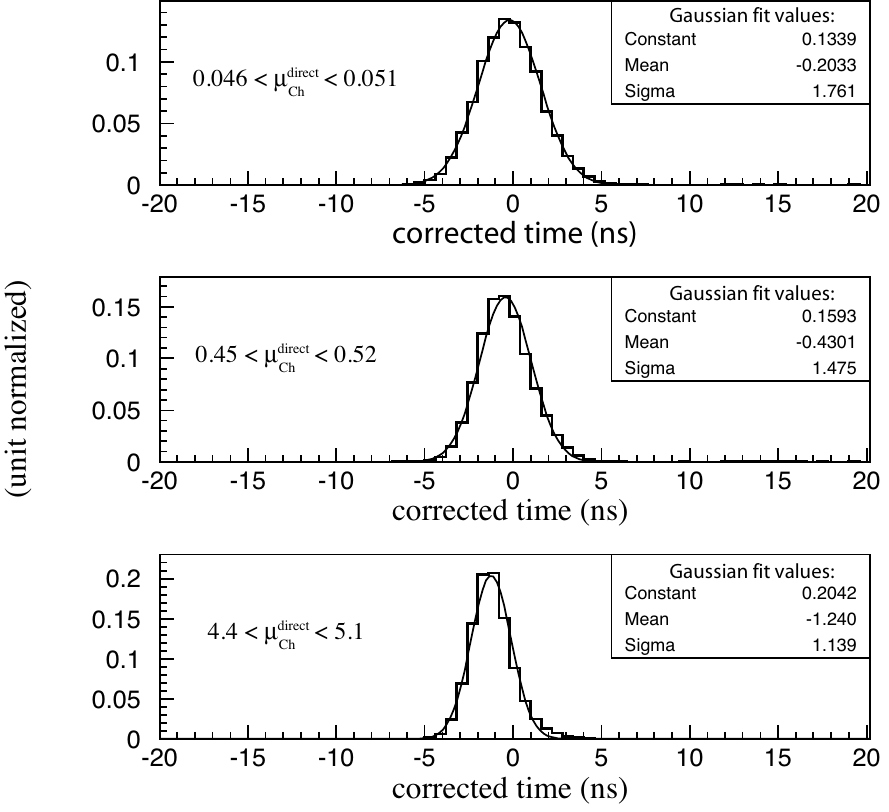}
\caption{\label{fig:tcmu} Distributions of $\tc$ for direct Cherenkov light for R5912 PMTs.  The primary difference between the two PMT types is that the R1480 PMTs show wider prompt timing peaks than do the newer R5912 PMTs.}
\end{figure}

\begin{figure}[t]
\centering
\includegraphics[width=12 cm] {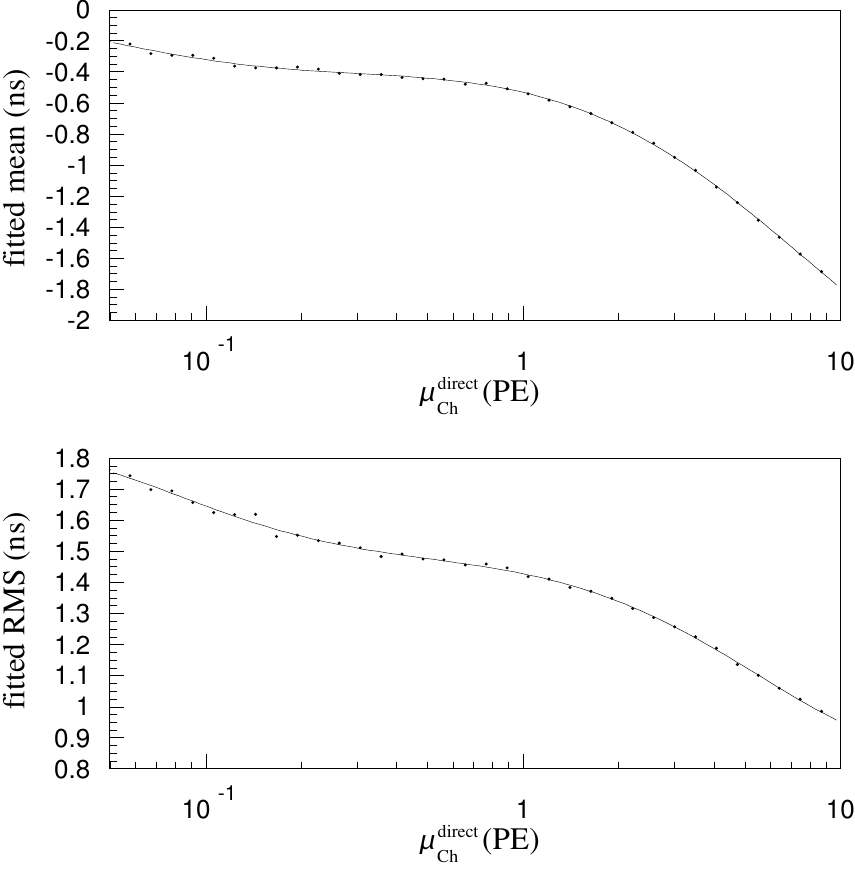}
\caption{\label{fig:gaussparmudirect2} For 300~MeV muons, the mean and width parameters from 
Gaussian fits like those in Figure~\ref{fig:tcmu} versus $\mu_{\cer}^{\direct}$. The dependence
is parameterized as a sixth-order polynomial in $\log(\mu_{\cer}^{\direct})$.}
\end{figure}

\begin{figure}[t]
\centering
\includegraphics[width=12 cm] {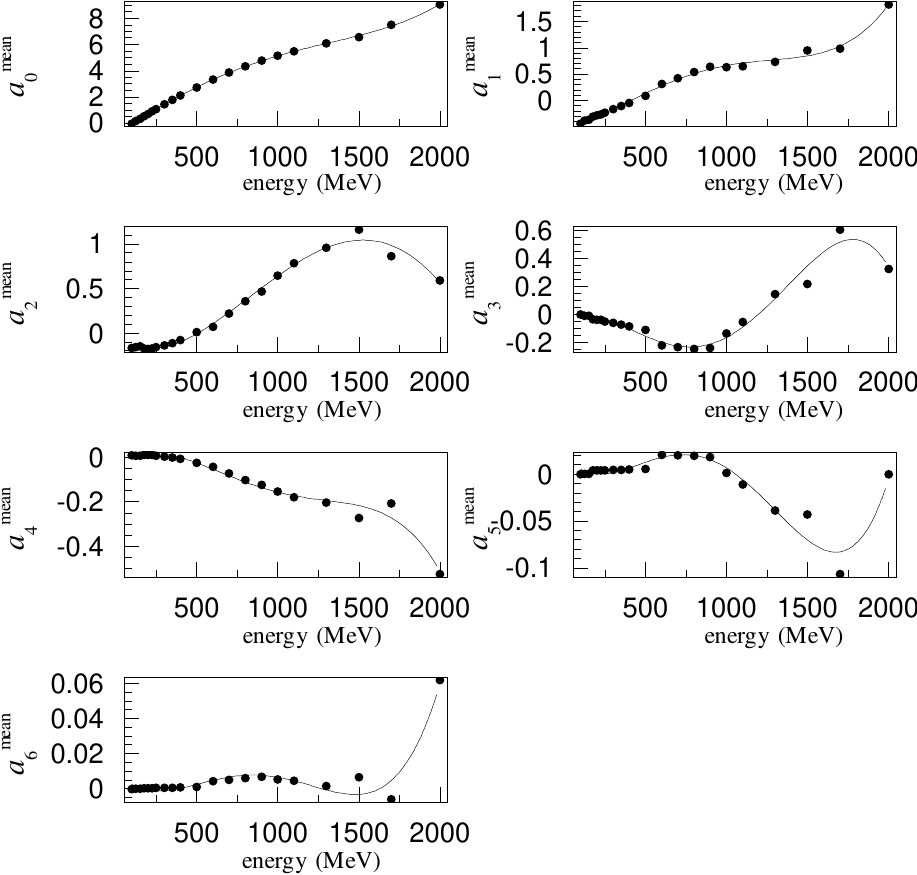}
\caption{\label{fig:gaussparmudirect3} Third-level fits parametrizing the energy dependence
of the second-level parameters describing the Gaussian mean and width parameters. }
\end{figure}

\begin{figure}[t]
\centering
\includegraphics[width=11cm] {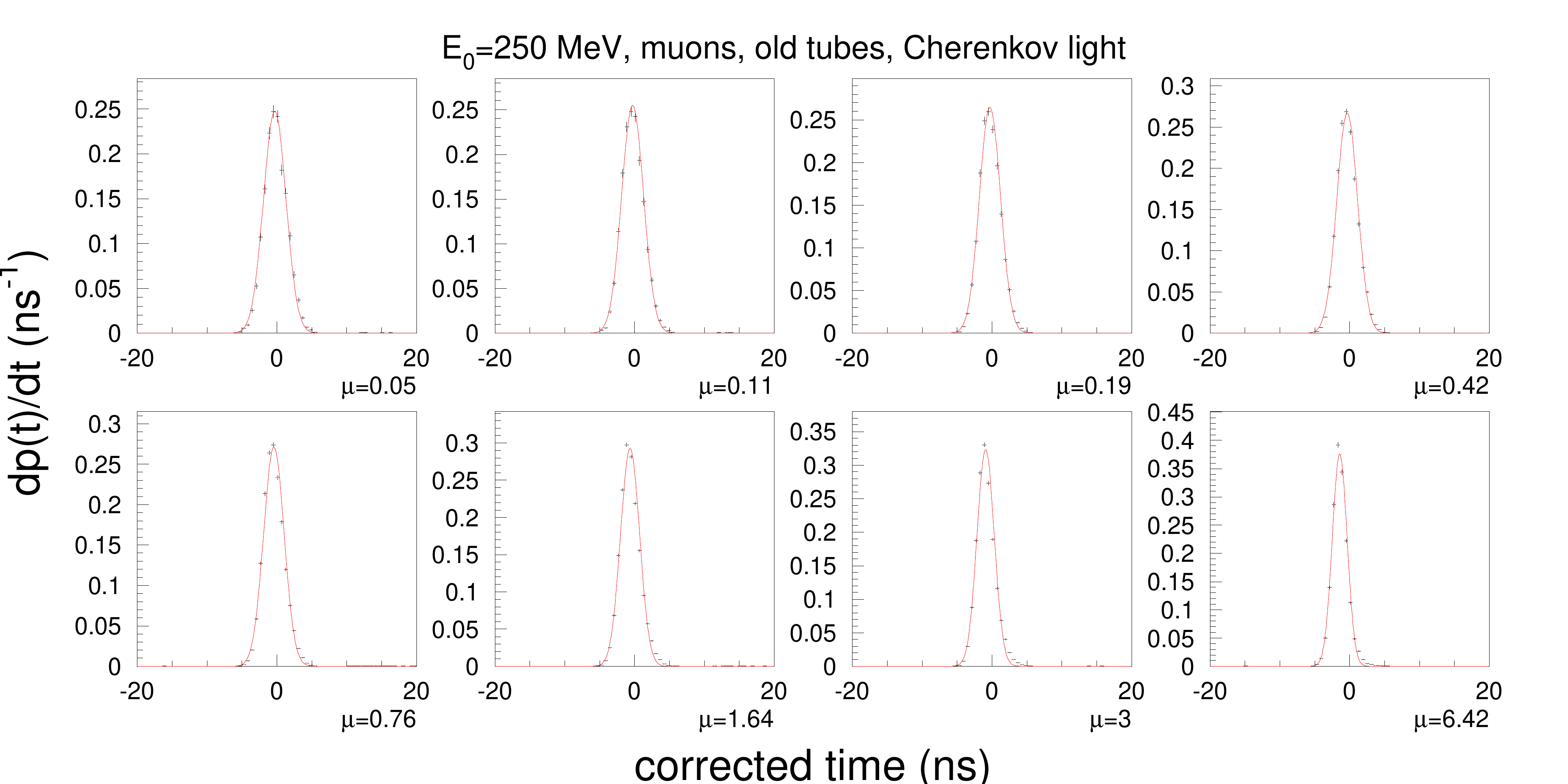} \\  \vspace{0.5 cm}
\includegraphics[width=11 cm] {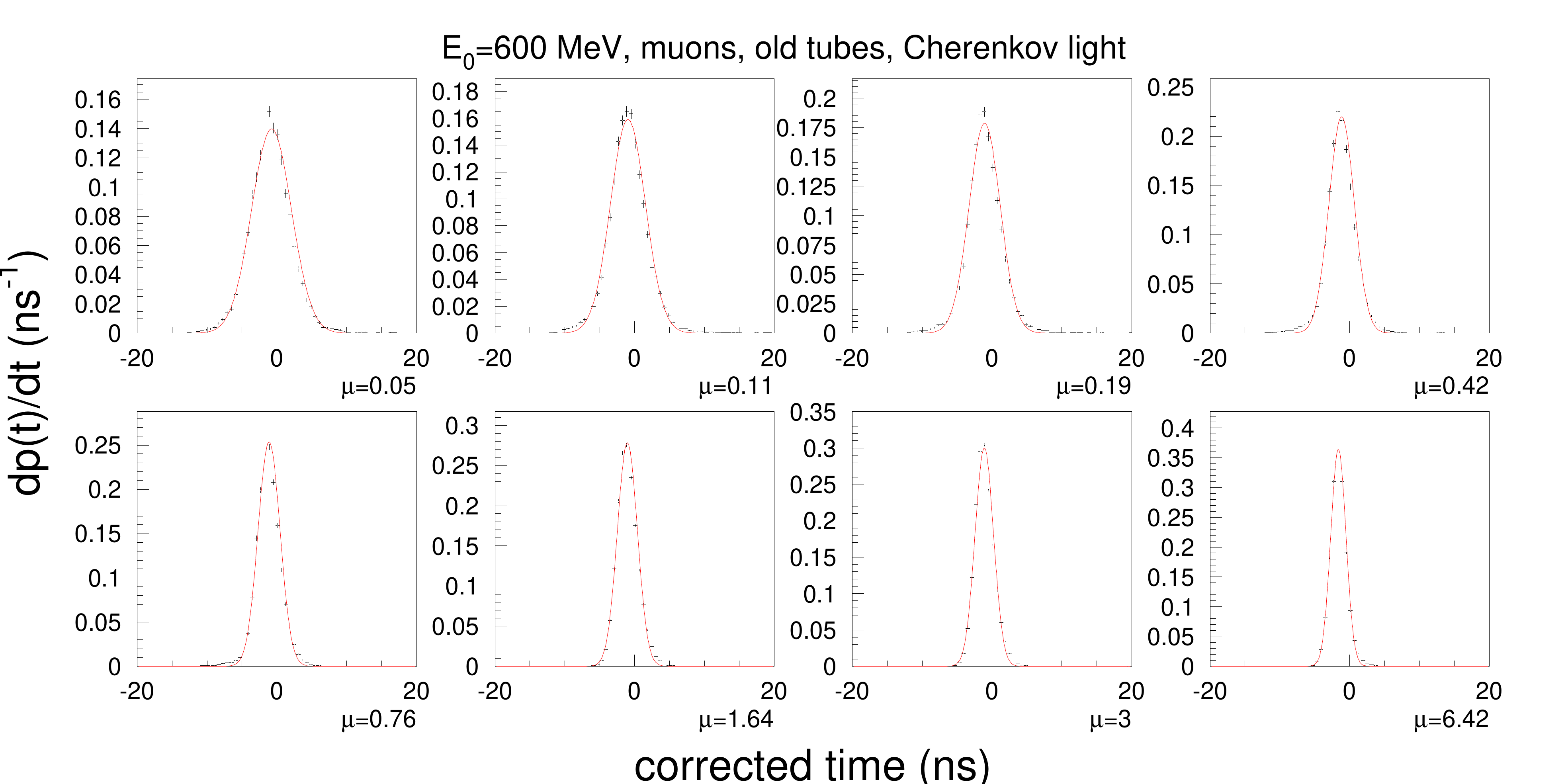} \\  \vspace{0.5 cm}
\includegraphics[width=11 cm] {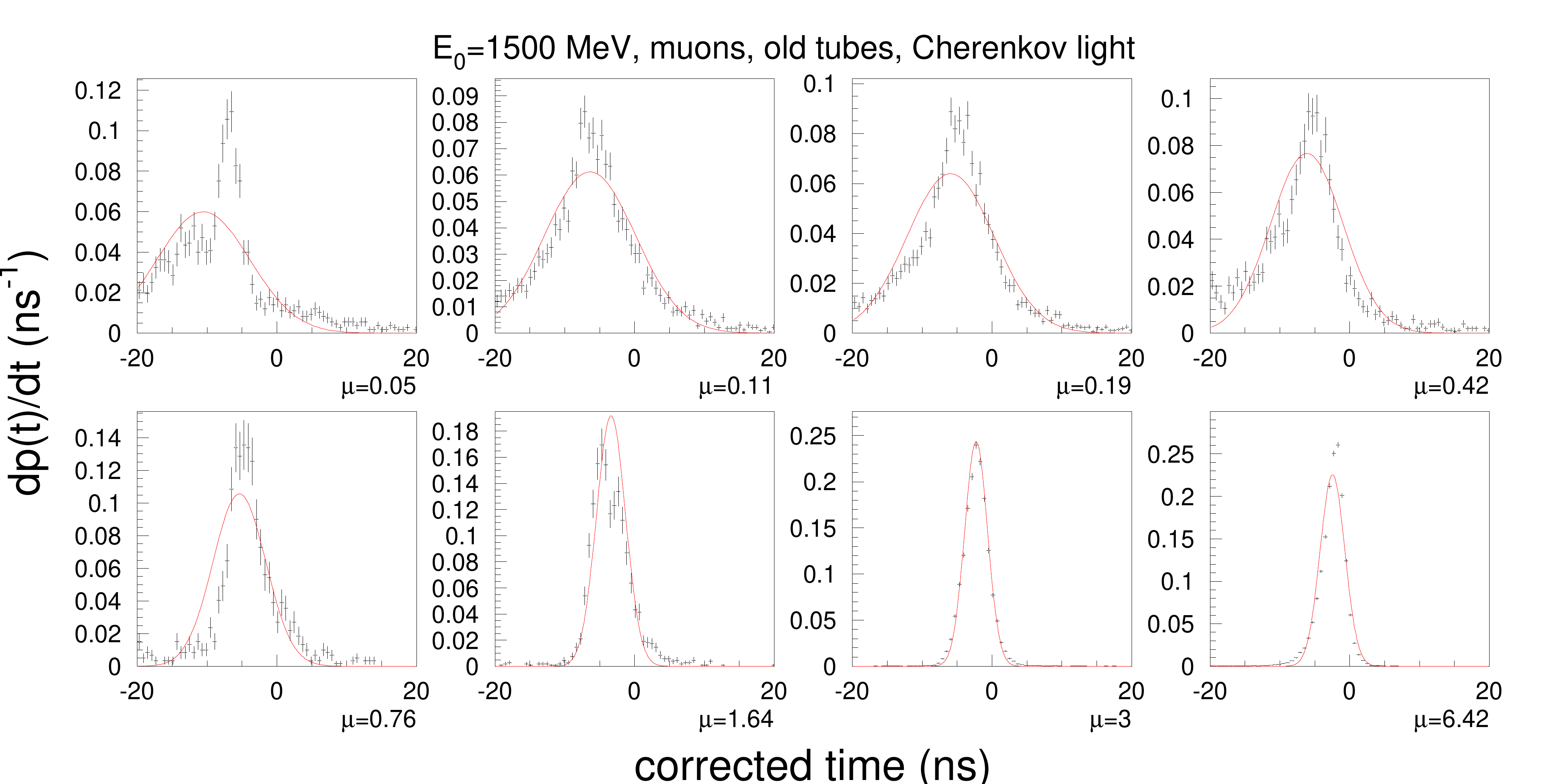}
\caption{\label{fig:tcercheck} Check of primitive distributions. The parametrized $\tc$ likelihood
distributions for Cherenkov light are compared against the actual $\tc$ distributions
for 250 (top), 600 (middle) and 1500 (bottom) $\mev$ muons as a function of predicted
charge.  (Note that the high-$E_0$, low-$\mu$ hits, which are less well modeled, are rare.)}
\end{figure}

The Cherenkov (prompt) primitive distributions are created by simulating particles
throughout the detector with isotropically chosen directions and fixed energy $E_0$. The particles
are created with direct Cherenkov light only; all other sources of light (scintillation, scattering, etc.)
are turned off. For each event,  the true track parameters are used
to evaluate the direct Cherenkov predicted charge $\mu^{\direct}_{\cer}$ for each hit. Histograms
of the corrected time $\tc$ of the hits are produced for various ranges of predicted charge. Figure \ref{fig:tcmu} shows three such $\tc$ distributions for $300\mev$ muons. Since only direct Cherenkov light
is present, the histograms show no late-time features.  The shapes of the time spectra depend
on the $\mu^{\direct}_{\cer}$ values involved, with the distributions becoming narrower and earlier with
increasing predicted charge due to the increasing probability that an early photon will be recorded
at the PMT.  The $\tc$
distribution in each predicted charge range is fit to a Gaussian distribution, and the resulting Gaussian parameters (mean and width) are subsequently parametrized across $\mu^{\direct}_{\cer}$ using a sixth-order polynomial in $\log(\mu^{\direct}_{\cer})$.
Figure \ref{fig:gaussparmudirect2} shows an example of these ``second-level'' fits for 
$300\mev$ muons. The procedure is repeated at many values of $E_0$, with the two second-level fits
providing seven parameters for the Gaussian mean and  seven parameters for the Gaussian width at each 
energy. The energy dependences of these 14 parameters are then fit as a function of $E_0$ to fourth-order
polynomials in a third-level parametrization, 
as shown in Figure \ref{fig:gaussparmudirect3} for the first seven of the second-level parameters. In addition
to conveniently summarizing the dependence of the time response on $\mu_{\cer}^{\direct}$ and $E_0$, the 
parametrizations provide a smooth  likelihood surface, as required by the minimization algorithm.

This completes the prompt primitive distributions $G_{\cer}(\tc;E_0,\mu^{\direct}_{\cer})$. They are calculated for a given predicted charge $\mu^{\direct}_{\cer}$ at  a PMT for a track with energy $E_0$ by evaluating the 14 third-level
curves which parametrize the $E_0$ dependence of the second-level functions. These 14 values then determine
the two functions which parametrize the dependence of the mean and width of the Gaussian parameters on
 $\mu^{\direct}_{\cer}$ at $E_0$. These are evaluated at the particular value of $\mu^{\direct}_{\cer}$ to determine
 the appropriate mean and width of the Gaussian distribution which describes the Cherenkov primitive distribution appropriate for the
 particular values of $\mu^{\direct}_{\cer}$ and $E_0$.  
  The validity of the assumptions and simplifications can be tested directly by comparing the calculated primitive distributions with those actually observed in the Monte Carlo simulated events. Figure 
\ref{fig:tcercheck} shows such a comparison. 

\begin{figure}[t]
\centering
\includegraphics[width=11cm] {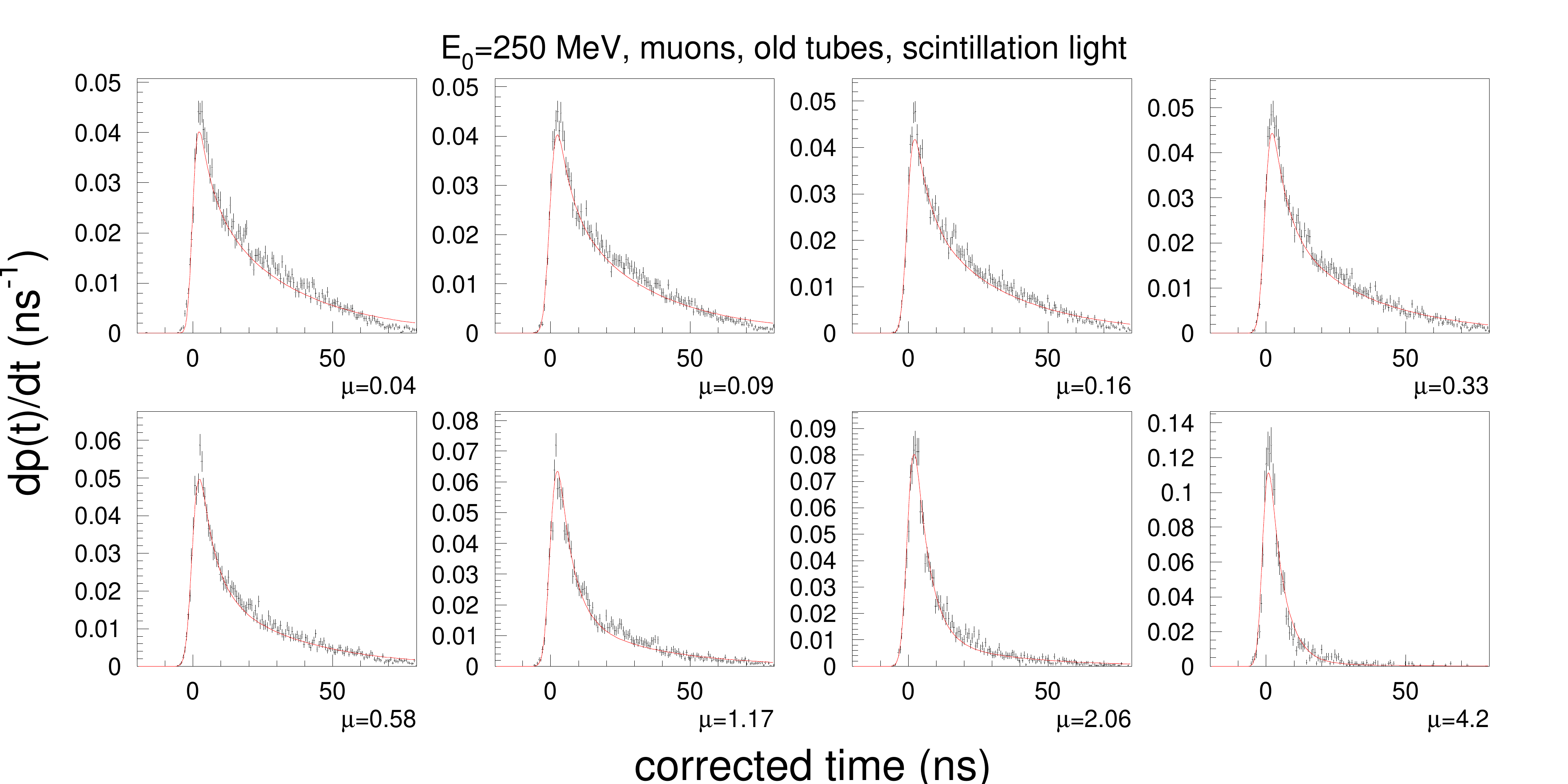} \\  \vspace{0.5 cm}
\includegraphics[width=11 cm] {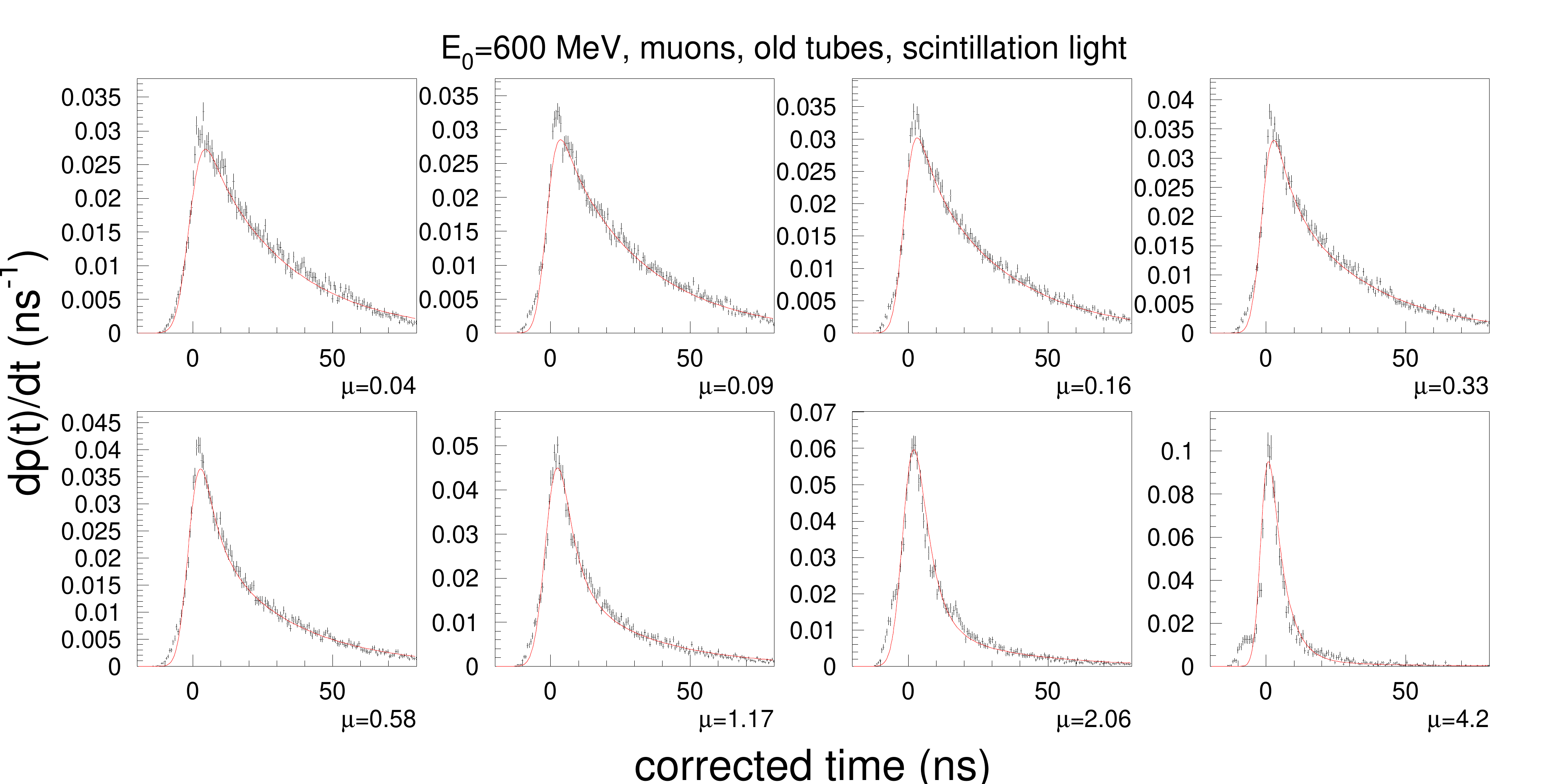} \\  \vspace{0.5 cm}
\includegraphics[width=11 cm] {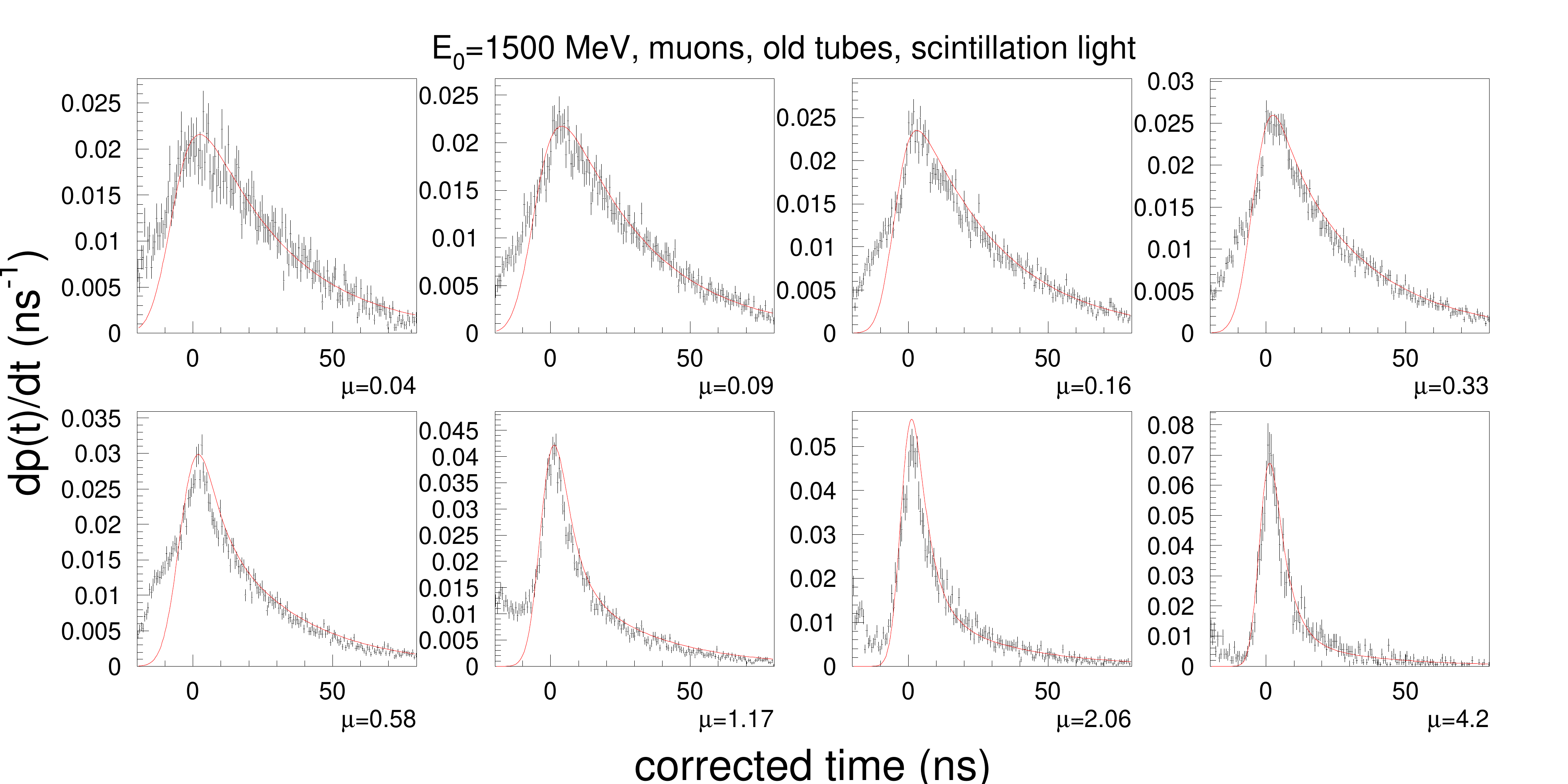}
\caption{\label{fig:tscicheck} Check of scintillation primitive distributions.
 The parametrized $\tc$ likelihood
distributions for scintillation light are compared against the actual $\tc$ distributions
for 250 (top), 600 (middle) and 1500 (bottom) $\mev$ muons as a function of predicted
charge.}
\end{figure}

For the scintillation (late) primitive distributions, events are generated with scintillation light only, and the $\tc$ 
histograms are created in bins of $\mu_{\sci}^{\direct}$. Each $\tc$ histogram is fit to a sum of two exponentials
convolved with a Gaussian distribution. The exponential decay constants are fixed at $\tau_1=5\ns$ and
$\tau_2=30\ns$, leaving three free parameters: the time origin, the Gaussian resolution, and the relative
weight of the two exponentials. At a given $E_0$, the three parameters are extracted in each histogram to obtain their
dependence on $\mu_{\sci}^{\direct}$, which is parametrized as a sixth-order polynomial in $\log(\mu_{\sci}^{\direct})$.
The dependence of the three sets of seven parameters from the polynomials are parametrized as a function of
$E_0$ by fourth-order polynomials, in analogy with the Cherenkov case. We now have $G_{\sci}(\tc;E_0, \mu^{\direct}_{\sci})$, the scintillation primitive distribution.
 Figure \ref{fig:tscicheck} shows the primitive distributions
compared with actual $\tc$ distributions from the Monte Carlo simulation for
muons at $E_0=250$, 600 and $1500\mev$. 

To obtain the complete PDF $f_{\tc}(\tc)$ for a given PMT hit, we first divide the predicted charge into ``prompt" and ``late'' components based on the sources:
\begin{equation}
\begin{array}{ll}
\mu_{\sprompt}& = 0.95 \mu_{\cer}^{\direct} \\
\mu_{\slate} & = 0.05 \mu_{\cer}^{\direct} + \mu_{\cer}^{\indirect} + \mu_{\sci}^{\direct} + \mu_{\sci}^{\indirect}
\end{array}
\label{eq:tlike_1t}
\end{equation}

These definitions of $\mu_{\sprompt}$ and $\mu_{\slate}$ encapsulate the following assumptions. First, all prompt
light is due to direct Cherenkov light; all other light (including indirect Cherenkov light) follows the late
scintillation-based time distribution. The $5\%$ of direct Cherenkov light which is included in
$\mu_{\slate}$ accounts for the late pulsing of PMTs which causes promptly arriving light to
appear at late times. The late time distribution is used to model indirect Cherenkov light since
the fluorescence, scattering, and reflection processes which lead to indirect light have a time
structure similar to that of scintillation.  Further, late timing is less critical to the reconstruction, as prompt light provides essentially all the useful timing information.

The $\mu_{\sprompt}$ and $\mu_{\slate}$ values are used to combine the prompt and late
primitive distributions to form the total PDF $f_{\tc}(\tc)$. In the process, one must account for
the fact that a PMT can register only one hit for a given track due to the latency period of the
electronics. Since the time reported by
the PMT  reflects the time of the first photoelectron, we assume that a PMT hit that includes a prompt 
photoelectron obeys the prompt primitive distribution regardless of how many late photoelectrons may follow. The Poisson distribution gives us the probabilities that no prompt or late photoelectrons
are produced based on their expected mean values:
\begin{equation}
\begin{array}{ll}
P(\mbox{no prompt PEs}) & = e^{-\mu_{\sprompt}} \\
P(\mbox{no late PEs})      & = e^{-\mu_{\slate}} 
\end{array}
\label{eq:propnope}
\end{equation}
With these probabilities, one can calculate the probability that a given hit contains at least
one prompt photoelectron:
\begin{equation}
P(\mbox{prompt PE present}|\mbox{hit}) =  \frac{ 1- P(\mbox{no prompt PEs})}{1-P(\mbox{no prompt PEs})P(\mbox{no late PEs})}
\label{eq:pprompt}
\end{equation}
We assign this as the weight $w_p$ for the prompt primitive distribution within the total PDF,
while $w_l=1-w_p$ is used as the weight for the late primitive distribution, yielding:
\begin{equation}
f_{\tc}(\tc;E_0, \mu_{\sprompt}, \mu_{\slate}) = w_p G_{\cer}(\tc;E_0,\mu_{\sprompt}) + w_l G_{\sci}(\tc;E_0,\mu_{\slate})
\label{eq:ftcweight}
\end{equation}
where $G_{\cer}$ and $G_{\sci}$ are the prompt Cherenkov-based and late scintillation-based timing distributions,
respectively.  Note that Eq.~(\ref{eq:pprompt}) removes the overall probability of a hit occurring. Thus,
even if the absolute probability of a prompt photoelectron is small, $w_p\sim 1$ if $\mu_{\slate} \ll \mu_{\sprompt}$.

\begin{figure}[t]
\centering
\includegraphics[width=13cm] {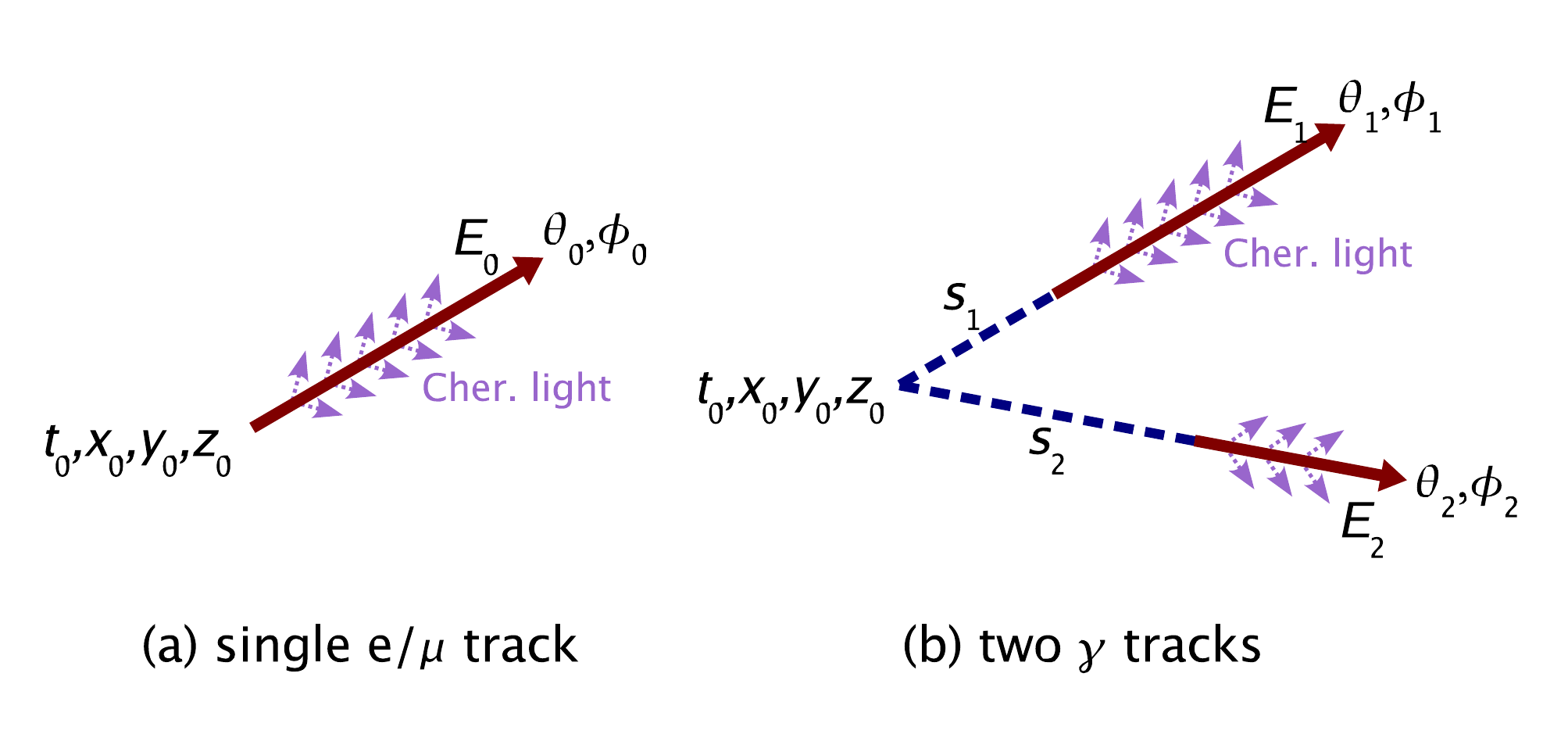} 
\caption{\label{fig:onetwotrack} Internal fit parameters for (a) a single muon or electron track and (b)
two photon tracks. Each photon track includes a conversion distance parameter $s$. The two-track
parameters can be constrained such that the invariant mass of the two photons $M_{\gamma\gamma}$
is always $M_{\piz}$.}
\end{figure}

%% file: twotrack.tex
\section{The Two-track Reconstruction}
\label{sec:twotrack}
The single track model with seven parameters is sufficient for reconstructing  $\num$ and $\nue$
CCQE events as well as cosmic muons or their decay electrons. As discussed in 
Section \ref{sec:nuint}, NC $\piz$ events require a two-track model. Figure \ref{fig:onetwotrack}
shows the 12 parameters needed to describe two photon tracks originating from a common
vertex. The electron track model serves double-duty as the photon track model, under the assumption that
a showering electron is indistinguishable from a showering photon apart
from the conversion distance.  (The mean conversion length in Marcol 7 is 67 cm.) While the two-track model is conceptually a straightforward extension of the one-track case, complexities of the likelihood space require special treatment in the likelihood maximization algorithm.

\subsection{Two-track Charge Likelihood}
Since the charge likelihood $F_q$ depends only on the measured charge $q$ and
the total predicted charge $\mu$ at each PMT, the two-track charge likelihood can be
obtained by adding together the predicted charges from the two tracks to obtain
the total predicted charge at each PMT:
\begin{equation}
\mu = \mu_{\mbox{\scriptsize{track 1}}} + \mu_{\mbox{\scriptsize{track 2}}}
\end{equation}

\subsection{Two-track time likelihood}
The one-track time likelihoods account for the ``first photoelectron only'' nature of the electronics by calculating the probability that a prompt photoelectron is present and by then weighting the prompt primitive distribution's influence on the full time PDF accordingly.  This scheme is extended to handle two tracks as follows.

The two single-track primitive distributions, $G^{i}_{\cer}$ and $G^{i}_{\sci}$ are formed for each track,
where $i$ labels the track number. In anticipation of aggregating all late light later, the late ``scintillation''
primitive distributions are averaged to form:
\begin{equation}
\hat{G}_{\sci} = \frac{1}{2}\left( G^1_{\sci} + G^2_{\sci} \right)
\end{equation}
Of the two tracks, one will have a midpoint that is nearer to the target PMT than the other.
This track's quantities are labeled with ``n'' (near); the other's, ``f'' (far).
In analogy with Eq.~(\ref{eq:tlike_1t}), we define:
\begin{equation}
\begin{array}{ll}
\mu_{\sprompt,\sn}& \equiv  0.95\; \mu_{\cer,\sn}^{\direct} \\
\mu_{\sprompt,\sf}& \equiv  0.95\; \mu_{\cer,\sf}^{\direct} \\
\mu_{\slate} & \equiv \mu_{\stot} -  \mu_{\sprompt,\sn} - \mu_{\sprompt,\sf}
\end{array}
\label{eq:tlike_2t}
\end{equation}
The probabilities of not obtaining a photoelectron from these sources are:
\begin{equation}
\begin{array}{lll}
\overline{P}_{\sn} \equiv  & P(\mbox{no prompt PE from near track}) & = e^{-\mu_{\sprompt,\sn}} \\
\overline{P}_{\sf} \equiv  & P(\mbox{no prompt PE from far track}) & = e^{-\mu_{\sprompt,\sf}}        \\
\overline{P}_{\sl} \equiv  & P(\mbox{no late PEs}) & = e^{-\mu_{\slate}}  \\
\end{array}
\end{equation}
from which the following weights are obtained:
\begin{equation}
\begin{array}{ll}
w_{\sn} & = \displaystyle{\frac{1-\overline{P}_{\sn}}{1 - \overline{P}_{\sn} \overline{P}_{\sf} \overline{P}_{\sl}} } \\
  & \\
w_{\sf} & = \displaystyle{\frac{1-\overline{P}_{\sf}}{1 - \overline{P}_{\sf} \overline{P}_{\sl}} (1-w_{\sn})   }\\
&  \\
w_{\sl} & = 1-w_{\sn} - w_{\sl}
\end{array}
\end{equation}
where $w_{\sn}$ is the probability that a prompt photoelectron from the near track exists given
that any photoelectron exists, {\em etc.} The weights are used to combine the two prompt
primitive distributions and the averaged late primitive distribution:
\begin{equation}
f_{\tc}(\tc) = w_{\sn} \; G_{\cer,\sn}(\tc) + w_{\sf} \; G_{\cer,\sf} (\tc) + w_{\sl} \; \hat{G}_{\sci}(\tc)
\end{equation}
resulting in the complete two-track corrected time PDF.

\section{Minimization of $-\log(\like)$}
\subsection{One-track}
With the likelihood $\like$ defined, the parameter set $\vecx$ that minimizes its negative
logarithm $F\equiv -\log(\like)$ must be found. In the single track minimization process,
two issues arise.

First, the energy parameter $E_0$ is tied to the geometry of the event via the track profiles
$\rho(s;E_0)$. If the spatial parameters ($x_0$, etc) are varied simultaneously
with $E_0$, the minimization algorithm can get confused by this correlation. The solution
is to iterate the minimization process, as described below.

A second issue arises from the discrete PMT lattice, which imprints a small fluctuating
signal on $\like(\vecx)$ despite the smooth parameterizations of the input tables.  Minimization
algorithms that rely on gradients or that do not have controllable step sizes are thus poorly
suited to the problem. Therefore, the SIMPLEX method in Minuit\cite{minuit} is chosen over the MIGRAD method.

To initiate the minimization process, a vector of seed parameters is obtained from a fast fitter. In the first iteration of the minimization, the energy $E_0$ is held at its seeded 
value while the six remaining parameters ($x_0$, $y_0$, $z_0$, $t_0$, $\theta_0$, $\phi_0$)
are varied via Minuit/SIMPLEX to find a temporary minimum of $F$.  These six parameters are then
fixed, while $E_0$ is freed for a second iteration. Finally, $E_0$ is fixed once more with 
the six other parameters varied (as in the initial iteration), to find the final minimum of F. The parameters
from the final SIMPLEX call are returned as the best-fit set $\bf{x}$.

\begin{figure}[t]
\centering
\includegraphics[width=13cm] {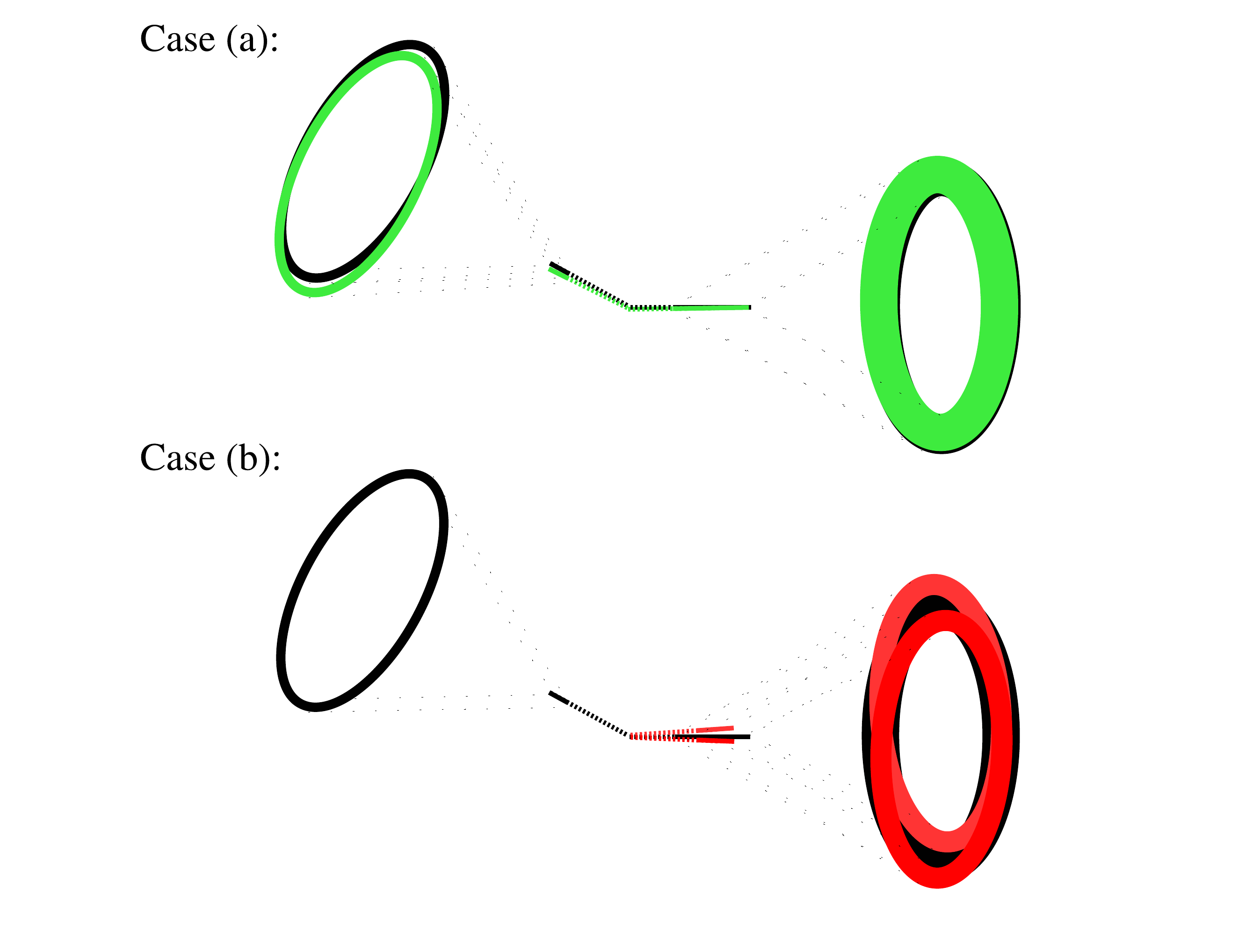} 
\caption{\label{fig:twotrackseed}The two $\pi^0$ fitter configurations discussed in the text.}
\end{figure}

\subsection{Two-track}
\label{sec:twotrackmin}
Figure \ref{fig:twotrackseed} shows two situations that may arise in fitting a $\piz$ event.
The event has two photon tracks whose Cherenkov rings are represented in black, while
green and red represent two ring configurations corresponding to different
starting values for the parameters. In case (a), the 12 parameters are near the correct
configuration. In case (b), the parameters are such that both tracks are directed toward the
dominant ring.  The latter case corresponds to a local minimum which may offer
a worse likelihood than case (a); however, the minimization algorithm may be trapped by the
fact that any small changes to the parameters result in an increased value of $F$. For example,
sweeping one track over to the smaller black ring involves passing through a region with little
detected Cherenkov light.  The intermediate states form a barrier of disfavored values of $F$
that the minimization algorithm code will have difficulty overcoming, especially
if several parameters must be adjusted simultaneously even to make the distant configuration
an improvement. 

The two-track fits require a minimization approach that avoids these traps in the likelihood
surface. Monte Carlo $\piz$ events were used to identify both common and rare
trapping scenarios, and the minimization algorithm addresses each of these through the following collection of seed configurations and minimization sequences:n
\begin{itemize}
\item The conversion lengths $s_1$ and $s_2$ are seeded with either 50 cm or 250 cm, leading
to four possible pairs.
\item $\theta_1$ and $\phi_1$ are seeded with the results from the electron single-track fit
or with one of the eight perturbations (see below).
\item$\theta_2$ and $\phi_2$ are seeded with the ``best'' directions from a full grid
of tested directions (see below).
\item The four-vertex of the event is seeded with the four-vertex from the electron
single-track fit shifted according to $s_1$, $\theta_1$ and $\phi_1$.
\item $E_1$ is seeded with approximately $E_0$ from the electron single-track
fit, while the seed for $E_2$ is, when possible, the energy needed to give an
invariant mass equal to $M_{\piz}$. 
\end{itemize}

For the last item, the energy seeds are based on an
empirical expression that accounts for the second ring's energy contribution to the one-track fit
energy as a function of the angle between the two $\gamma$ tracks. The function is obtained by
simulating NC $\piz$ events and reconstructing them with the single track electron fit. Using
the true photon energies ($E_1>E_2$) and opening angle 
$\theta_{\gamma\gamma}$, the relationship between $(E_0 - E_1)/E_2$ and 
$\cos\theta_{\gamma\gamma}$ is parametrized by a linear function. This function is used
to determine the seed $E_1$ and $E_2$ values for a given $E_0$ and 
$\cos\theta_{\gamma\gamma}$.

The nine ($1\mathord{+}8$) possible seed directions for track 1 are created as follows. The spatial distribution
of PMT charge is projected onto a plane perpendicular to the electron fit direction. The 
major axis of the covariance ellipse of the resulting two-dimensional distribution is found.
The nine possible seed directions are rotations of the best electron-fit direction by
$0$, $\pm 0.159$, $\pm 0.450$ and $\pm 0.644$ radians parallel to the major axis
and $\pm 0.159$ radians perpendicular to the major axis.

The seed directions for track 2 come from either a ``coarse'' $24\mathord{\times} 12$ $(\theta,\;\phi)$ grid or a ``fine'' $50\mathord{\times} 25$ $(\theta,\;\phi)$ grid. The grid type is specified at run time.

Of the $9\mathord{\times}(24\mathord{\times}12)\mathord{=}2592$ or $9\mathord{\times}(50\mathord{\times}25)\mathord{=}11250$ possible combinations of track 1 and track 2 seed directions, only six are used to seed actual minimization sequences.  They are the six combinations with (1) the best total likelihood (time and charge), (2) the best charge likelihood when the estimated track energies\footnote{Recall the energy seed procedure discussed above.} are similar ($0.5<\frac{E_1}{E_2}<2$), (3) the best charge likelihood when the estimated track energies are dissimilar, and (4)$-$(6) the second-best likelihood in each of these three cases.  A small exception: combinations involving one of the eight perturbed track 1 directions {\em and} having estimated track energies that differ by more than a factor of five are not considered.  In these energetically asymmetric cases, the electron fit direction would be a good representation of track 1 since the influence of track 2 would be small, so there is no reason to consider alternate track 1 directions.

The six best and second-best track 1 and track 2 direction and energy combinations are chosen anew for each permutation of $(s_1, s_2)$, leaving 24 complete parameter sets in total.  Each set seeds two Minuit sequences.  In sequence 1, the SIMPLEX minimization routine is first run with the energy and angle parameters held fixed; once a minimum is found, SIMPLEX is called again with all parameters free.  In sequence 2, only the energy parameters are held fixed in the initial SIMPLEX call.  This results in 48 fit sequences, each of which ends with a completely free SIMPLEX minimization.  The best parameter set ({\em i.e.}, the one that results in the lowest $F$) from the forty-eight final SIMPLEX calls is reported as the best parameter set ${\bf x}$ for the two-track fit. During all minimizations, the photon conversion points are constrained to be within the main detector volume. This prevents the fit from removing the influence of a track through an inflated conversion length. The event vertex itself, however, is not constrained.

The SIMPLEX minimizations can be performed with a constraint on the invariant mass. This is accomplished by
removing $E_2$ as a free parameter and setting it via the
relation:
\begin{equation}
E_2 = \frac{M^2_{\piz}}{2 E_1 (1-\cos\theta_{\gamma\gamma})}
\end{equation}
where $\theta_{\gamma\gamma}$ is the angle between the photon tracks. This fixed-mass mode is the actual 
$\piz$ hypothesis, while the free-mass mode allows for mass reconstruction. Both are used for $\piz$
identification, with the former lending its maximized likelihood $\like_{\piz}$ and the latter providing
a reconstructed mass $M_{\gamma\gamma}$.  

%% file: performance.tex
\section{Performance}
\label{sec:performance}
\begin{figure}[hbtp]
  \begin{center}
  \includegraphics[width=0.45\textwidth]{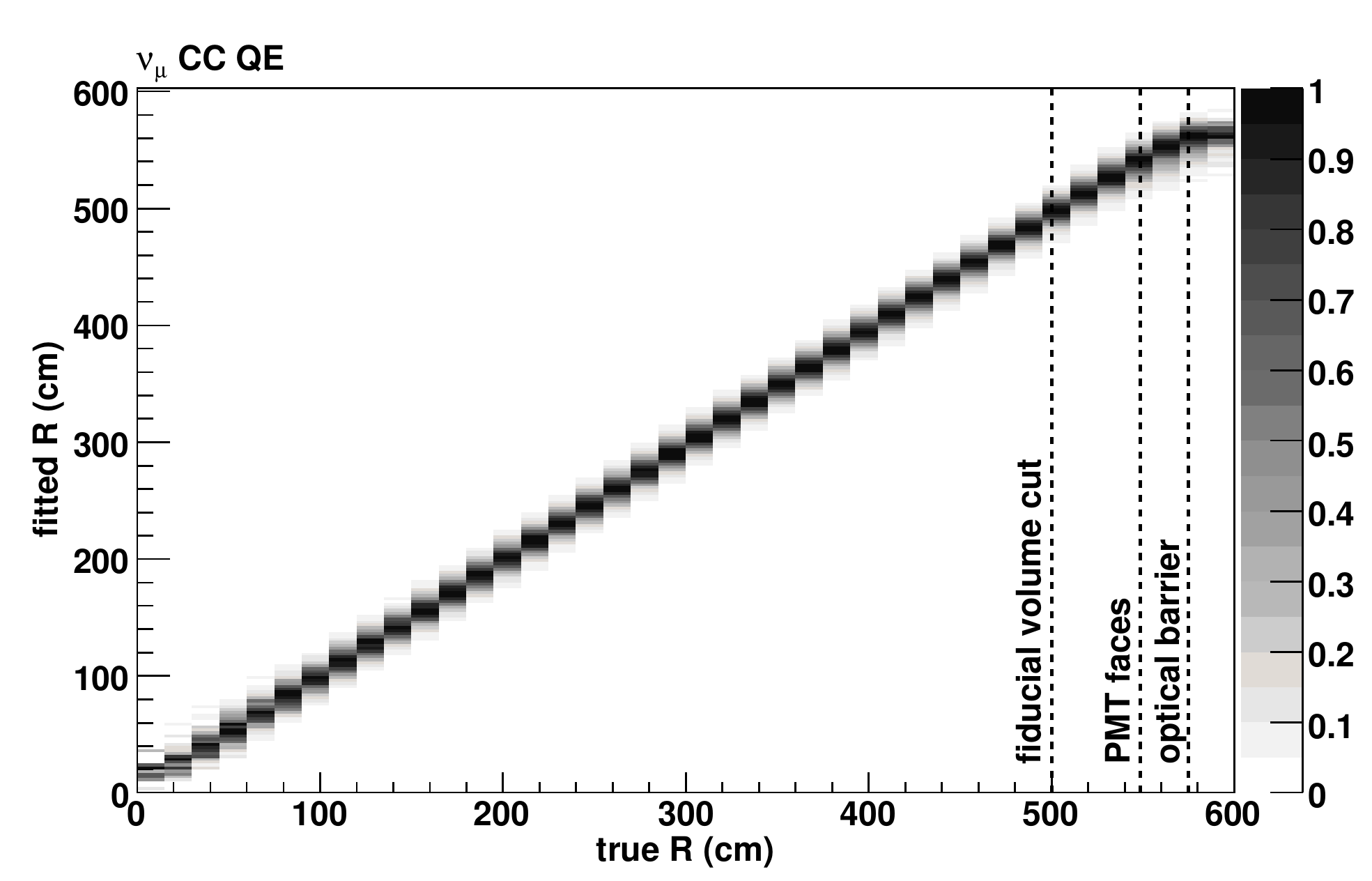}\hspace{0.05\textwidth}
  \includegraphics[width=0.45\textwidth]{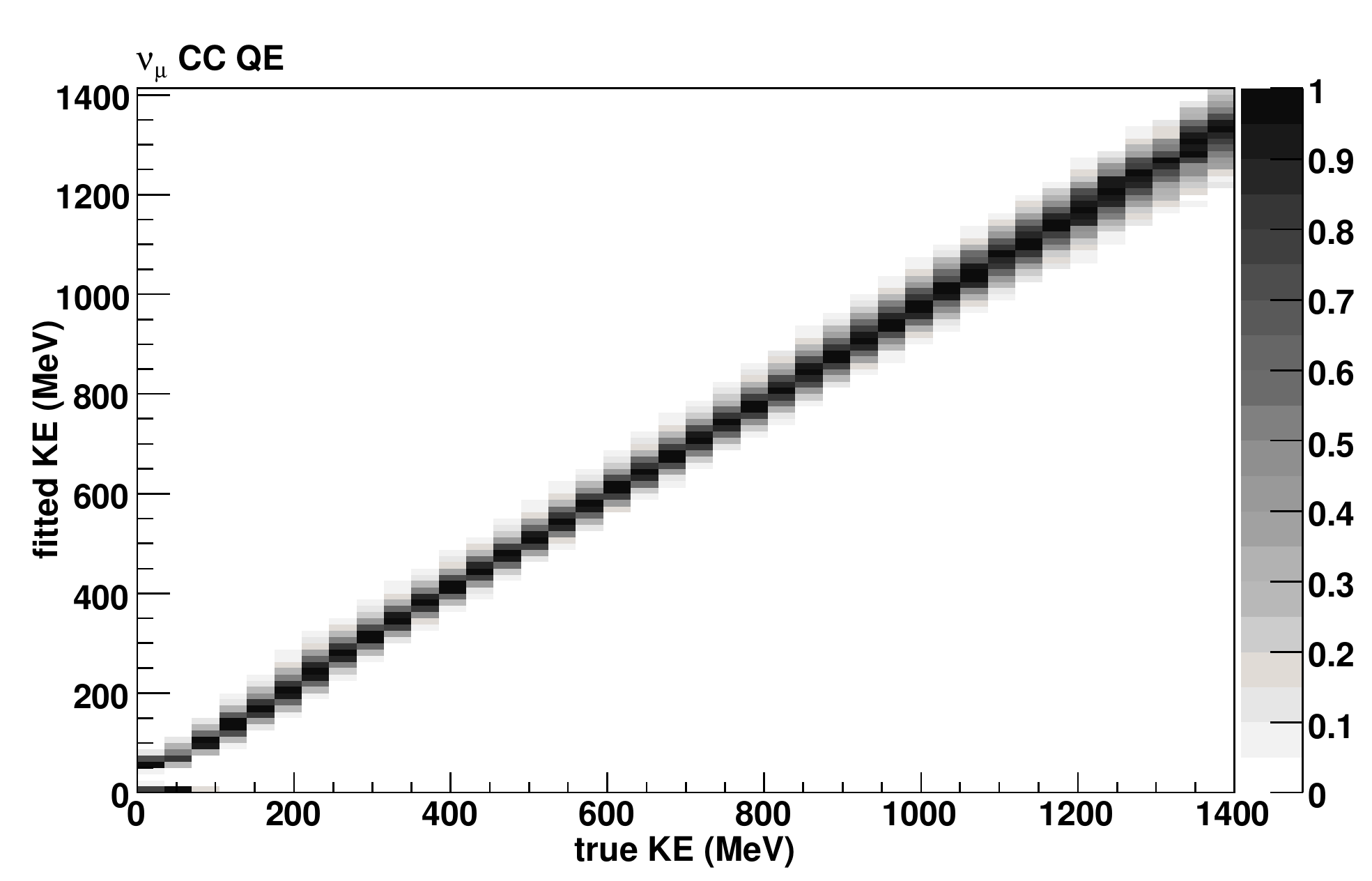}\\
  \includegraphics[width=0.45\textwidth]{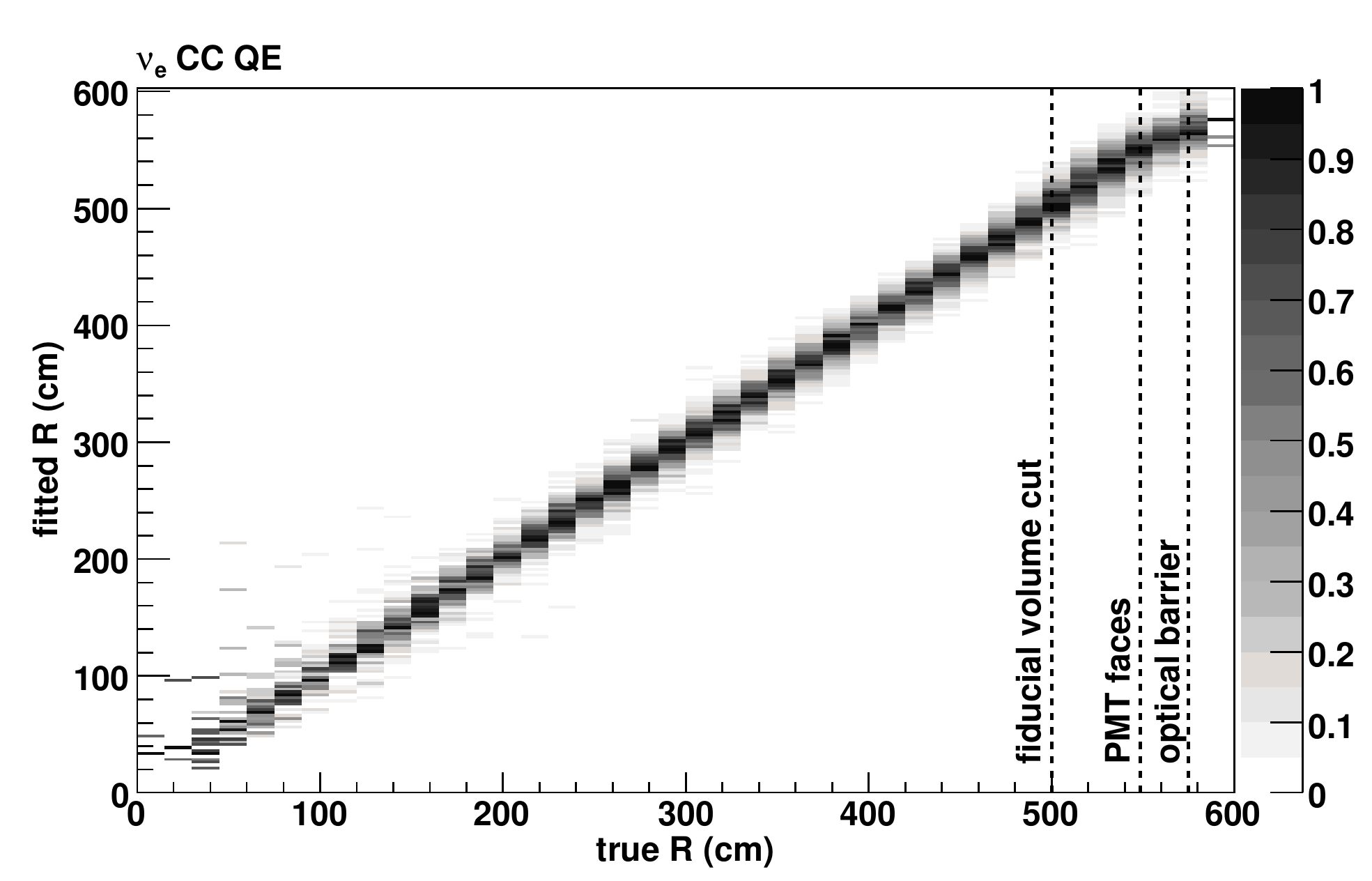} \hspace{0.05\textwidth}
  \includegraphics[width=0.45\textwidth]{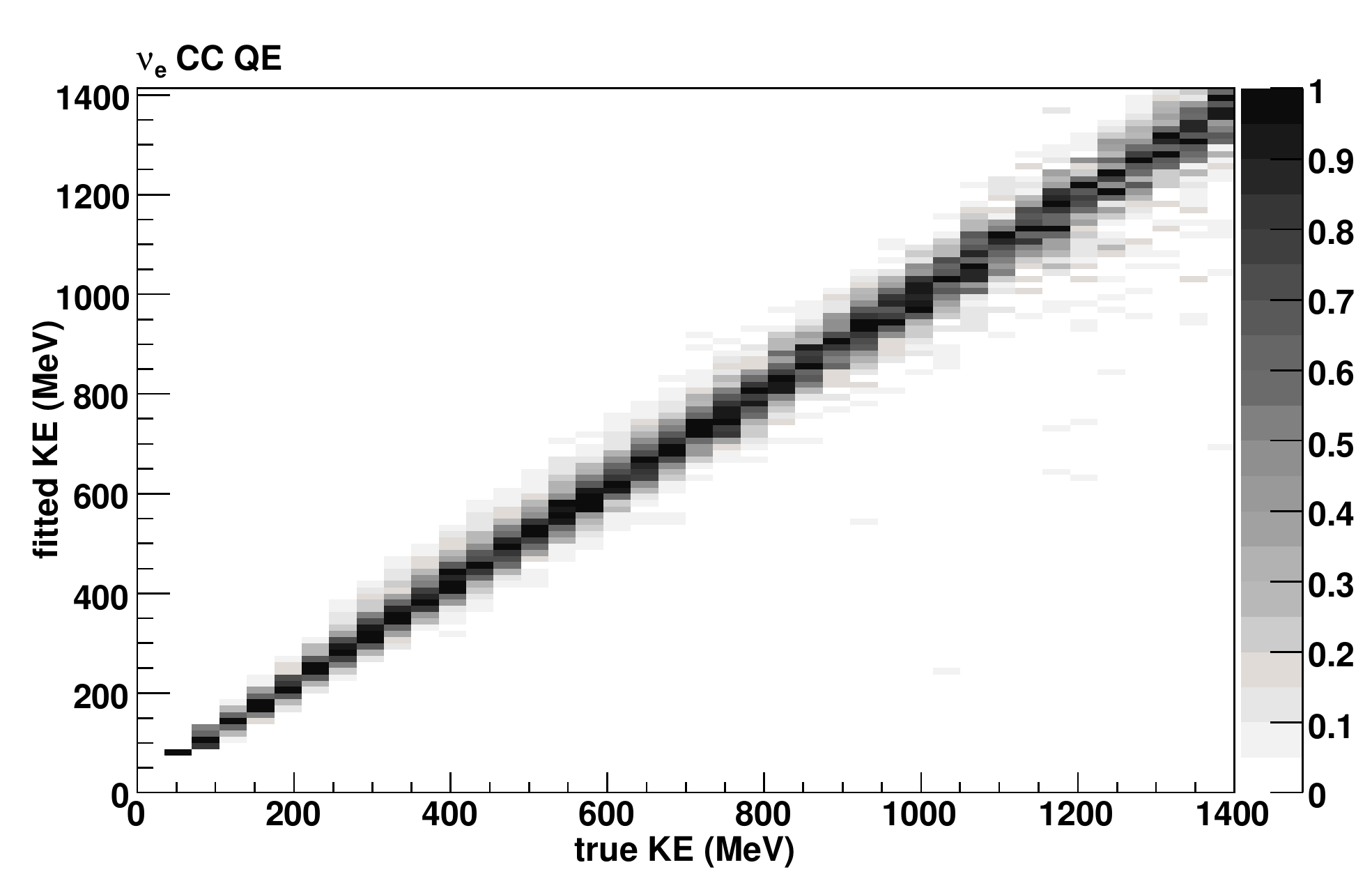} \\
  \includegraphics[width=0.45\textwidth]{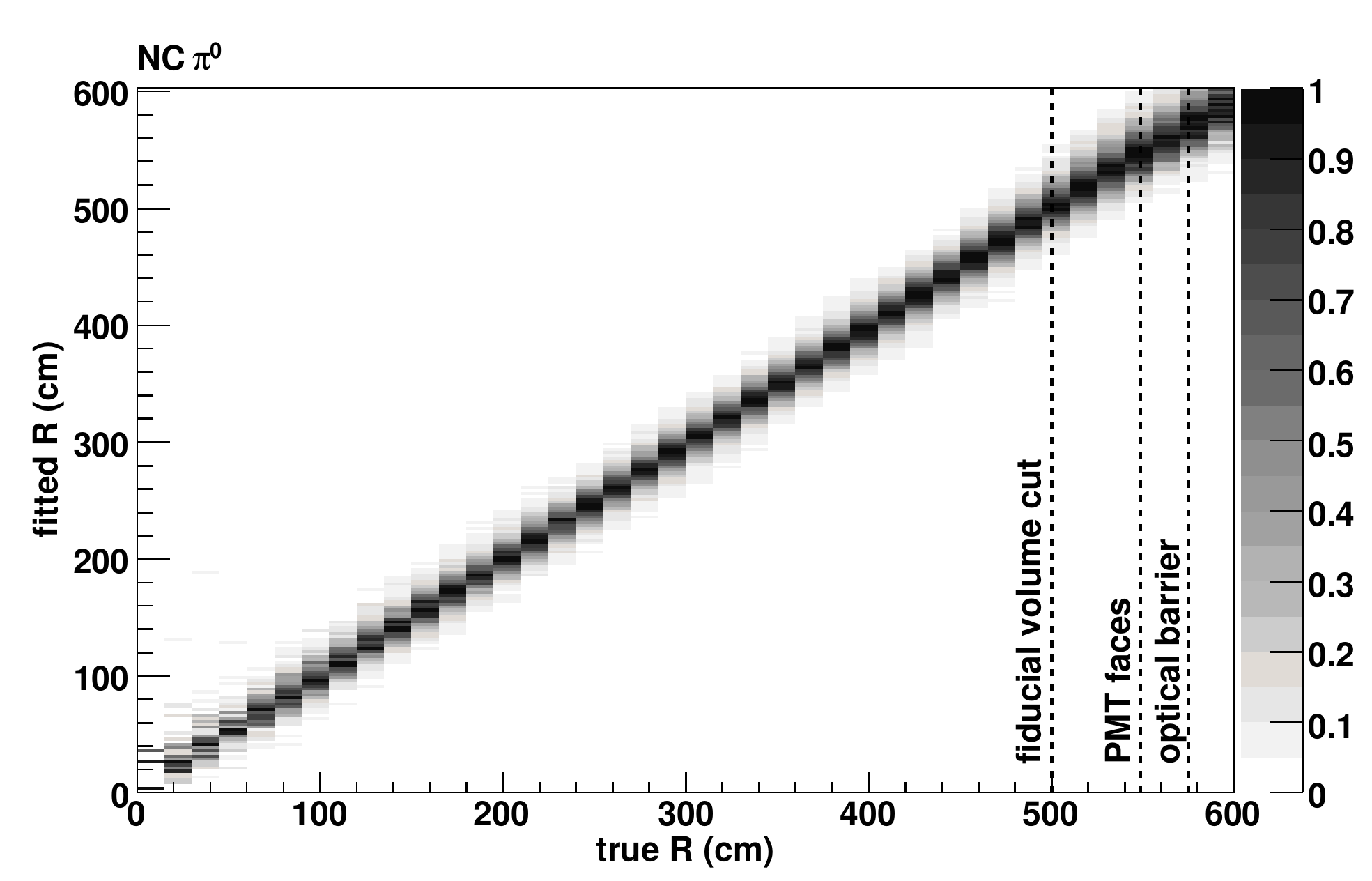}\hspace{0.05\textwidth}
  \includegraphics[width=0.45\textwidth]{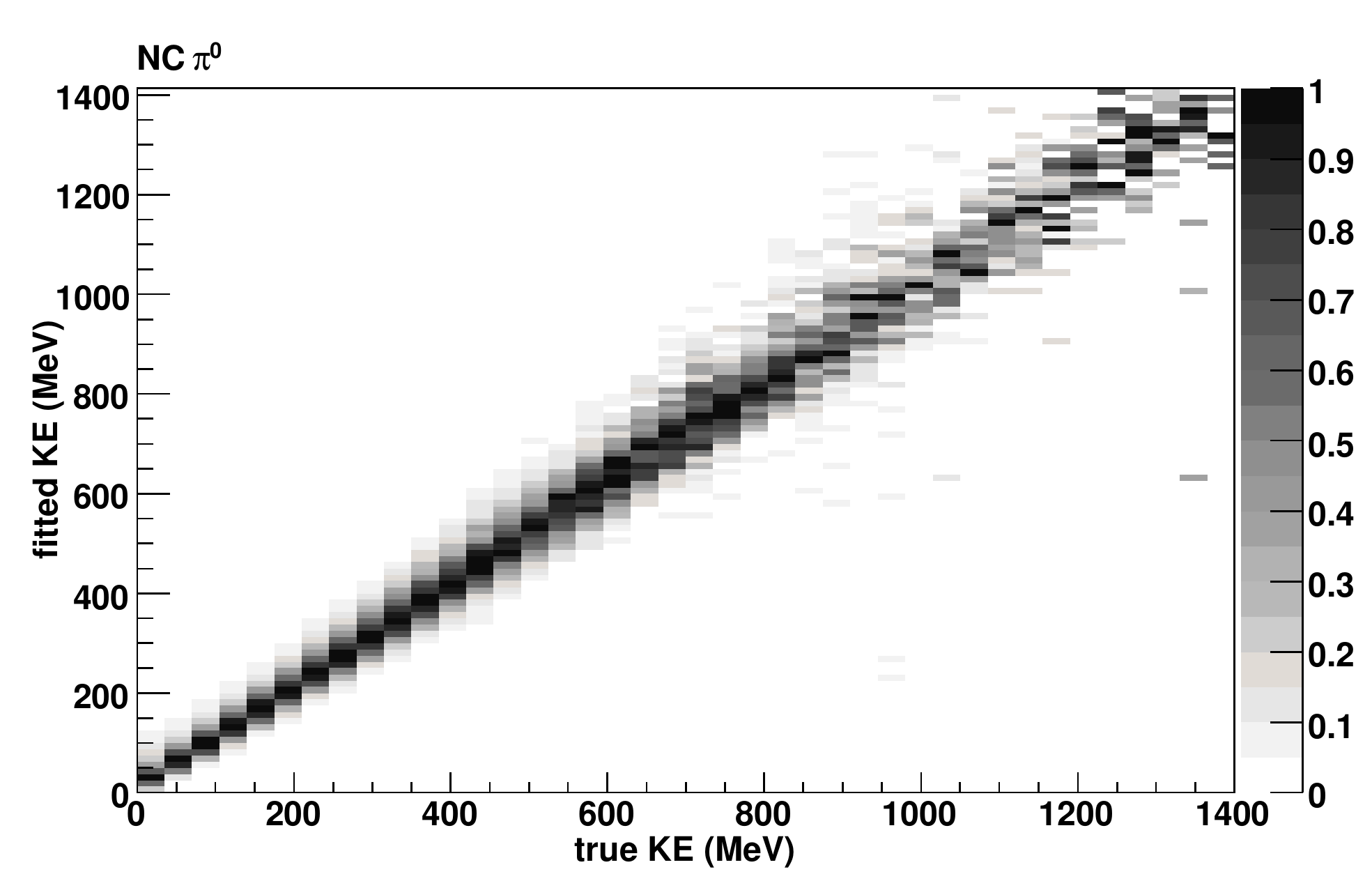} \\

  \caption{\label{fig:rVr} Reconstructed radial event position (left) and kinetic energy (right) plotted against their true values. The top, middle, and bottom panels show $\num$ CCQE, $\nu_e$ CCQE, and NC $\pi^0$ events.}
  \end{center}
\end{figure}

\begin{figure}[hbtp]
  \begin{center}
  \includegraphics[width=0.7\textwidth]{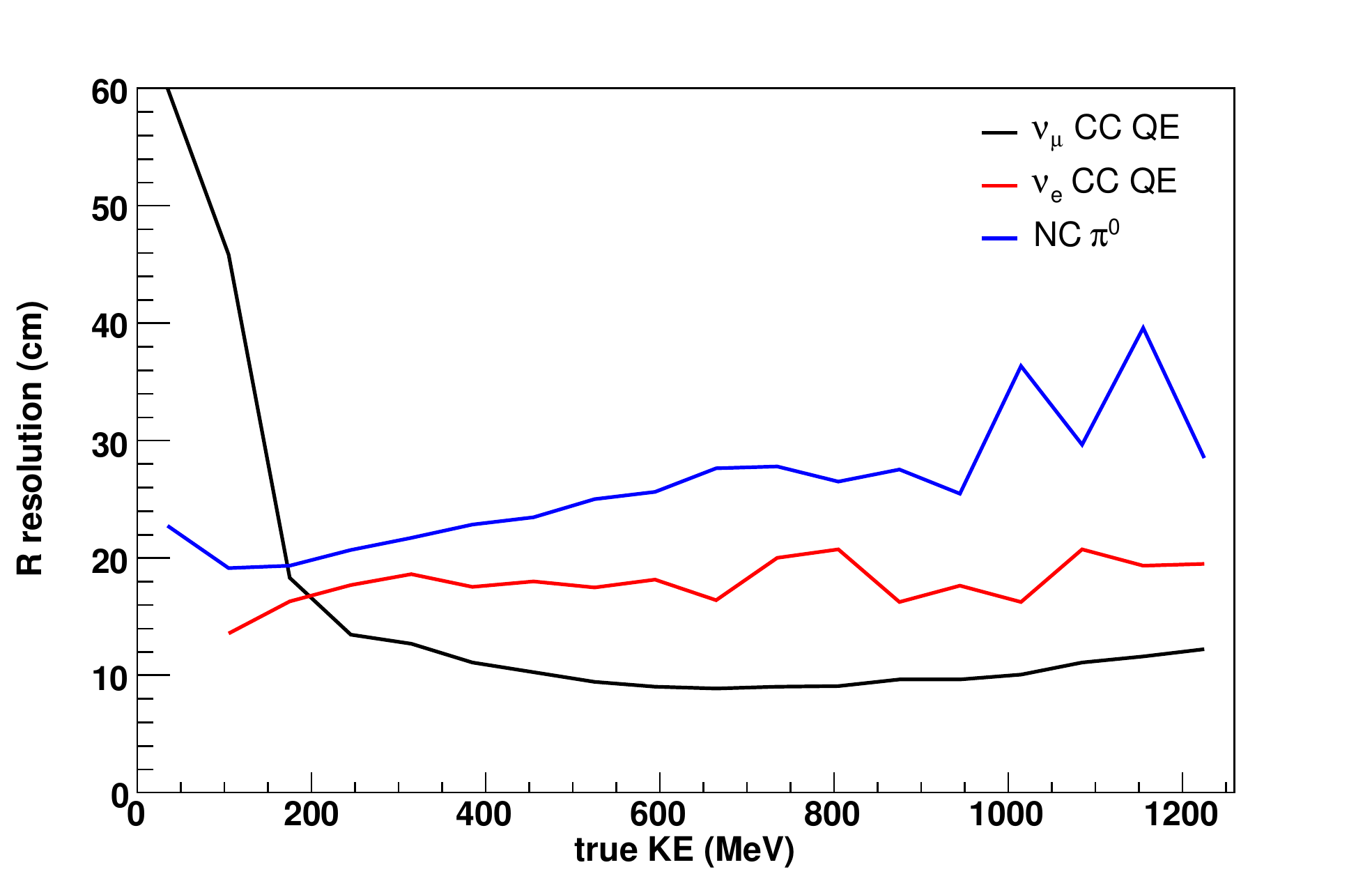}
  \includegraphics[width=0.7\textwidth]{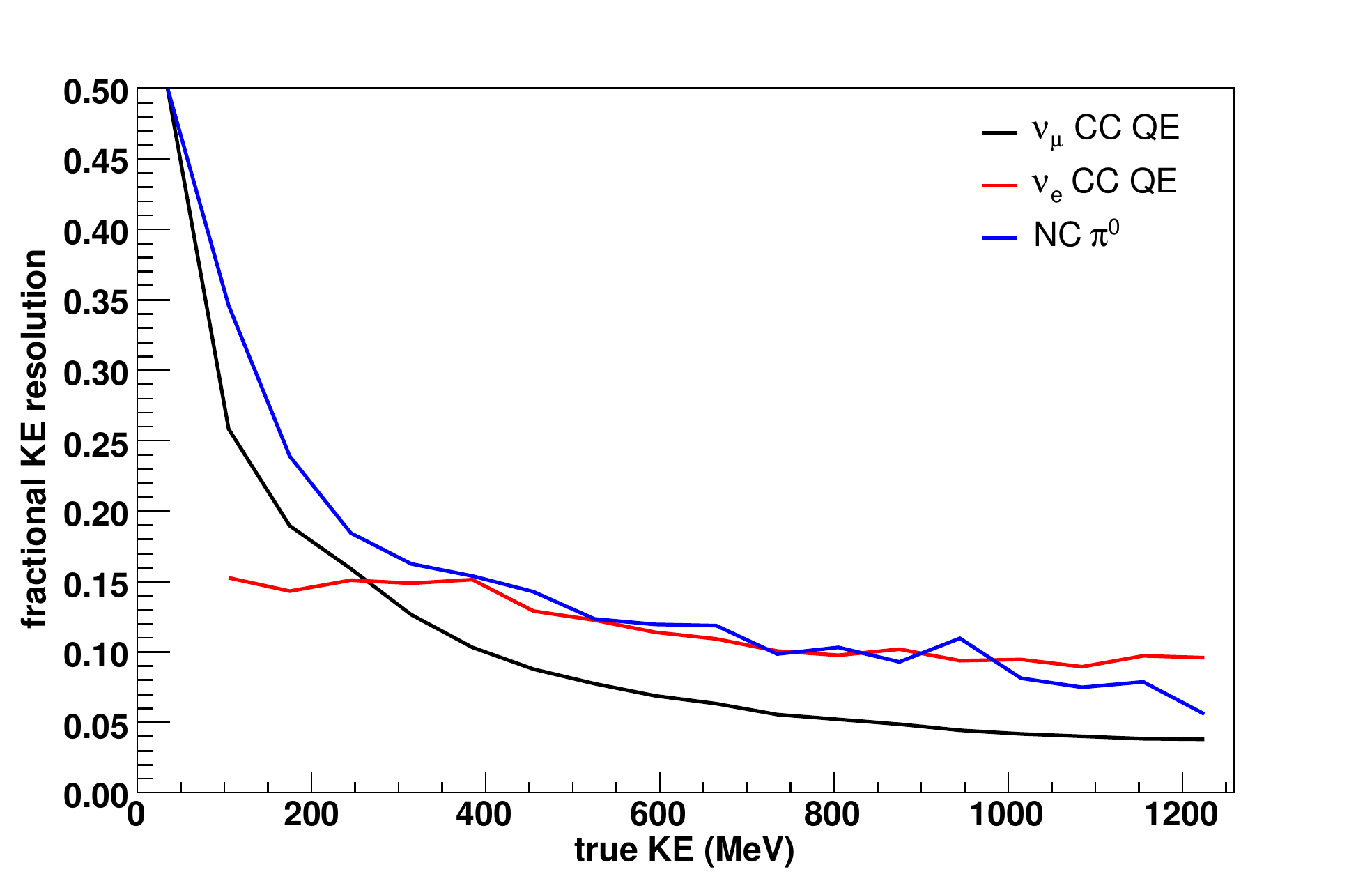}
  \includegraphics[width=0.7\textwidth]{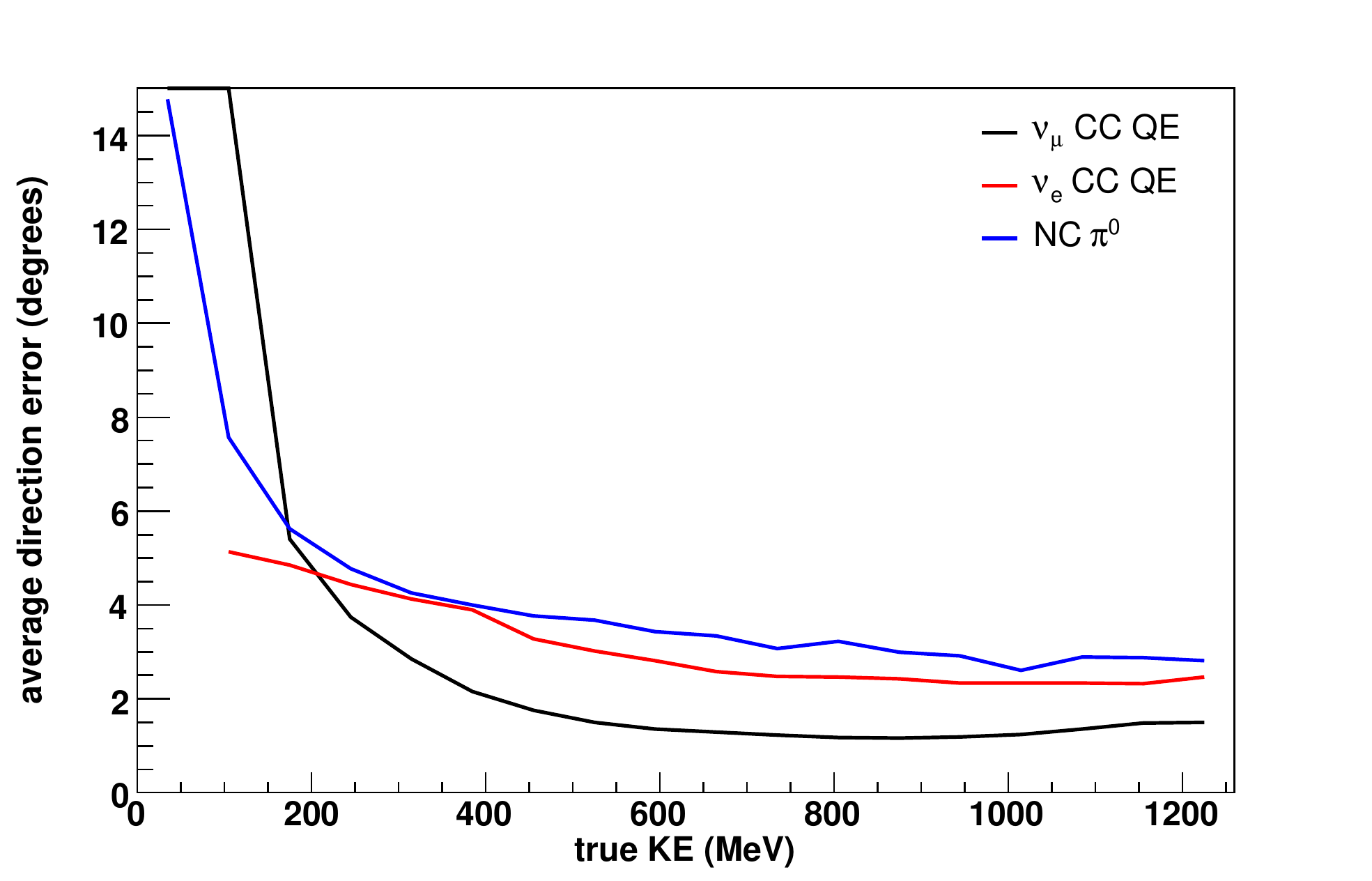}
  \caption{\label{fig:resolutions} Radius, kinetic energy, and direction resolutions.  The jitter is due to limited test sample statistics in some regions.  The \nue\ curves cut off at 100~MeV due to an unrelated upstream cut on the Monte Carlo sample.}
  \end{center}
\end{figure}

\begin{figure}[hbtp]
  \begin{center}
  \includegraphics[width=0.45\textwidth]{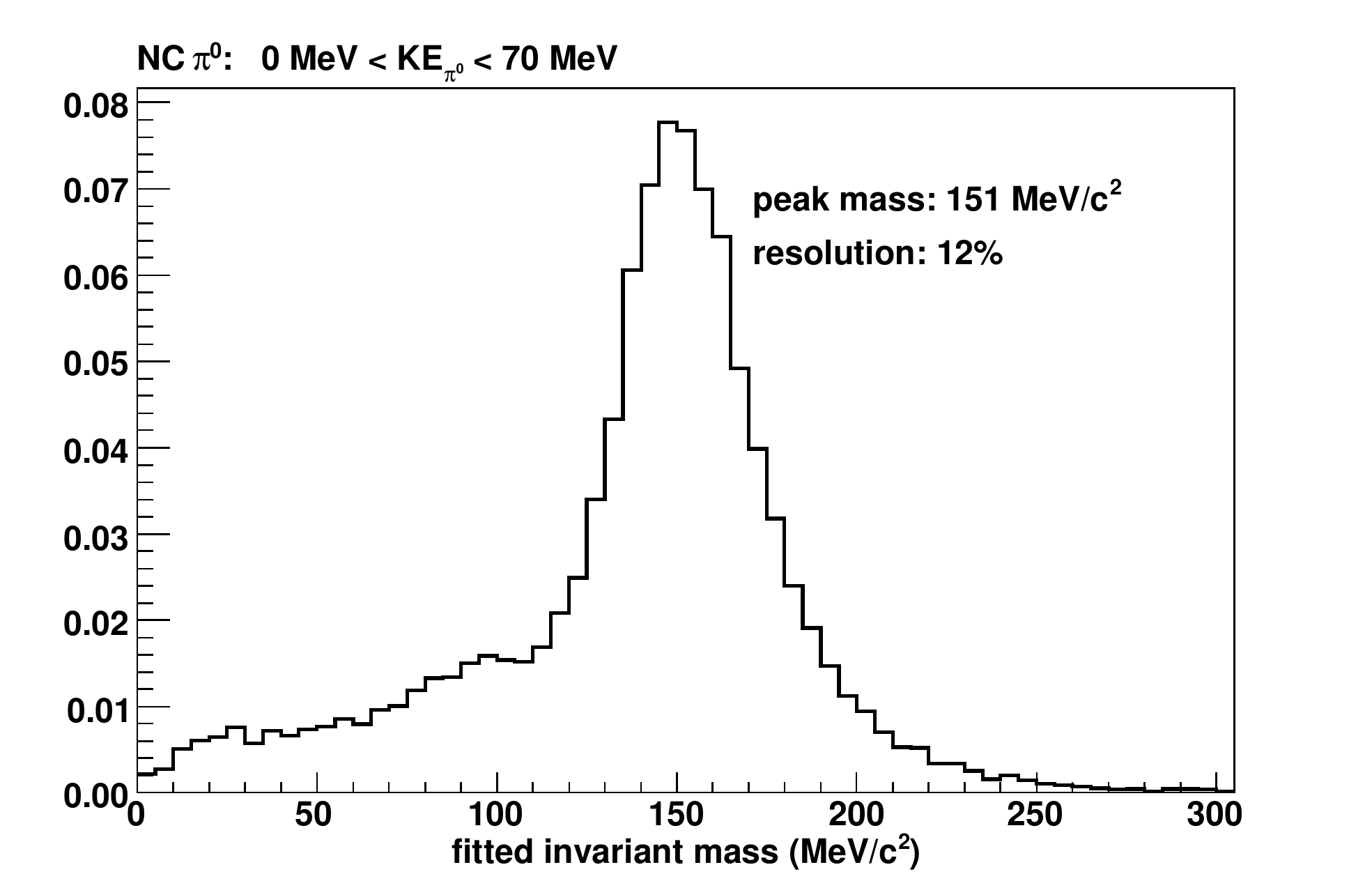} \hspace{0.05\textwidth}
  \includegraphics[width=0.45\textwidth]{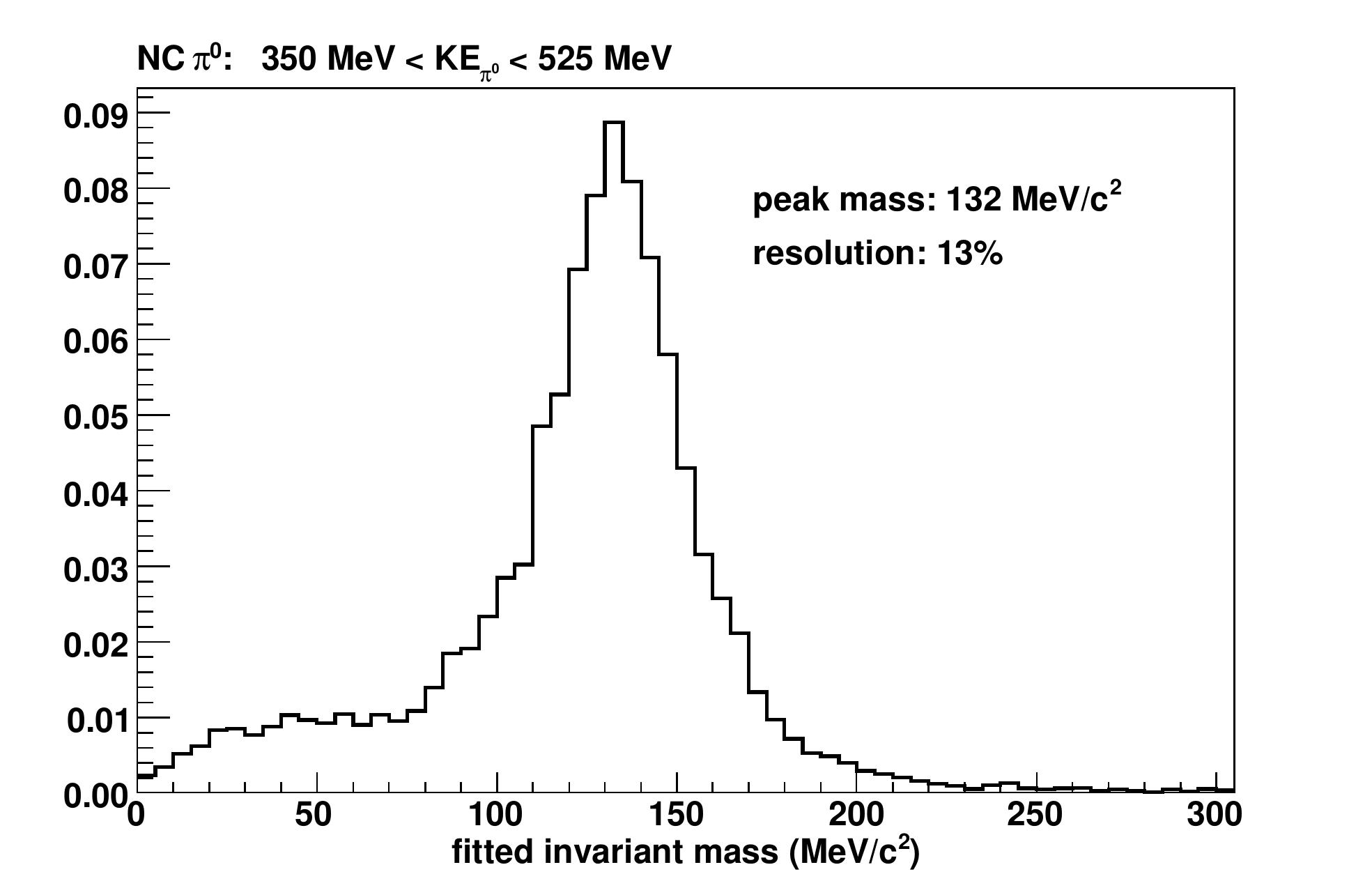} \\

  \includegraphics[width=0.45\textwidth]{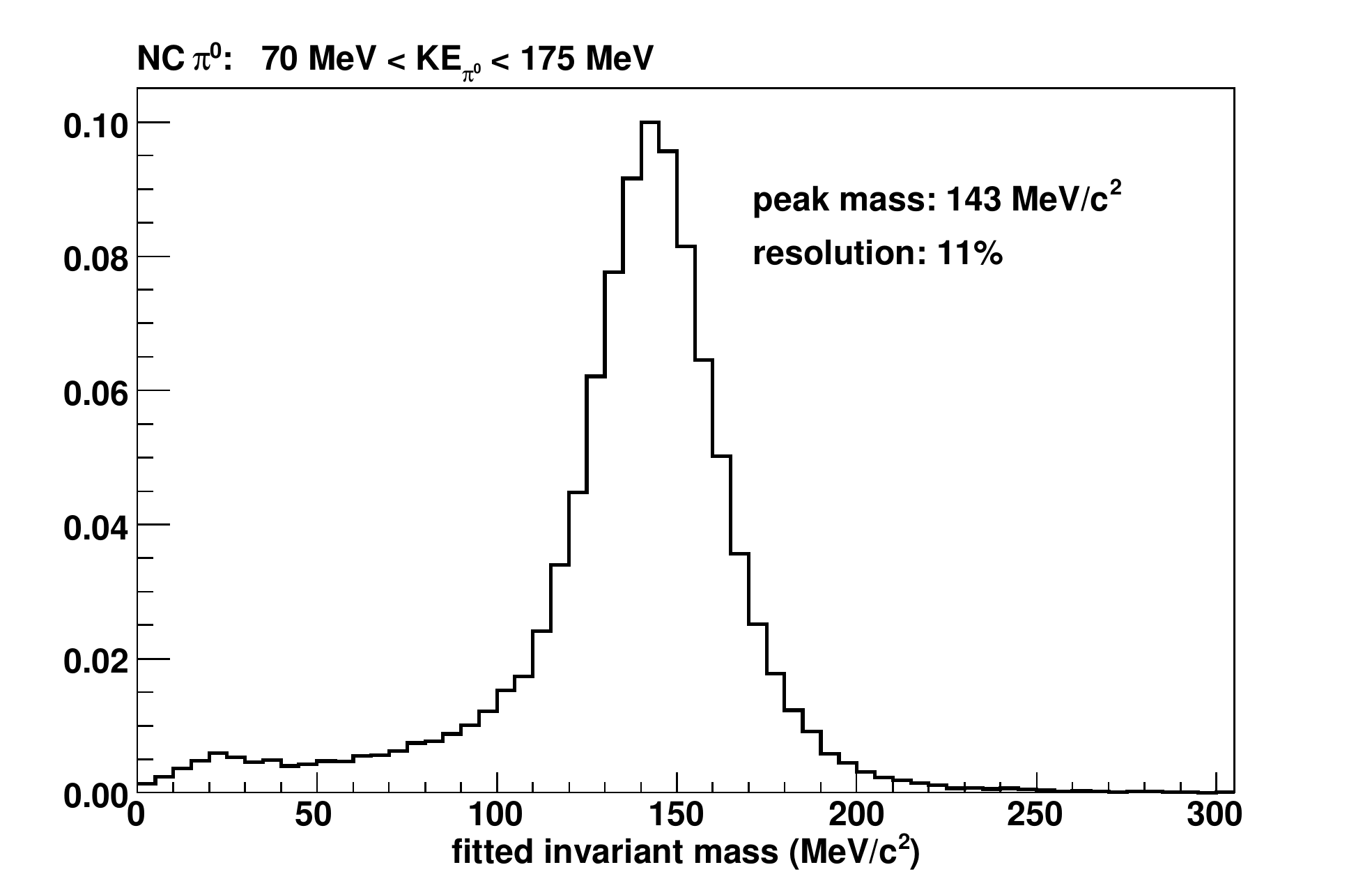}\hspace{0.05\textwidth}
  \includegraphics[width=0.45\textwidth]{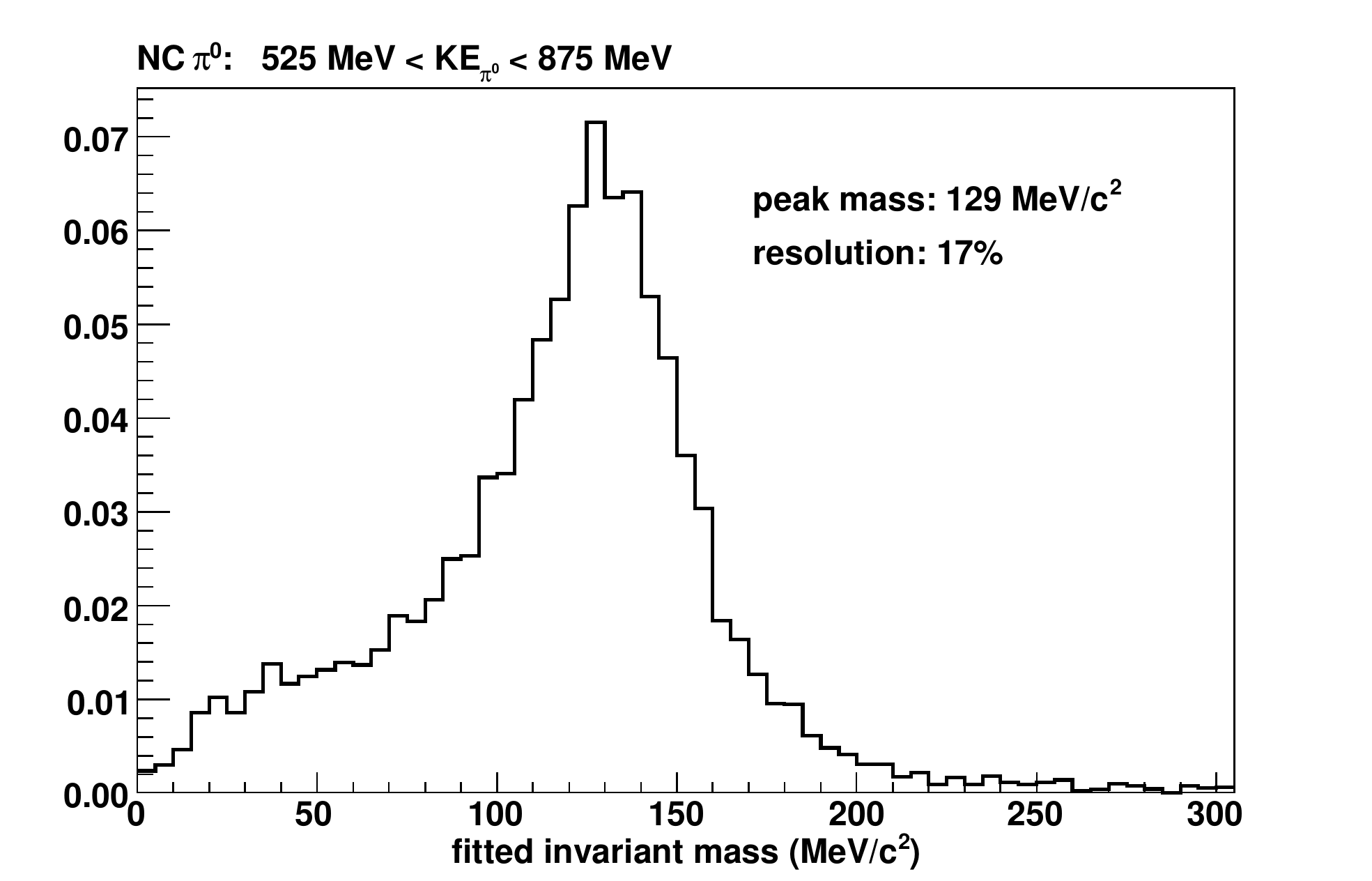} \\

  \includegraphics[width=0.45\textwidth]{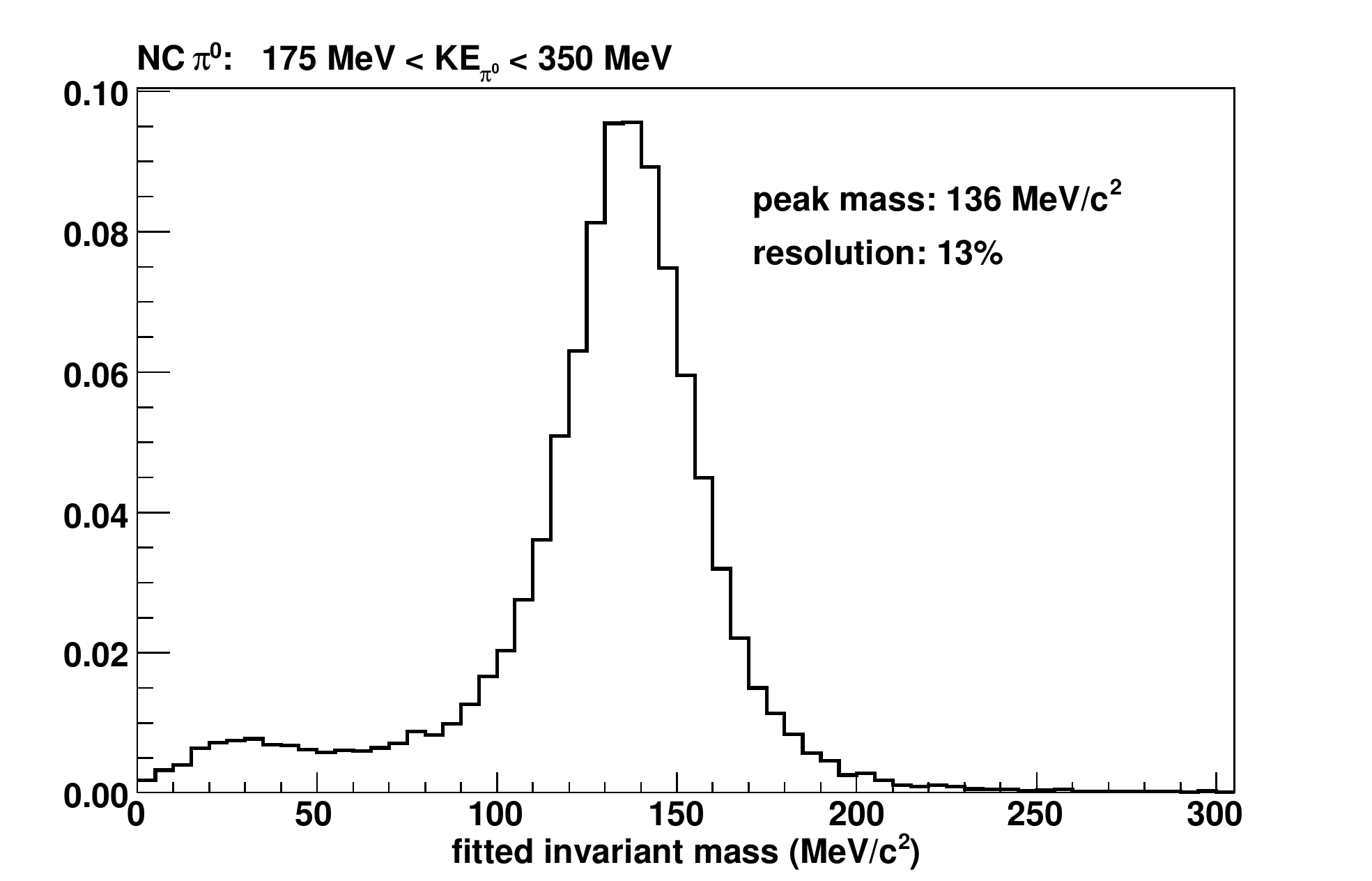}\hspace{0.05\textwidth}
  \includegraphics[width=0.45\textwidth]{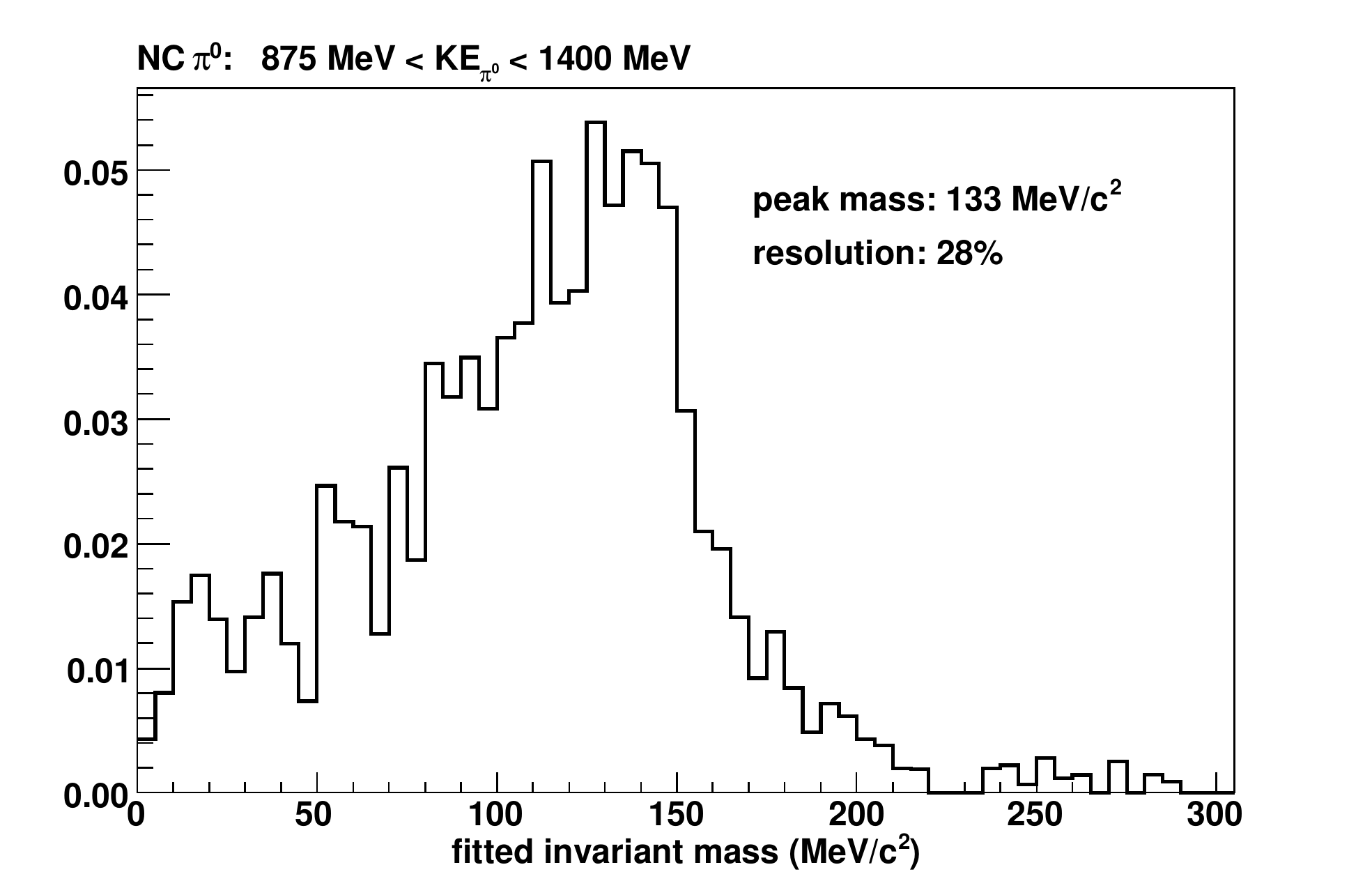} \\

  \caption{\label{fig:mcpires}Fitted invariant mass $M_{\gamma\gamma}$ for Monte Carlo-simulated NC $\pi^0$ events in three low energy regions (left) and three high energy regions (right).}
  \end{center}
\end{figure}

\begin{figure}[hbtp]
  \begin{center}
  \includegraphics[width=0.49\textwidth]{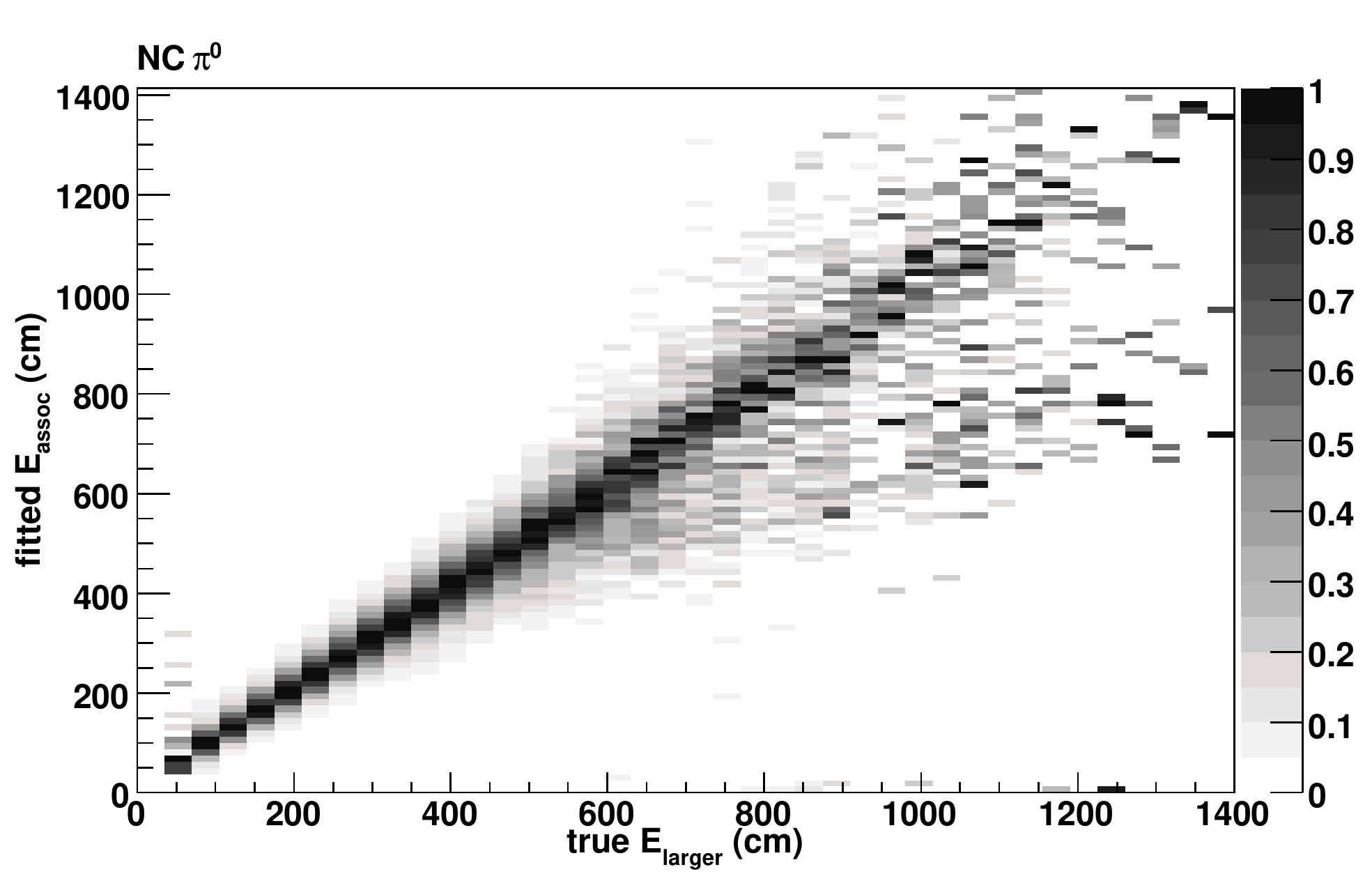}
  \includegraphics[width=0.49\textwidth]{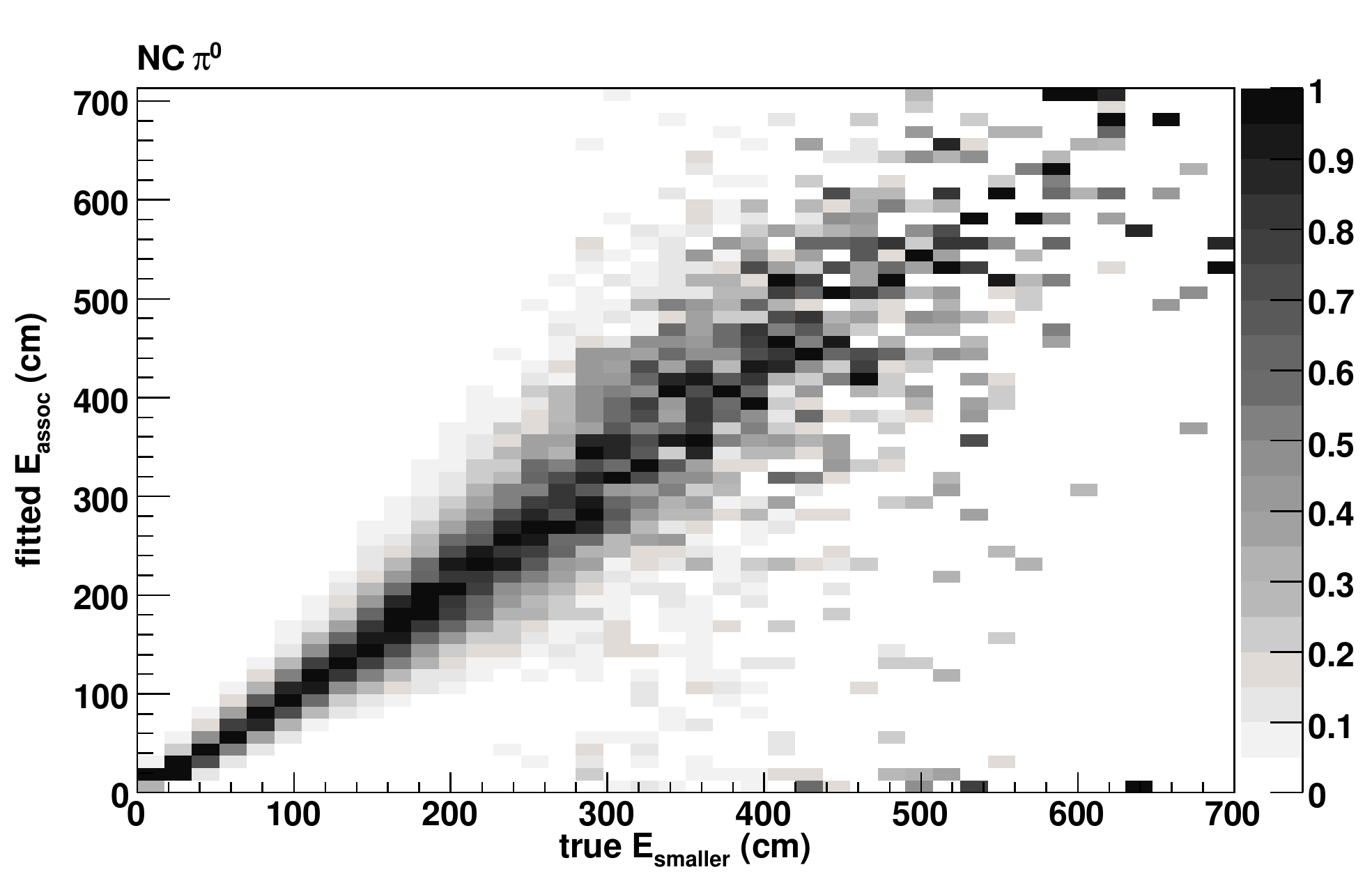}
  \caption{\label{fig:geVe} Reconstructed $\gamma$ energies in Monte Carlo-simulated NC $\pi^0$ events plotted against true
 values.  The left (right) panel shows the higher (lower) energy $\gamma$ from each event.  The association 
of each fitted track to the underlying true $\gamma$'s is, in general, ambiguous.  For these plots, we choose the assignment 
that gives the smaller combined energy and direction discrepancy.}
  \end{center}
\end{figure}

\begin{figure}[hbtp]
  \begin{center}
  \includegraphics[width=0.49\textwidth]{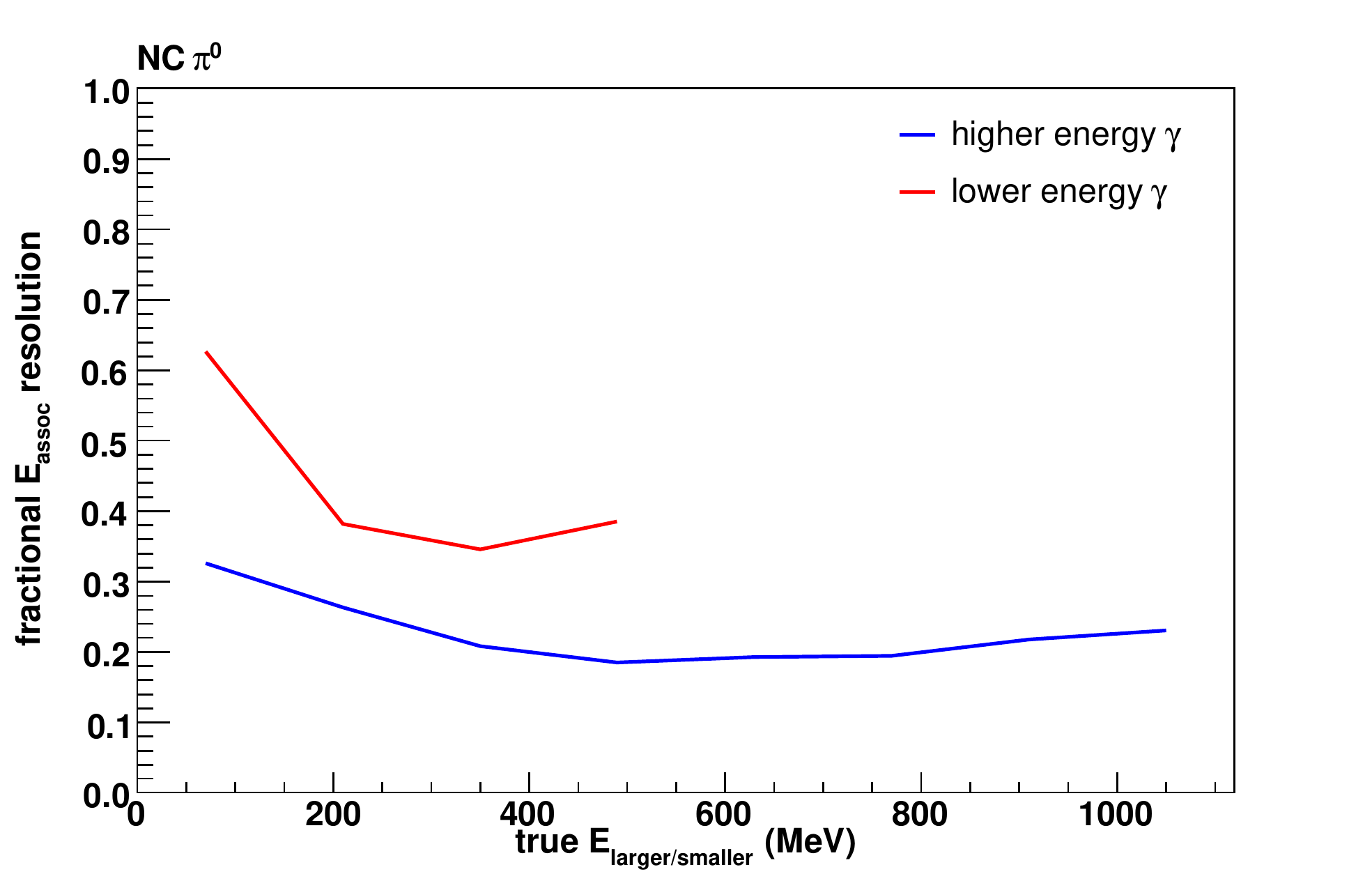}
  \includegraphics[width=0.49\textwidth]{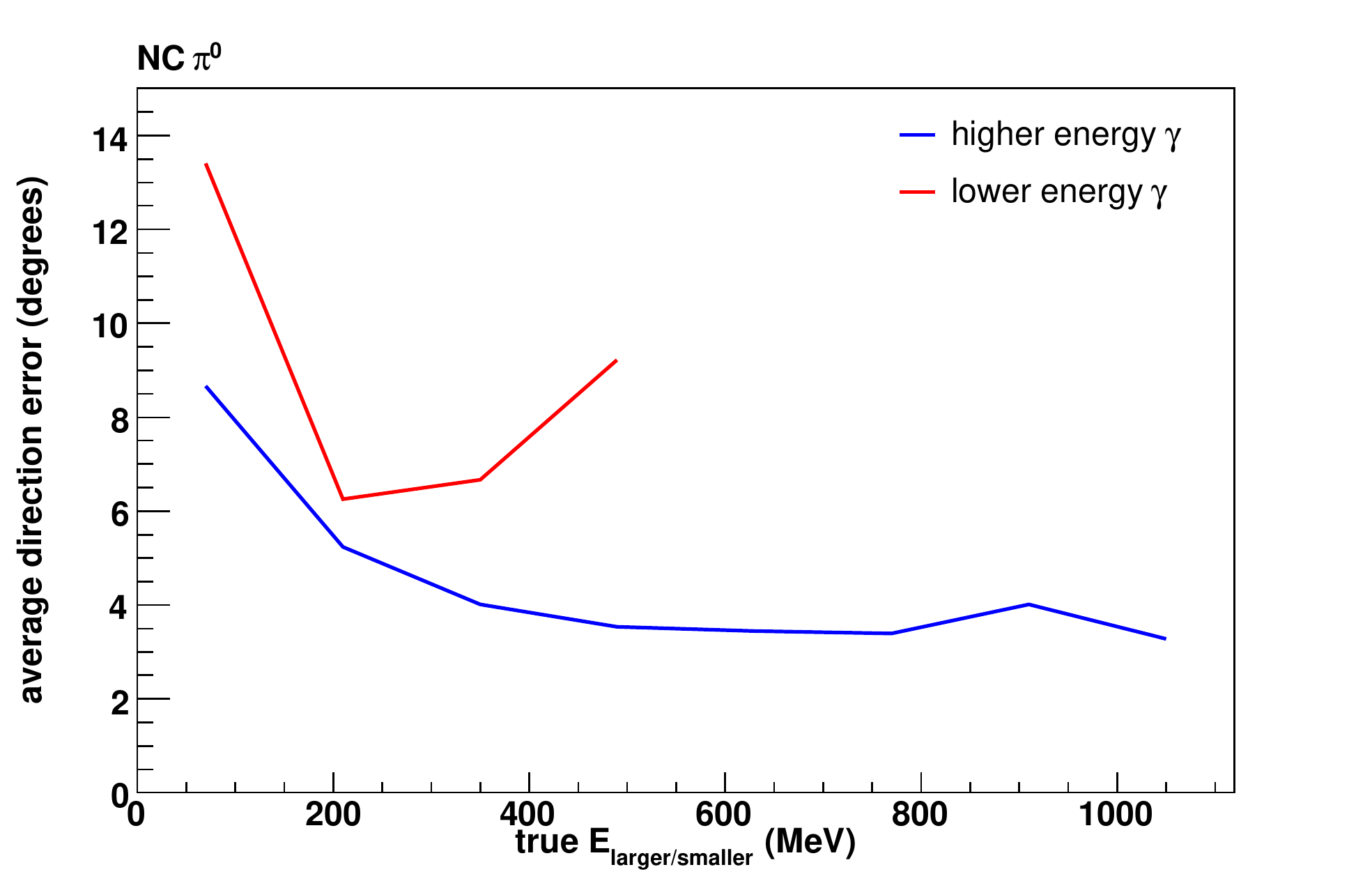}
  \caption{\label{fig:gresolutions} Energy (left) and direction (right) resolutions for $\gamma$'s in simulated NC $\pi^0$ events.  For a given true $\gamma$ energy, a dominant $\gamma$ has expectedly better resolution than does a subordinate $\gamma$.}
  \end{center}
\end{figure}

The performance of the reconstruction algorithm on simulated neutrino interactions
($\num$ CCQE, $\nue$ CCQE and NC $\piz$ events) generated with NUANCE~\cite{nuance} and
propagated through the detector Monte Carlo simulation is shown in Figures \ref{fig:rVr}$-$\ref{fig:gresolutions}.\footnote{For comparisons of Monte Carlo to data and for discussion of the development and validation of the Monte Carlo, see Refs.~\cite{pattersonthesis} and \cite{coherentpi0}.}  The electron and muon one-track fits are applied to the $\num$ and $\nue$ CCQE events, respectively,
while the two-track fits, with and without the mass constraint, are applied to the NC $\piz$ events.
A minimal selection is applied to ensure that the events are contained within the main detector region and, for plots not directly related to vertex reconstruction, within the fiducial volume
of the analysis. 

A few observations about the figures, particularly Figure~\ref{fig:resolutions}: (1) The resolutions for $\num$ CCQE
events degrade as the energy approaches the Cherenkov threshold from above, as most of the information in the event comes from the (rapidly decreasing) prompt Cherenkov light. (2) A low-energy $\piz$ offers little information on its direction, as all that is observed is the isotropically produced $\gamma\gamma$ final state.  Thus, the angular
resolution (but not the vertex resolution) becomes poor.  (3) The vertex resolution for $\piz$
events is worse than $\nue$ CCQE events since the former has spatially
displaced light production thanks to the $\nearsim 70\cm$ average photon conversion length.
(4) Typical performance
numbers for $\nue$ CCQE events: 20 cm vertex resolution, $12\%$ kinetic energy resolution, and
$4^{\circ}$ angular resolution.  (5) Typical performance numbers for $\num$ CCQE events: 10 cm vertex resolution,
$8\%$ kinetic energy resolution and $2^\circ$ angular resolution.  (6) Figure~\ref{fig:mcpires} shows a $\piz$ mass resolution of $\nearsim 13\%$.

%% file: pid.tex
\section{Particle Identification}
\label{sec:likelihood}
The maximum likelihoods returned from each fit can be used for hypothesis testing.  The two quantities
\begin{equation}
\likeemu \equiv \log\frac{\like_e}{\like_\mu} = \log\like_e - \log\like_\mu
\label{eq:logemu}
\end{equation}
and
\begin{equation}
\likeepi \equiv \log\frac{\like_e}{\like_\pi} = \log\like_e - \log\like_\pi
\label{eq:logepi}
\end{equation}
are used to determine for a given event whether the electron hypothesis is preferred over the muon and $\piz$ hypotheses. In these expressions, $\like_e$, $\like_\mu$ and $\like_\pi$ are the maximized likelihoods returned by the electron, muon, and (fixed-mass) two-track fits, respectively.

\begin{figure}[hbtp]
  \begin{center}
  \includegraphics[width=0.9\textwidth]{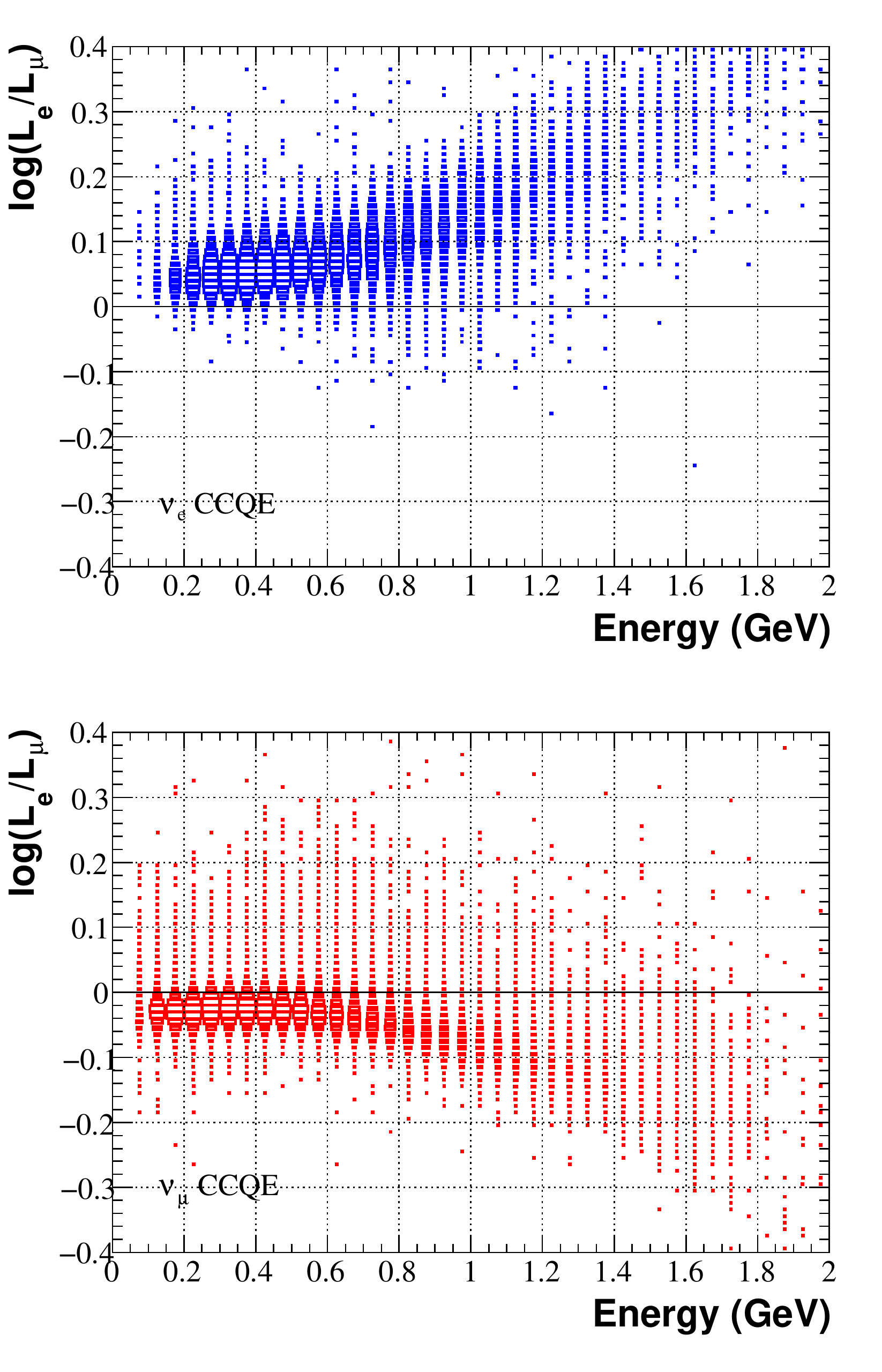}
  \caption{\label{fig:logemu} Distribution of $\likeemu\mathord{=}\log(\like_e/\like_\mu$) for Monte Carlo simulated $\nue$ CCQE
  events (top) and $\num$ CCQE events (bottom) as a function of reconstructed
  energy (from the electron hypothesis fit).}
    \end{center}
\end{figure}

\begin{figure}[hbtp]
  \begin{center}
  \includegraphics[width=0.45\textwidth]{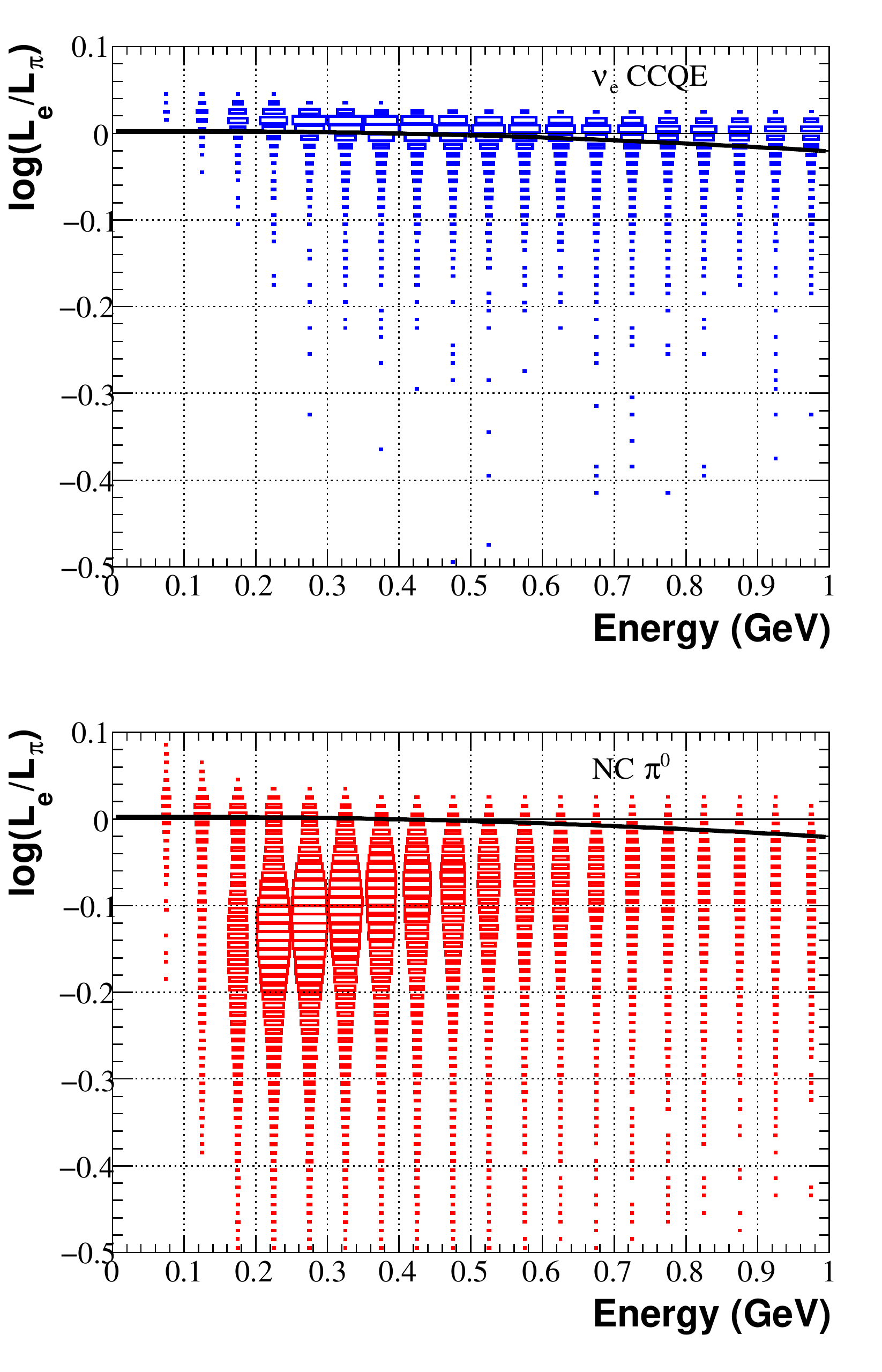}\hspace{0.05\textwidth}
  \includegraphics[width=0.45\textwidth]{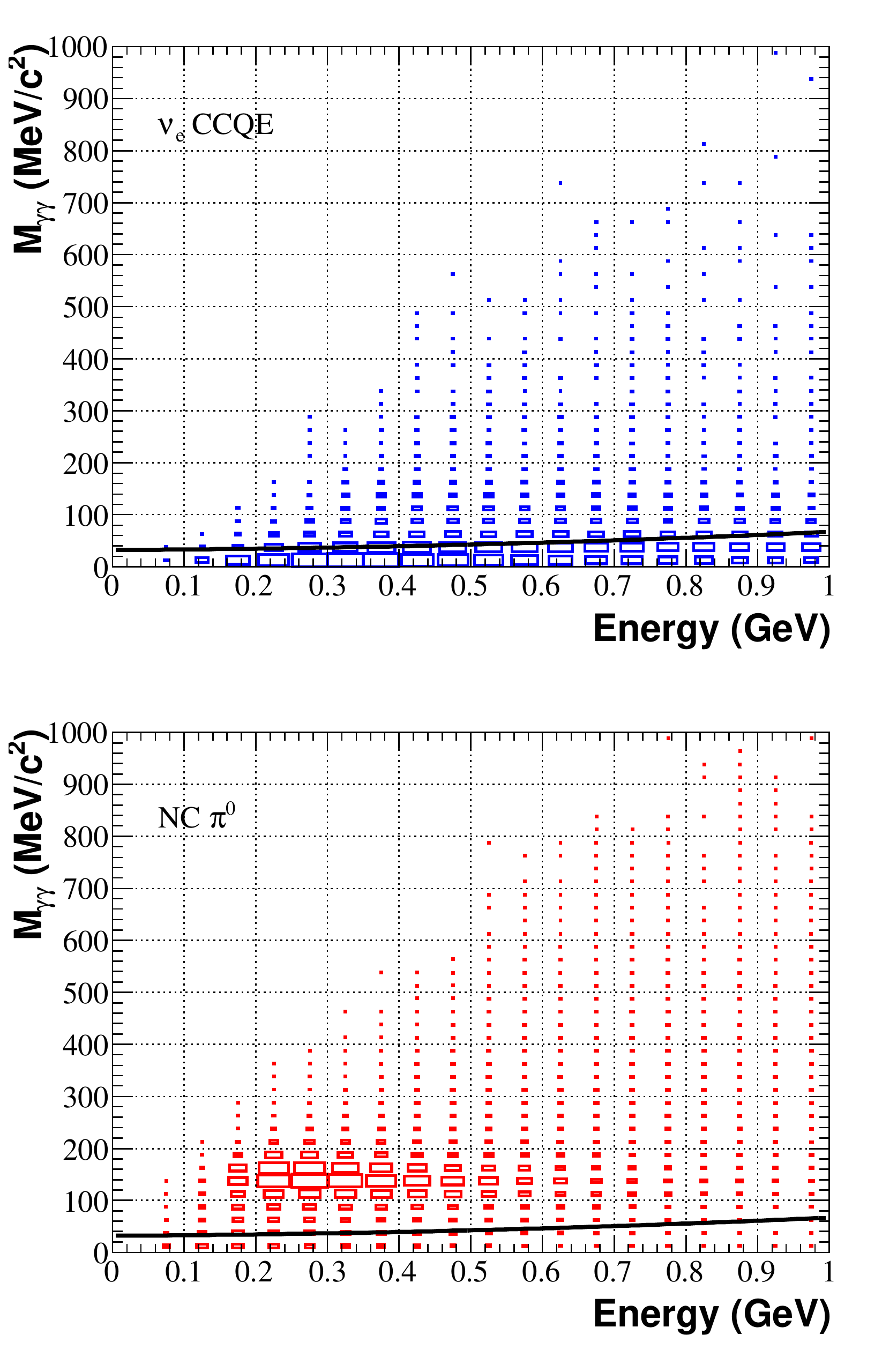}
  \caption{\label{fig:pipid} Left: Distribution of $\likeepi\mathord{=}\log(\like_e/\like_\pi)$ for Monte Carlo simulated $\nue$ CCQE
  events (top) and NC $\piz$ events (bottom). Right: same for $M_{\gamma\gamma}$. The black lines on each distribution
show the selection used in the MiniBooNE $\num\to\nue$ oscillation analysis \cite{prl}.}
    \end{center}
\end{figure}

\subsection{Electron/Muon separation}

Figure \ref{fig:logemu} shows for simulated $\nue$ CCQE events and
$\num$ CCQE events the distribution of $\likeemu$ as a function
of the energy reconstructed by the electron fit. For each sample, the
events have been subject to a set of preselection criteria used in the
$\nue$ appearance analysis. These require that there
is only one time-cluster of PMT hits in the event, eliminating obvious $\num$ CC
events that produce decay electrons.
Cosmic backgrounds are eliminated by requiring the minimum 
number of PMT hits in the main region  to be $\mathord{>}200$  and
the number of veto hits to be $\mathord{<}6$. Furthermore,
the average time of hits is required to lie within the expected beam
delivery window from the Booster. 
 
 For the $\nue$ CCQE events, $\likeemu$ tends to take positive values, indicating that the
fit to the event with the electron hypothesis is favored over the muon hypothesis.
Likewise, $\likeemu$ tends to be negative for the $\num$ CCQE events, indicating that the
muon hypothesis fits these events better than the electron hypothesis.
At high energies ($\mathord{>}1\gev$), the electron/muon separation is aided by the fact that the muon
pathlength grows approximately linearly with energy, while the electron shower grows
more slowly.

\subsection{Electron/$\piz$ separation}

The left plots in Figure \ref{fig:pipid} show the distributions of $\likeepi$
as a function of energy reconstructed by the single-track electron reconstruction for
simulated $\nue$ CCQE events and NC $\piz$ events. The quantity $\likeepi$ tests
whether a given event fits better as a single electron track or as a $\piz$.  Overall,
electron/$\piz$ separation becomes more difficult at high energies as the energies of the
decay photons in $\piz$ events become more asymmetric and the shower fluctuations
in a single electron event get large enough to mimic the presence of a second photon.  Also, $\piz$ events in which one of the two photons leaves the detector unconverted present an essentially irreducible background at low energy.

Another quantity that can be used for electron/$\piz$ separation is $\mgg$, the invariant
mass of the two photons returned by the free-mass two-track fit. The NC $\piz$ events peak at
the $\piz$ mass, whereas the $\nue$ CCQE
events peak toward zero. This is seen in the right plots in Figure \ref{fig:pipid},
where the $\mgg$ distribution is shown as a function of reconstructed energy. The use of the true $\piz$ mass in the fit's seeding procedure (Section~\ref{sec:twotrackmin}) induces no appreciable high-mass peak in the $\nue$ CCQE events while helping considerably in the correct identification of NC $\piz$ events.

The $\nue$ appearance analysis at MiniBooNE utilizes both $\mgg$ and $\likeepi$ to separate $\nue$ CCQE events from NC $\piz$.



\subsection{Particle Identification Performance}
\begin{figure}[hbtp]
  \begin{center}
  \includegraphics[width=0.95\textwidth]{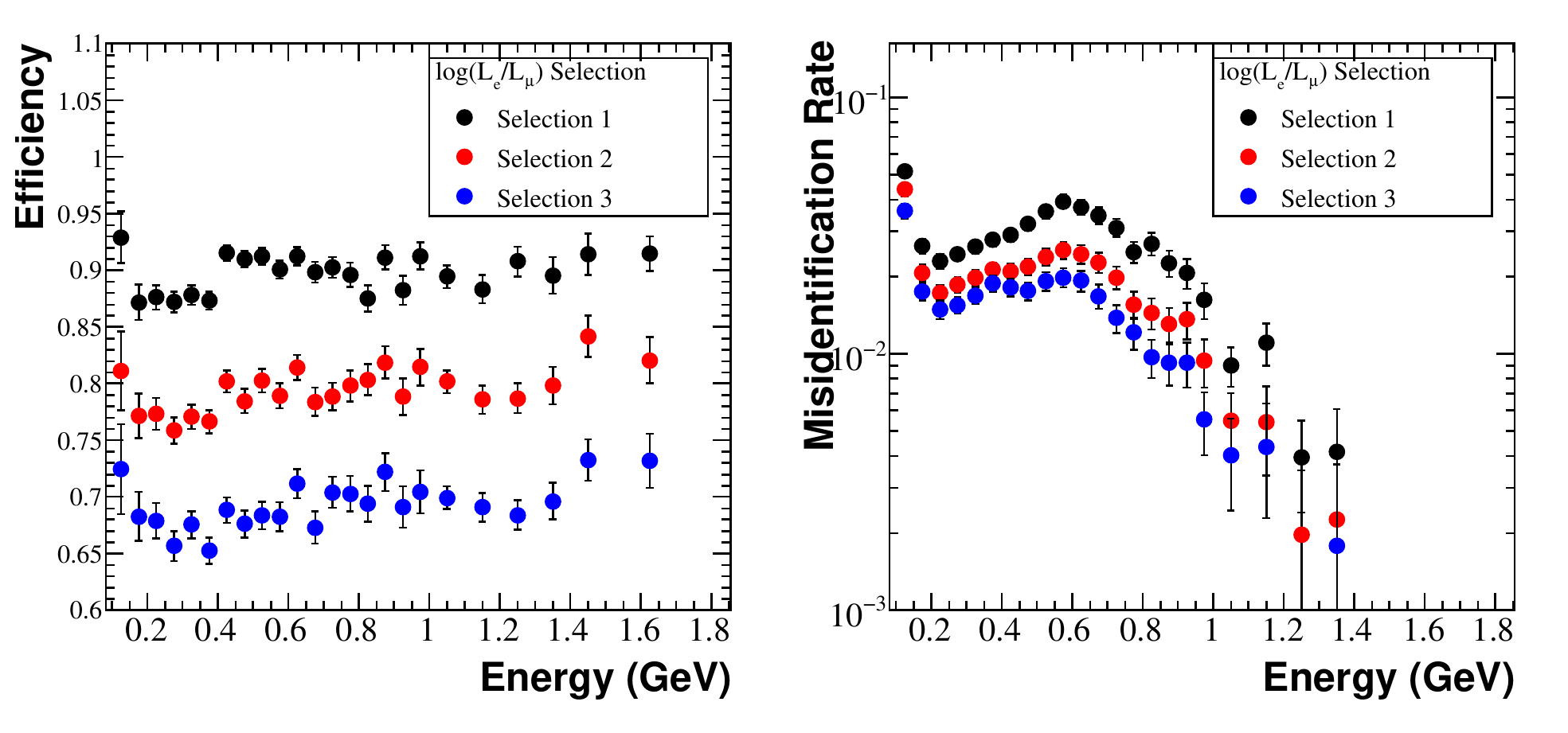}
  \caption{\label{fig:logemu_perf} Performance of three prototype $\likeemu\mathord{=}\log(\like_e/\like_\mu)$ cuts for the $\nue$ selection. Left: Signal ($\nue$ CCQE) efficiency versus the electron reconstruction's energy.  Right: Misidentification rate of $\num$ CCQE events versus the electron reconstruction's energy.}
  \end{center}
\end{figure}

\begin{figure}
  \begin{center}
  \includegraphics[width=0.95\textwidth]{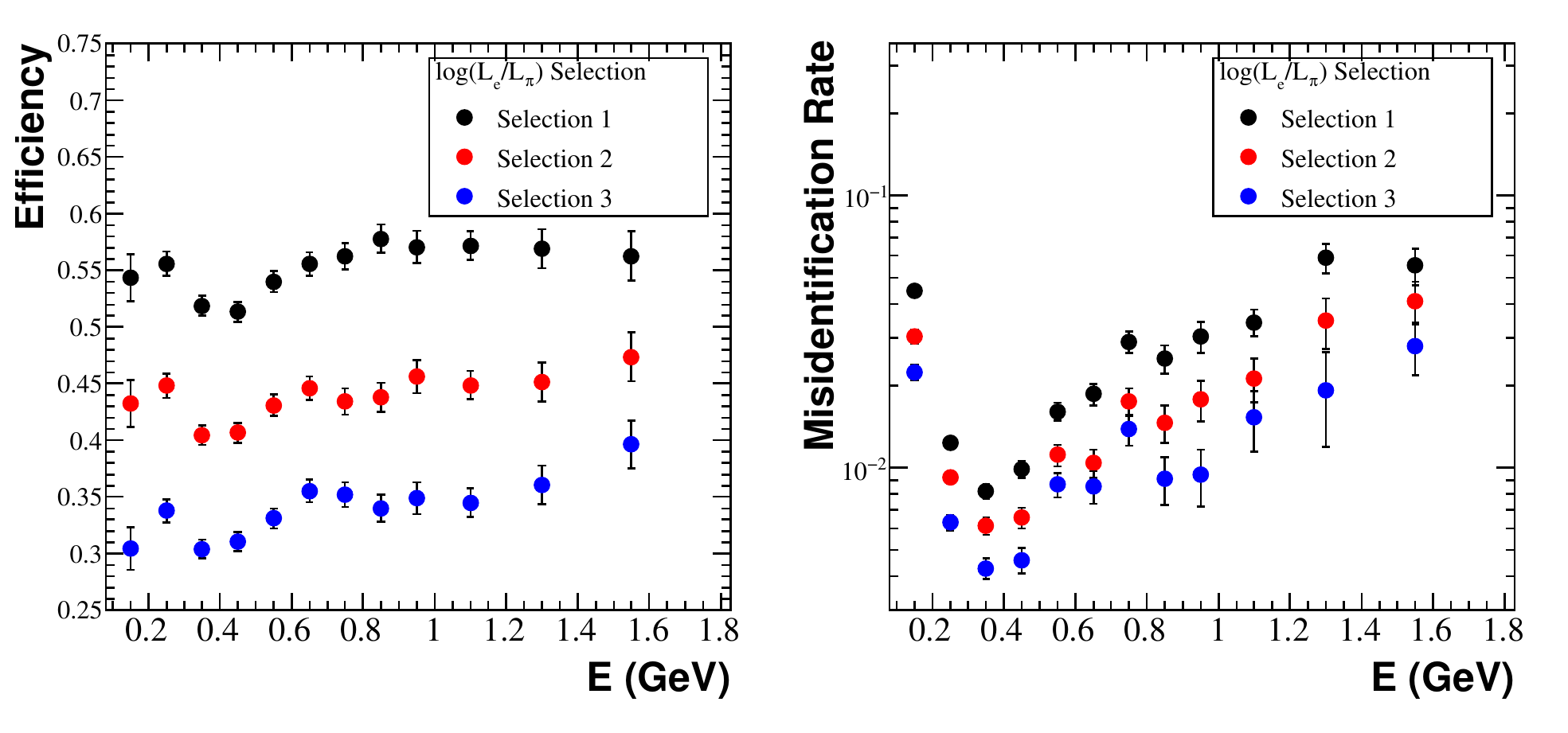}\\
  \includegraphics[width=0.95\textwidth]{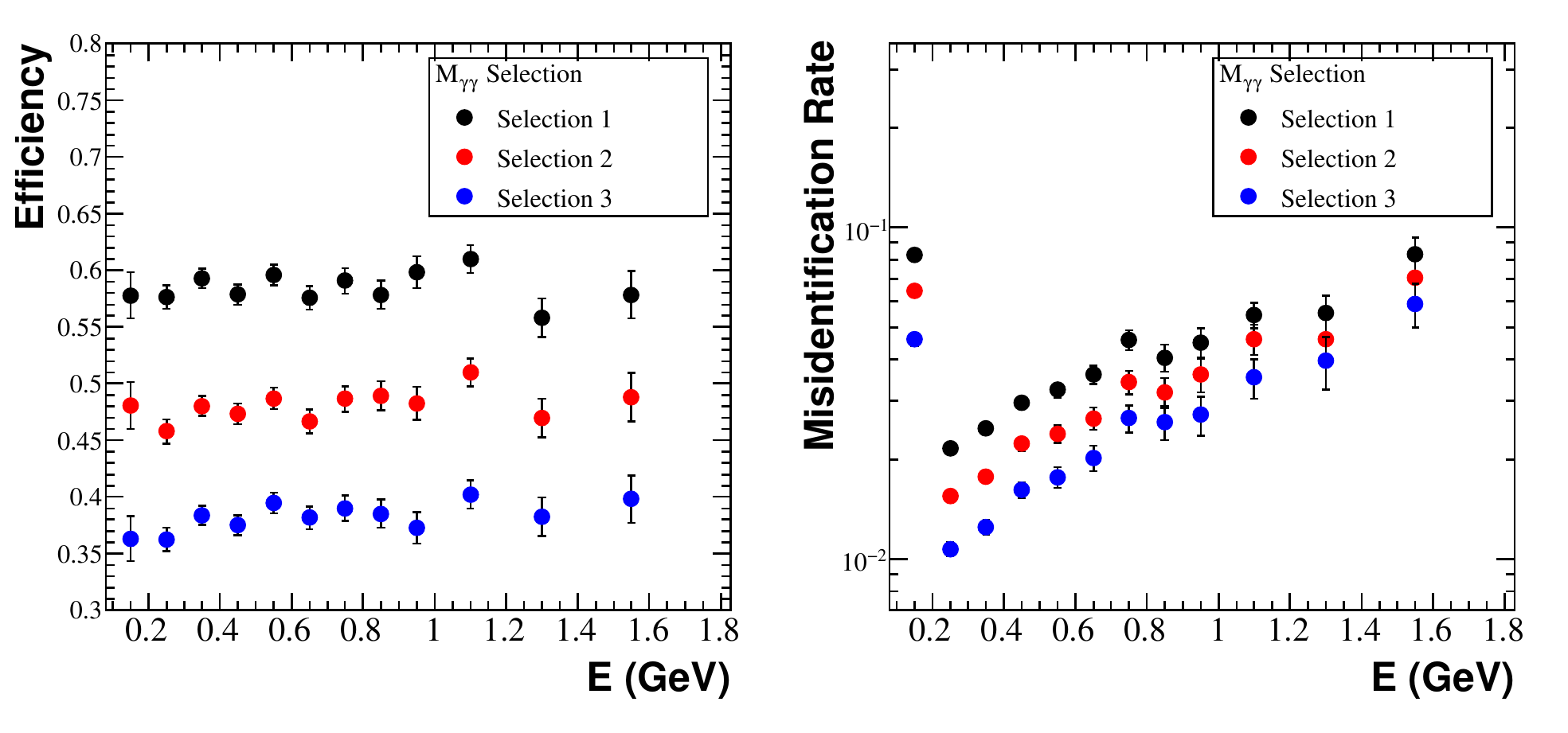}
  \caption{\label{fig:pi_perf}  Performance of three prototype $\likeepi\mathord{=}\log(\like_e/\like_\pi)$ (top) and $\mgg$ (bottom) cuts for the $\nue$ selection. Left: Signal ($\nue$ CCQE) efficiency versus the electron reconstruction's energy.  Right: Misidentification rate of NC $\piz$ events versus the electron reconstruction's energy.}
  \end{center}
\end{figure}

The background rejection levels reachable with these fitter-derived quantities are shown for Monte Carlo events in Figures~\ref{fig:logemu_perf} and~\ref{fig:pi_perf}.   Figure~\ref{fig:logemu_perf} shows the efficiency, after fiducial volume and cosmic rejection cuts, for selecting signal $\nue$ CCQE events alongside the efficiency (or misidentification rate) for $\num$ CCQE events.  Three prototype selections are used, each chosen to give an approximately fixed signal efficiency at all energies.  Note that the misidentification rate is shown with a logarithmic vertical scale. Figure~\ref{fig:pi_perf} shows similar plots for $\piz$ rejection, one for the $\likeepi$ cut and one for the $\mgg$ cut.

Misidentification rates of a few percent for $\num$ CCQE events are seen for signal efficiencies from $70-90\%$.  (Note that $\num$ CCQE events are further reduced by about an order of magnitude by recognizing the delayed muon decay.)  The two $\piz$ variables indicate that the misidentification rate increases with energy; this is expected, since the two decay photons become highly asymmetric energetically and since the Cherenkov rings can achieve a greater degree of overlap.  The rise in the $\piz$ misidentification rate at low energies is due to events with an escaping photon (corresponding to the cluster of $\piz$
events with positive $\likeepi$ at 50$-$200~$\mev$ in Figure~\ref{fig:pipid}).  Between 200~$\mev$ and 600~$\mev$,
where most of the NC $\piz$ events lie, a misidentification rate of less than 1 percent
can be achieved with $40\%$ efficiency using only the $\likeepi$ cut, while rates of a few percent can be achieved with only the $\mgg$ cuts.  In the MiniBooNE $\num\to\nue$ oscillation analysis, 50\% signal efficiency and 1\% $\pi^0$ misidentification rate are obtained by requiring both $\mgg<\mgg^{\mathrm{max}}(E)$ and $\likeepi>\likeepi^{\mathrm{max}}(E)$, where the ``max'' functions are quadratic in the energy $E$ obtained from the single-track electron fit.  These quadratic selection boundaries are indicated by the solid black lines in Figure~\ref{fig:pipid}.  These cuts, along with an analogous $\likeemu$ cut, were tuned to optimize sensitivity to LSND-like oscillations.


\section{Summary}
A maximum likelihood reconstruction algorithm is used at MiniBooNE to
determine track parameters under electron, muon, and $\piz$ hypotheses.  Underlying the likelihood is a track and detector model 
that calculates for a given parameter set the charge and time PDFs expected
for each PMT, accounting for the spatially extended production of
Cherenkov and scintillation light as well as the effects of indirect light from subsequent
optical processes.

The maximized likelihoods for a given
event obtained under different event hypotheses are used for event
selection. In particular, the ratio of the likelihoods under the electron and 
muon models is used to suppress $\num$ CC events in the MiniBooNE
$\num\to\nue$ oscillation search; the ratio of the likelihoods under the electron
and two-track $\piz$ fit (with fixed invariant mass) and the
invariant mass obtained from the free-mass two-track fit are used to
suppress NC $\piz$ events. While the reconstruction has been developed
within the context of the MiniBooNE $\num\to\nue$ oscillation search,
other experiments employing similar Cherenkov detection techniques should
be able to use the techniques discussed here.